\documentclass[iop]{emulateapj}
\usepackage{natbib}
\usepackage{mathrsfs}

\begin{document}
\title{Molecular Clouds in the North American and Pelican Nebulae: Structures}
\author{Shaobo Zhang\altaffilmark{1,2}, Ye Xu\altaffilmark{1}, Ji Yang\altaffilmark{1}}
\altaffiltext{1}{Purple Mountain Observatory, \& Key Laboratory for Radio Astronomy, Chinese Academy of Sciences, Nanjing 210008, China; shbzhang@pmo.ac.cn}
\altaffiltext{2}{Graduate University of the Chinese Academy of Sciences, 19A Yuquan Road, Shijingshan District, Beijing 100049, China}

\begin{abstract}

We present observations of 4.25 square degree area toward the North American and Pelican Nebulae in the $J = 1-0$ transitions of $^{12}$CO, $^{13}$CO, and C$^{18}$O. Three molecules show different emission area with their own distinct structures. These different density tracers reveal several dense clouds with surface density over 500~$M_\sun$~pc$^{-2}$ and a mean H$_2$ column density of 5.8, 3.4, and 11.9$\times10^{21}$~cm$^{-2}$ for $^{12}$CO, $^{13}$CO, and C$^{18}$O, respectively. We obtain a total mass of $5.4\times10^4~M_\odot$ ($^{12}$CO), $2.0\times10^4~M_\odot$ ($^{13}$CO), and $6.1\times10^3~M_\odot$ (C$^{18}$O) in the complex. The distribution of excitation temperature shows two phase of gas: cold gas ($\sim$10~K) spreads across the whole cloud; warm gas ($>$20~K) outlines the edge of cloud heated by the W80 H~II region. The kinetic structure of the cloud indicates an expanding shell surrounding the ionized gas produced by the H~II region. There are six discernible regions in the cloud including the Gulf of Mexico, Caribbean Islands and Sea, Pelican's Beak, Hat, and Neck. The areas of $^{13}$CO emission range within 2-10~pc$^2$ with mass of (1-5)$\times10^3~M_\odot$ and line width of a few km~s$^{-1}$. The different line properties and signs of star forming activity indicate they are in different evolutionary stages. Four filamentary structures with complicated velocity features are detected along the dark lane in LDN~935. Furthermore, a total of 611 molecular clumps within the $^{13}$CO tracing cloud are identified using the ClumpFind algorithm. The properties of the clumps suggest most of the clumps are gravitationally bound and at an early stage of evolution with cold and dense molecular gas.

\end{abstract}


\section{Introduction}

The study of massive star formation is limited. The molecular clouds within a few hundred parsecs of the sun provide an ideal environment for improving our knowledge of star forming process. Among these clouds, low-mass star-forming regions constitute the majority of the population, while regions with massive clumps and dense clusters like the Orion nebula are infrequent. The North American (NGC~7000) and Pelican (IC~5070) Nebulae (referred to as the ``NAN complex" hereafter) are together one of the nearby ($\sim$600~pc, \citealt{lau02}) star forming regions with large numbers of massive stars. This is the next closest region showing signs of massive star formation after Orion, but has been rarely studied to-date.

The studies of molecules \citep{bal80,dob94} and near-infrared extinction \citep{cam02} all confirm substantial quantities of molecular gas along the Lynds Dark Nebula (LDN) 935 \citep{lyn62} which lies between the North American and Pelican nebulae. All three objects (NGC~7000, IC~5070, and LDN~935) are thought to be a part of W80, a large H~II region mainly in the background. \citet{com05} identified an O5V star, 2MASS J205551.25+435224.6, hidden behind the LDN~935 cloud to be the ionizing star of the H~II region. Mid-infrared observations as Mid-course Space Experiment (MSX, \citealt{ega98}) have found several Infrared dark clouds (IRDCs) in LDN~935 which indicates the existence of a cold, dense environment in the molecular cloud. Other signposts of on-going star formation, such as HH objects, and H$\alpha$ emission-line stars (e.g., \citealt{bal03, com05, arm11}, etc.), are also found in the NAN complex. However, studies of molecules in the NAN complex, which can reveal both the spatial and velocity structures, have only been conducted in a few small regions or are limited by resolution.

In this work, we use molecular data tracing different environments to study the properties of the individual regions, filamentary structures, and clumps in the NAN complex. There is a divergence in the distance estimation of the complex as discussed by \citet{wen83, str93, cer07}, etc and reviewed by \citet{rei08}. In our calculation, we adopt a commonly used distance of 600~pc based on multi-color photometric results for hundreds stars \citep{lau02, lau07}.

\section{Observations and Data Reduction}\label{sec:obs}

We observed the NAN complex in $^{12}$CO~(1$-$0), $^{13}$CO~(1$-$0), and C$^{18}$O~(1$-$0) with the Purple Mountain Observatory Delingha (PMODLH) 13.7~m telescope as one of the scientific demonstration regions for Milky Way Imaging Scroll Painting (MWISP) project\footnote{\url{http://www.radioast.nsdc.cn/yhhjindex.php}} from May 27 to June 3, 2011. The three CO lines were observed simultaneously with the 9-beam superconducting array receiver (SSAR) working in sideband separation mode and with the fast Fourier transform spectrometer (FFTS) employed \citep{sha12}. The typical receiver noise temperature ($T_{\rm rx}$) is about 30~K as given by status report\footnote{\url{http://www.radioast.nsdc.cn/zhuangtaibaogao.php}} of PMODLH.

\begin{figure}
\centering
\includegraphics[width=0.4\textwidth]{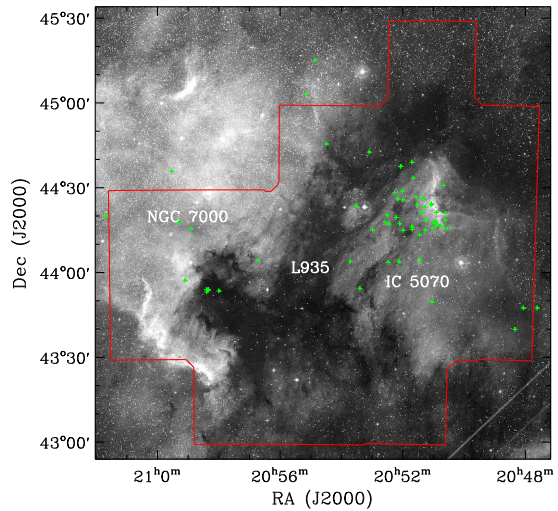}
\caption{The location of observation coverage, superimposed on the second Palomar Observatory Sky Survey (POSS II) red image. Green crosses mark the T-Tauri type stars identified by \citet{her58}}
\label{fig:guide}
\end{figure}

Our observations were made in 17 cells of dimension 30\arcmin$\times$30\arcmin\ and covered an area of total 4.25~deg$^2$ (466~pc$^2$ at the distance of 600~pc) as shown in Figure~\ref{fig:guide}. The cells were mapped using the on-the-fly (OTF) observation mode, with the standard chopper wheel method for calibration \citep{pen73}. In this mode, the telescope beam is scanned along lines in RA and Dec directions on the sky at a constant rate of 50\arcsec/sec, and receiver records spectra every 0.3~sec. Each cell was scanned in both RA and Dec direction to reduce the fluctuation of noise perpendicular to the scanning direction. Further observations were made toward the regions with C$^{18}$O detection to improve their signal to noise ratios. The typical system temperature during observations was $\sim$280~K for $^{12}$CO and $\sim$185~K for $^{13}$CO and C$^{18}$O.

After removing the bad channels in the spectra, we calibrated the antenna temperature ($T_a^*$) to the main beam temperature ($T_{\rm mb}$) with a main beam efficiency of 44\% for $^{12}$CO and 48\% for  $^{13}$CO and  C$^{18}$O. The calibrated OTF data were then re-gridded to 30\arcsec pixels and mosaicked to a FITS cube using the GILDAS software package \citep{gui00}. A first order baseline was applied for the spectra. The resulting rms noise is 0.46~K for $^{12}$CO at the resolution of 0.16~km~s$^{-1}$, 0.31~K for $^{13}$CO and 0.22~K for C$^{18}$O at 0.17~km~s$^{-1}$. Such noise level corresponds to a typical integration time of $\sim$30~sec in each resolution element. A summary of the observation parameters is provided in Table~\ref{tbl:observation}

\begin{deluxetable*}{ccccccc}
\centering
\tabletypesize{\scriptsize}
\tablecolumns{7}
\tablewidth{0pc}
\tablecaption{Observation Parameters\label{tbl:observation}}
\tablehead{
\colhead{Line} &
\colhead{$\nu_0$} &
\colhead{HPBW} &
\colhead{$T_{\rm sys}$} &
\colhead{$\eta_{\rm mb}$} &
\colhead{$\delta v$} &
\colhead{$T_{\rm mb}$ rms noise} 
\\
\colhead{($J=1-0$)} &
\colhead{(GHz)} &
\colhead{(\arcsec)} &
\colhead{(K)} &
\colhead{} &
\colhead{(km~s$^{-1}$)} &
\colhead{(K)}
}
\startdata
$^{12}$CO & 115.271204 & 52$\pm$3 & 220-500 & 43.6\% & 0.160 & 0.46 \\
$^{13}$CO & 110.201353 & 52$\pm$3 & 150-310 & 48.0\% & 0.168 & 0.31 \\
C$^{18}$O & 109.782183 & 52$\pm$3 & 150-310 & 48.0\% & 0.168 & 0.22 \\
\enddata
\tablecomments{The columns show the line observed, the rest frequency of the line, the half-power beam width of the telescope, the system temperature, main beam efficiency, velocity resolution and rms noise of main beam temperature. The beam width and main beam efficiency are given by status report of the telescope.}
\end{deluxetable*}

\section{Result}

\subsection{General Distribution of Molecular Cloud}

\begin{figure}
\centering
\includegraphics[width=0.4\textwidth]{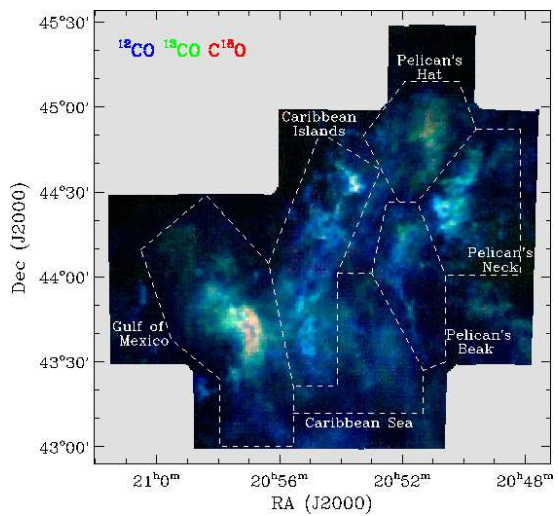}
\caption{A composite color image of the NAN complex made from the integrated intensity map, with $^{12}$CO in blue, $^{13}$CO in green, and C$^{18}$O in red, respectively. The spectra are integrated over $-20 \sim 20$~km~s$^{-1}$ for $^{12}$CO and $^{13}$CO, and $-10 \sim 10$~km~s$^{-1}$ for C$^{18}$O. We also overlay outlines of the six regions with their names on the plot.}
\label{fig:RGB}
\end{figure}

\begin{figure}[b]
\centering
\includegraphics[width=0.4\textwidth]{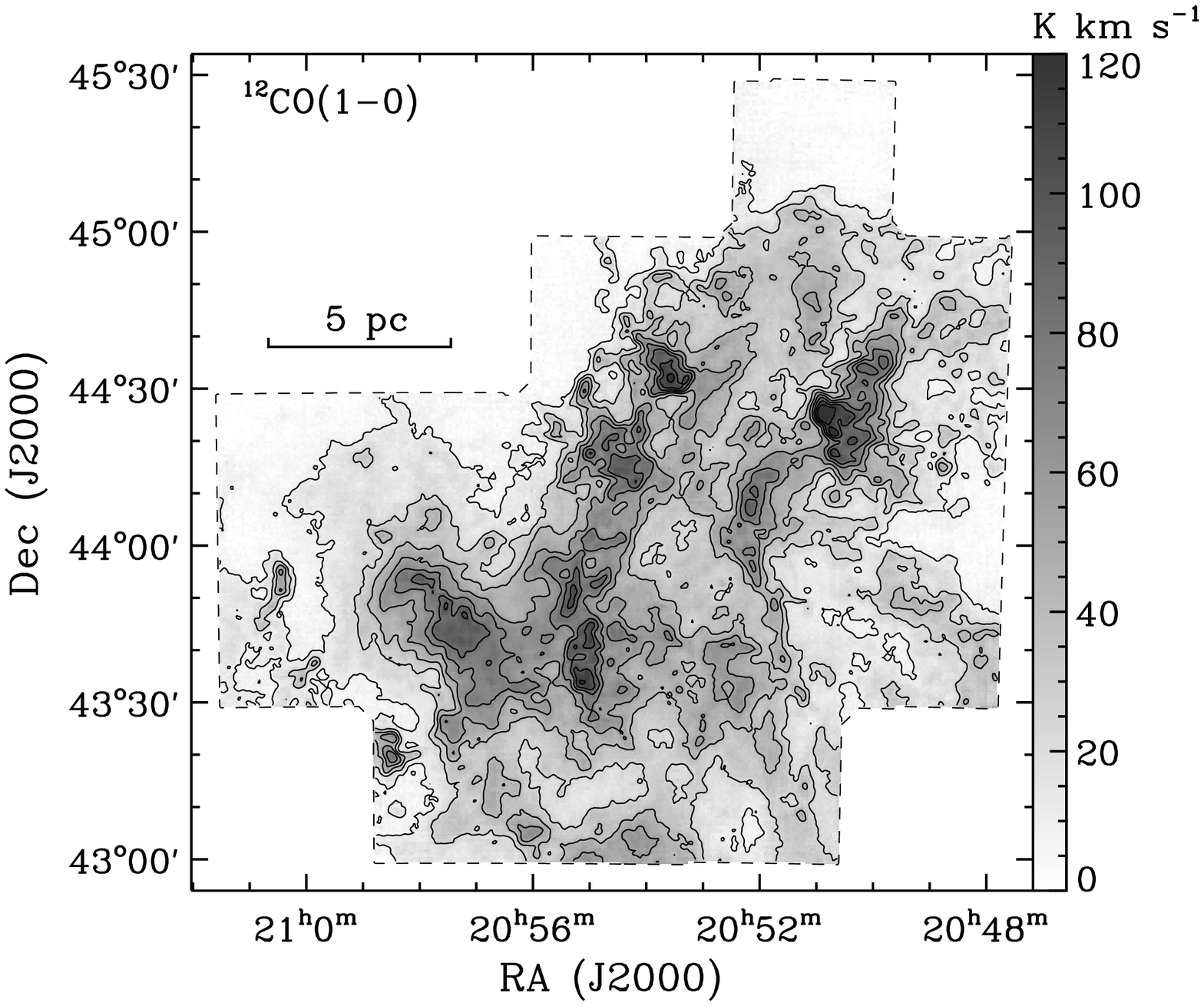}
\caption{Integrated intensity contours and gray-scale map of $^{12}$CO. The spectra are integrated over $-20 \sim 20$~km~s$^{-1}$. The contours are from 10~K~km~s$^{-1}$($\sim9\sigma$) at intervals of 15~K~km~s$^{-1}$, and the gray-scale colors correspond to a linear stretch of integrated intensity.}
\label{fig:12map}
\end{figure}

\begin{figure}[b]
\centering
\includegraphics[width=0.4\textwidth]{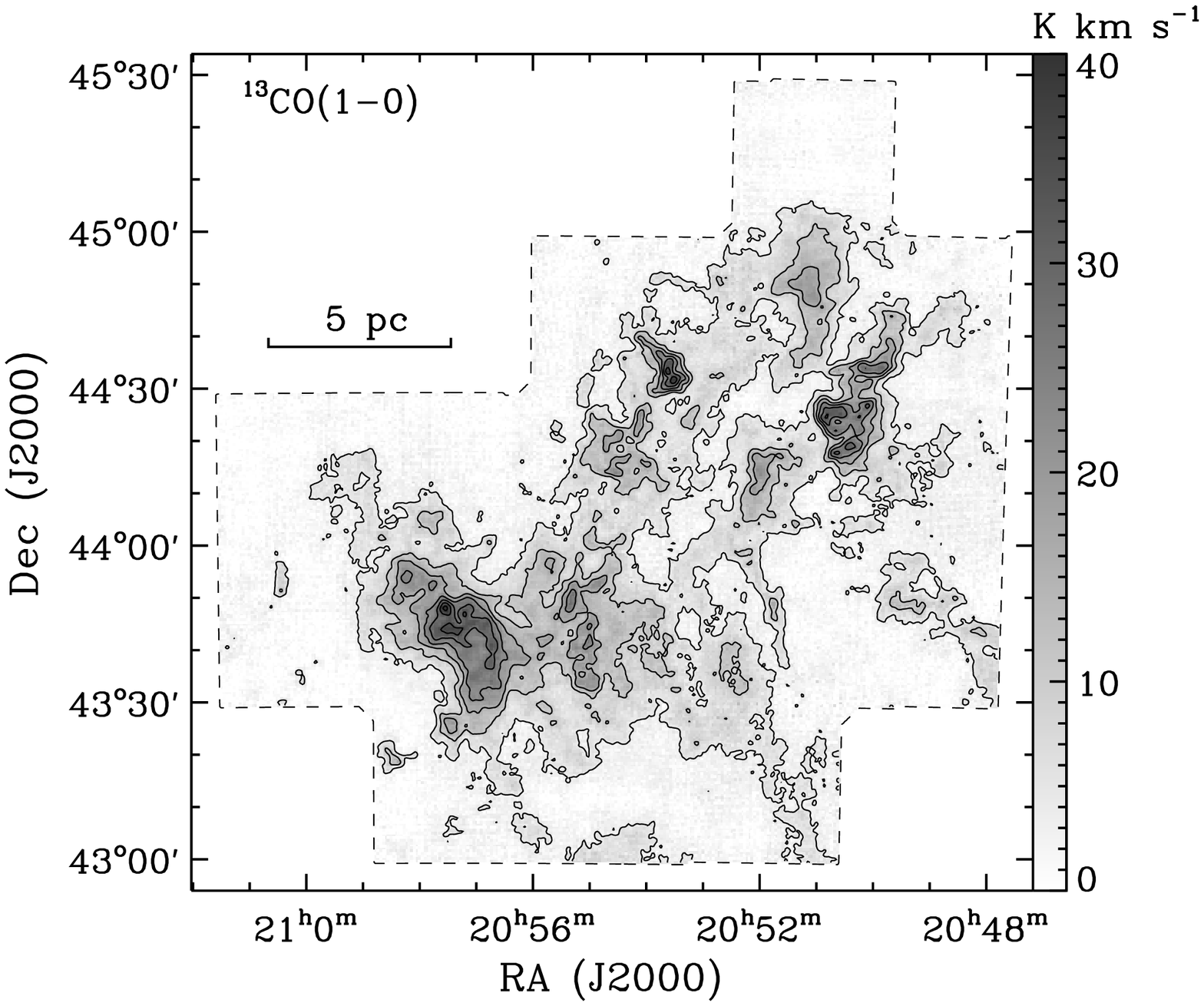}
\caption{Integrated intensity contours and gray-scale map of $^{13}$CO. The spectra are integrated over $-20 \sim 20$~km~s$^{-1}$. The contours are from 4~K~km~s$^{-1}$($\sim5\sigma$) at intervals of 6~K~km~s$^{-1}$, and the gray-scale colors correspond to a linear stretch of integrated intensity.}
\label{fig:13map}
\end{figure}

\begin{figure}[b]
\centering
\includegraphics[width=0.4\textwidth]{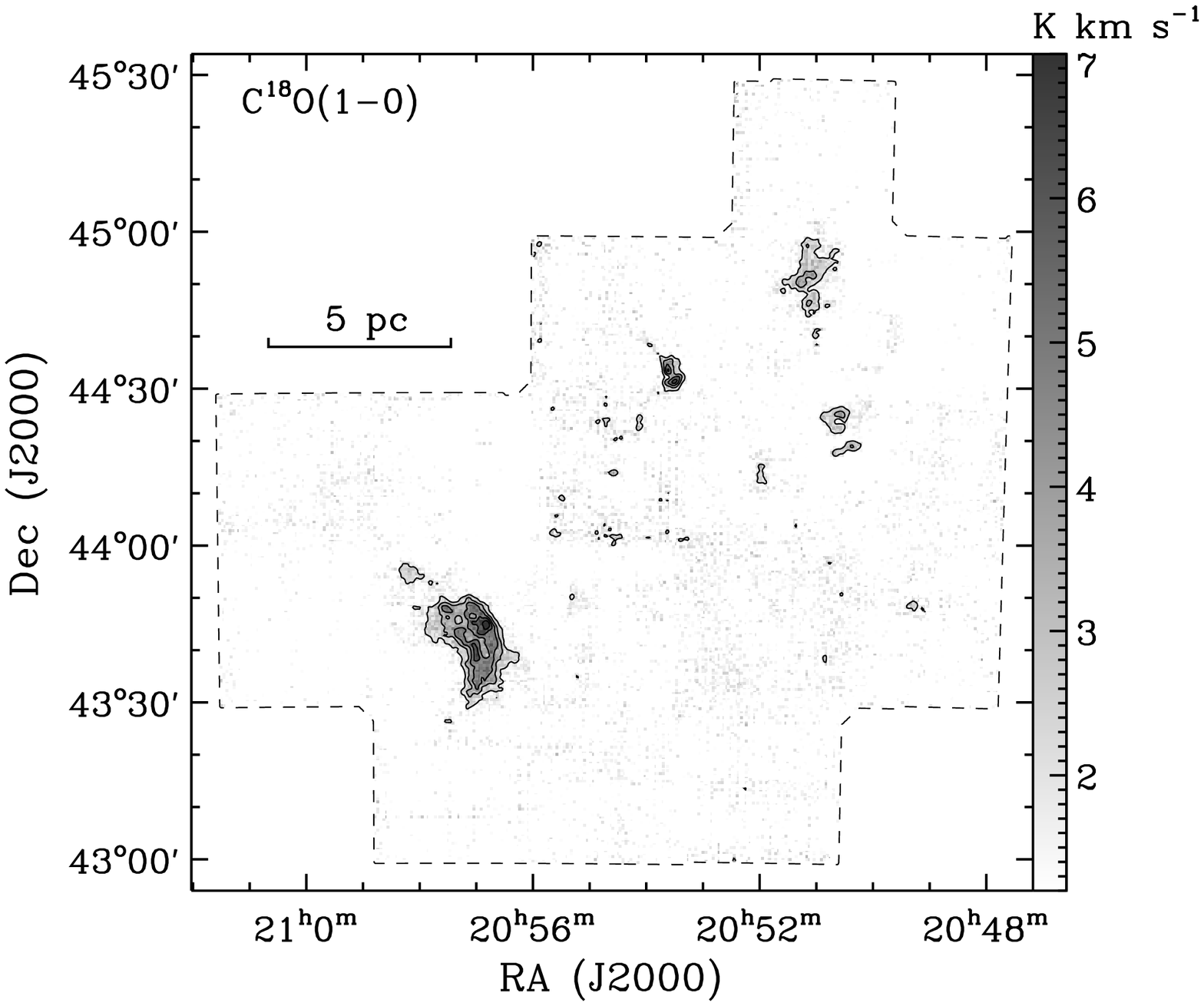}
\caption{Integrated intensity contours and gray-scale map of C$^{18}$O. The spectra are integrated over $-10 \sim 10$~km~s$^{-1}$. The contours are from 2~K~km~s$^{-1}$($\sim5\sigma$) at intervals of 1.2~K~km~s$^{-1}$, and the gray-scale colors correspond to a linear stretch of integrated intensity.}
\label{fig:18map}
\end{figure}

Figures~\ref{fig:RGB}-\ref{fig:18map} show the distributions of $^{12}$CO, $^{13}$CO and C$^{18}$O emissions. The distributions are elongated in the southeast-northwest direction along the dark lane. $^{12}$CO presents bright, complex, extended emission throughout the mosaic, while $^{13}$CO presents several condensations, and C$^{18}$O only appears at those brightest parts. From the distribution of molecules, we distinguish by eye six regions and designated their names following \citet{reb11}. Positions of these regions are indicated on the composed image in Figure~\ref{fig:RGB}. The brightest portions in all three lines are the Gulf of Mexico to the southeast, and the Pelican to the northwest. Between these, there are filamentary structures (the Caribbean Islands) and extended feature to the south (the Caribbean Sea) with few pixels of C$^{18}$O detection. The Caribbean Islands and Sea regions are spatially coincident along the line of sight but are separate in the velocity dimension.

\begin{figure}[b]
\centering
\includegraphics[width=0.4\textwidth]{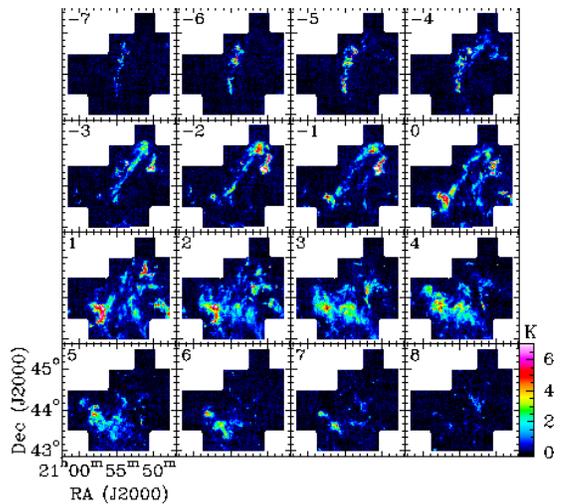}
\caption{Channel map of $^{13}$CO in the NAN complex. The central velocity of each channel, in km~s$^{-1}$, is marked on the top left corner of each map.}
\label{fig:channel}
\end{figure}

The channel map in Figure~\ref{fig:channel} illustrates the velocity structure of the molecules in the NAN complex. Three $^{13}$CO filaments are clearly presented in the velocity ranges of $-$7 to $-$4, $-$3 to $-$2, and $-$1 to 0~km~s$^{-1}$. The latter two filaments connect the Gulf of Mexico and the Pelican's Hat regions. The emissions in the Gulf of Mexico indicate an arc feature from 0 to 2~km~s$^{-1}$. Along with the Caribbean Sea, they show complicated structures in the following positive velocity panels. There is another filamentary structure near the Pelican's Beak, in the velocity range 3 to 4~km~s$^{-1}$.

\begin{figure}[t]
\centering
\includegraphics[width=0.4\textwidth]{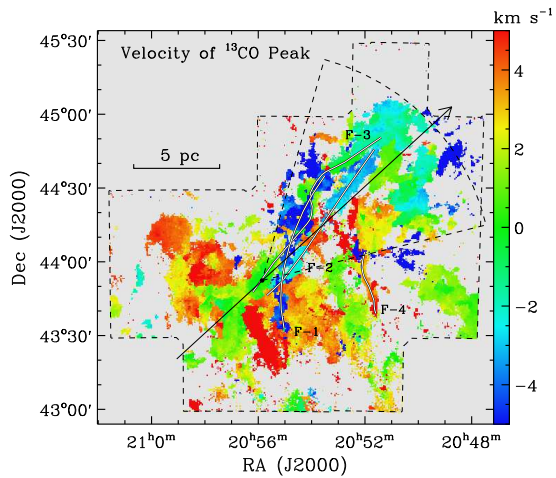}
\caption{Velocity-coded $^{13}$CO map. The color image shows the velocity distribution of the emission peak of $^{13}$CO. The black long arrow indicates the axis of the position-velocity map in Figure~\ref{fig:pvfull}. The black dot on the axis indicate the position of the ionizing star. A azimuthally averaged, around the ionizing star within the sector region, position-velocity map is given in Figure~\ref{fig:pvaz}. The positions of four filamentary structures are shown with white lines.}
\label{fig:velopeak}
\end{figure}

\begin{figure}[b]
\centering
\includegraphics[width=0.3\textwidth]{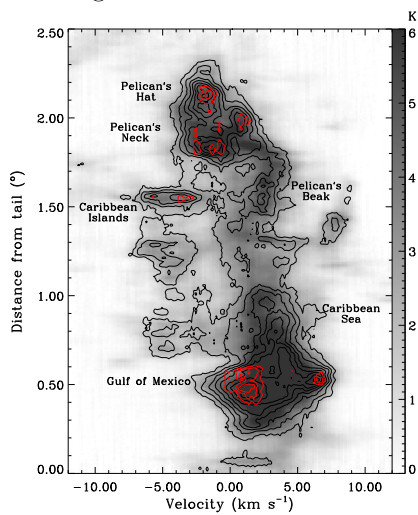}
\caption{Position-velocity map along the axis shown in Figure~\ref{fig:velopeak}. The spectrum on each position is averaged along a 1\arcdeg\ width line perpendicular to the axis. The gray-scale background indicates $^{12}$CO, black contours indicate $^{13}$CO, and red contours indicate C$^{18}$O. The lowest contour is 10$\sigma$ and the contour interval is 10$\sigma$ (0.28~K) for $^{13}$CO and 5$\sigma$ (0.1~K) for C$^{18}$O. Projected positions of six regions are marked.}
\label{fig:pvfull}
\end{figure}

The velocity-coded image shown in Figure~\ref{fig:velopeak} indicates the velocity distribution of the emission peak of $^{13}$CO. Near the center of the whole complex, there are several velocity components with high peak separation, and three filamentary structures showing with different color overlapping each other. The velocity components of the Pelican region in the northwest are relatively simple, while the peak velocities in the southeast show a component  around 0~km~s$^{-1}$, which outlines the Gulf of Mexico region, and another separated extended components at 3-4~km~s$^{-1}$. Such velocity structure could also be seen on the position-velocity map in Figure~\ref{fig:pvfull} along the axis through the full length of the complex in Figure~\ref{fig:velopeak}. In the center region of the whole complex, the molecular emission near $-$1~km~s$^{-1}$ is lacking and forms a cavity structure. \citet{bal80} pointed out that the molecular gas in the northwest part of the NAN complex belongs to an expanding shell surrounding the ionized gas produced by the W80 H~II region.

\begin{figure}[t]
\centering
\includegraphics[width=0.3\textwidth]{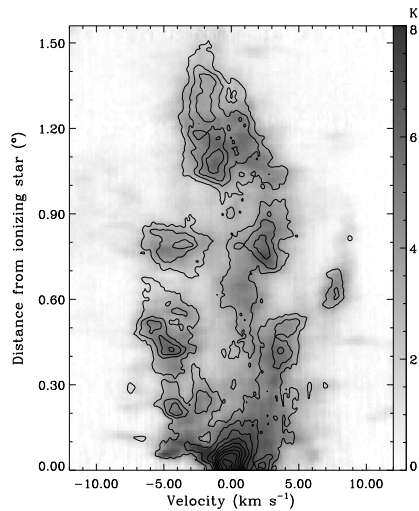}
\caption{Azimuthally average position-velocity map around the ionizing star within the sector region shown in Figure~\ref{fig:velopeak}. The gray-scale background indicates $^{12}$CO, and black contours indicate $^{13}$CO. The lowest contour is 5$\sigma$ (0.4~K) and the contour interval is 5$\sigma$.}
\label{fig:pvaz}
\end{figure}

In Figure~\ref{fig:pvaz}, we illustrate the kinematic of molecular shell near the Pelican region in detail. We could derive a expansion velocity of $\sim$5~km~s$^{-1}$. The Pelican's Hat at the far end is $\sim$14~pc away from the center of the H~II region. The cloud near the ionizing star at $\sim$0~km~s$^{-1}$ connects to the Gulf of Mexico region. Its velocity is close to the rest velocity of the whole complex, which is probably because the molecular gas in these region has not been penetrated by the shock of H~II region \citep{bal80}. In Figure~\ref{fig:pvfull}, we could further find there is a velocity gradient of $\sim$0.2~km~s$^{-1}$~pc$^{-1}$ within the complex along the axis of the position-velocity map.

\begin{figure}[b]
\centering
\includegraphics[width=0.4\textwidth]{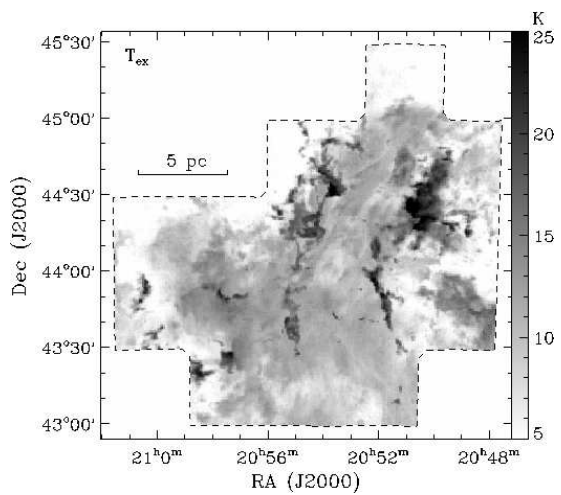}%
\caption{A gray-scale map of excitation temperature in the NAN complex. The excitation temperature is derived from the radiation temperature of $^{12}$CO within the velocity range of $-20 \sim 20$~km~s$^{-1}$.}
\label{fig:Tex}
\end{figure}

Our mapping region contains total areas of 403~pc$^2$ with $^{12}$CO detection, 225~pc$^2$ with $^{13}$CO detection, and 18~pc$^2$ with C$^{18}$O detection over 3$\sigma$ at the distance of 600~pc. Under the assumption of local thermodynamic equilibrium (LTE), we derive the excitation temperature from the radiation temperature of $^{12}$CO. The distribution of excitation temperature shown in Figure~\ref{fig:Tex} indicates gases of two different temperatures within the NAN complex: localized warm gas ($>$20~K) in the Caribbean Islands, Pelican's Neck and Beak, and in some small clouds to the southeast; and extended cold gas ($\sim$10~K) distributed throughout the whole of the dark nebula. The warm gas clearly matches the edge of the whole cloud, suggesting the warm clouds are heated by the background H~II regions.

We further calculate the column density and LTE mass with the $^{13}$CO data following the process given by \citet{nag98} and adopting a $^{13}$CO abundance of $N({\rm H_2})/N({\rm ^{13}CO})=7\times10^5$. We obtain a total mass of $2.0\times10^4~M_\odot$ in the NAN complex. Using the abundance of $N({\rm H_2})/N({\rm C^{18}O})=7\times10^6$ \citep{cas95}, a LTE mass based on C$^{18}$O data can also be derived as  $6.1\times10^3~M_\odot$. If we simply use the CO-to-H$_2$ mass conversion factor $X$ of $1.8\times10^{20}~{\rm cm^{-2} (K~km~s^{-1})^{-1}}$ given in the CO survey of \citet{dam01}, a mass of $5.4\times10^4~M_\odot$ can be derived for the complex. The mass of inner denser gas traced by $^{13}$CO accounts for 36\% of the mass in a larger area traced by $^{12}$CO, while the mass in a few small dense cloud traced by C$^{18}$O accounts for 11\% of the total mass. 

\begin{figure}[t]
\centering
\includegraphics[width=0.18\textwidth]{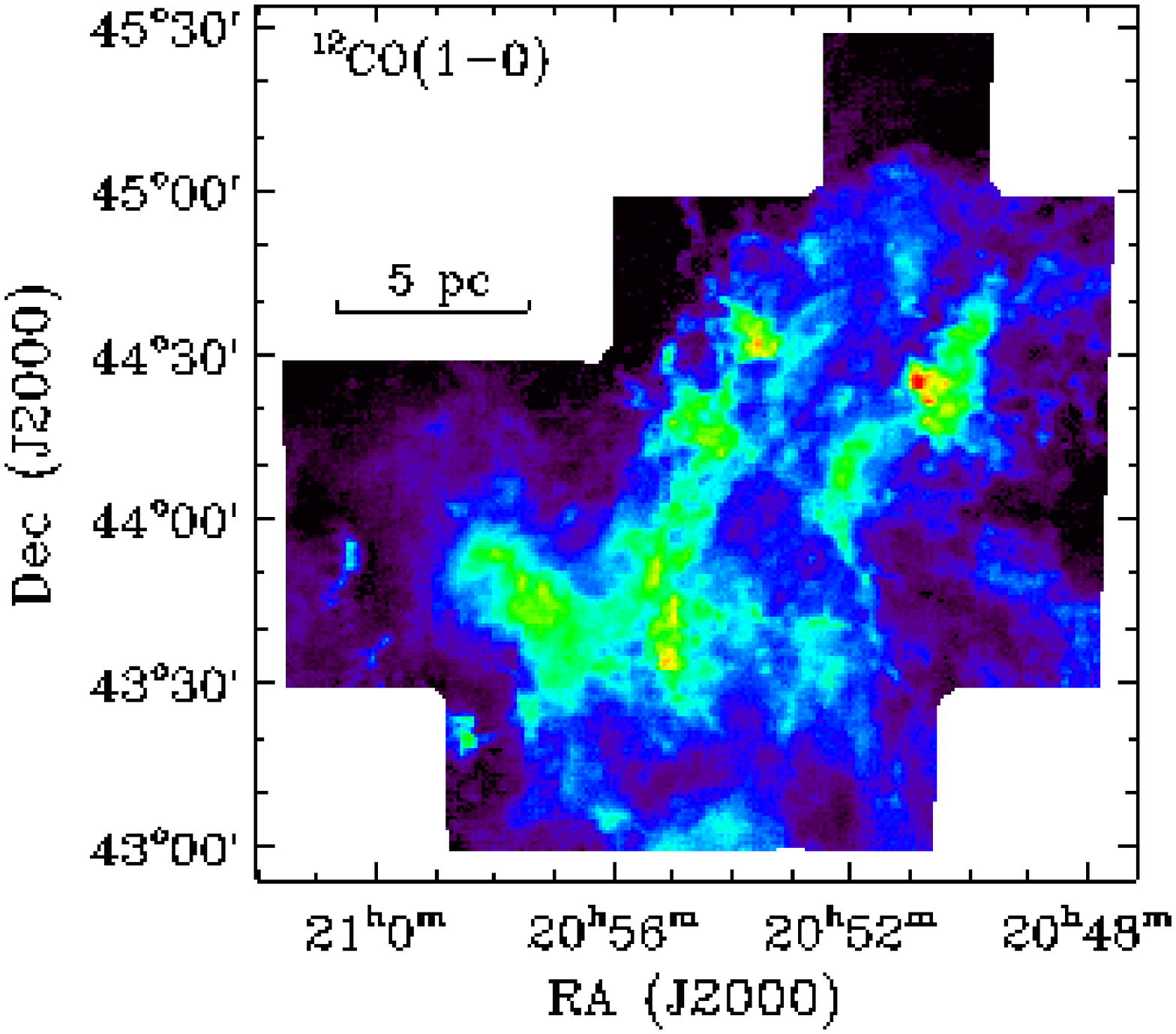}\ %
\includegraphics[width=0.18\textwidth]{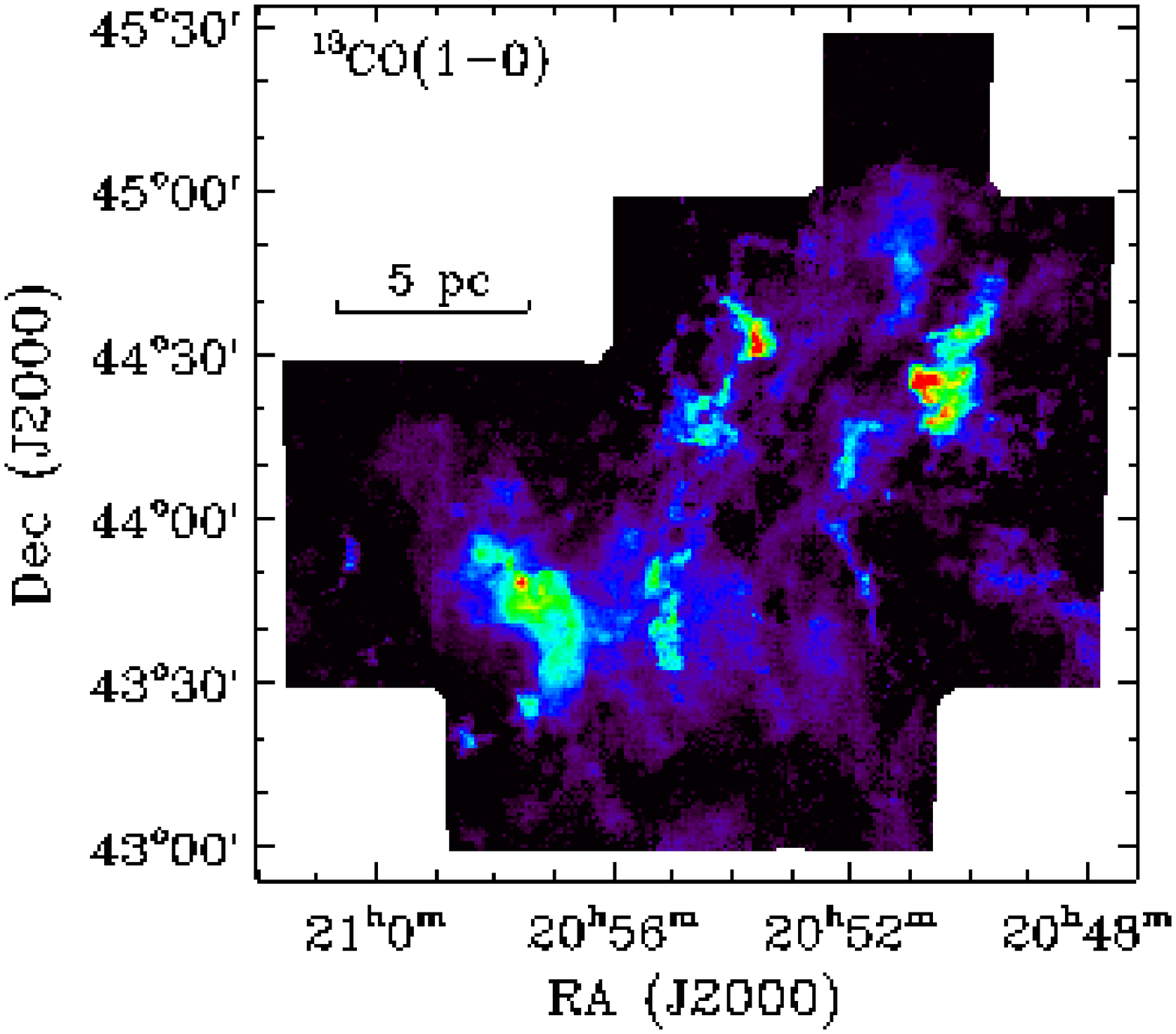}\ %
\includegraphics[width=0.18\textwidth]{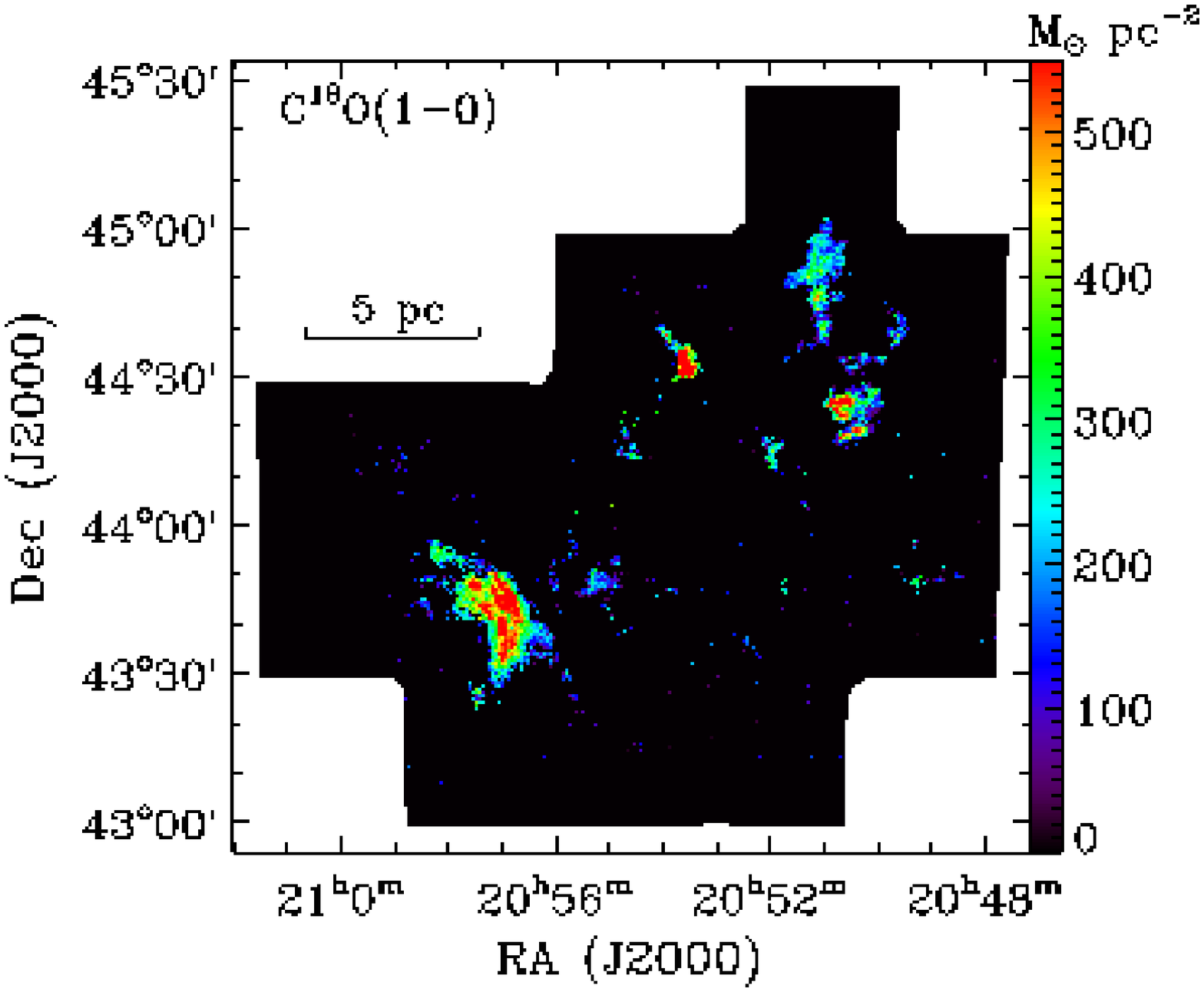}\ %
\caption{The surface density of H$_2$ traced by $^{12}$CO, $^{13}$CO, and C$^{18}$O with the same dynamical range. The abundance $N({\rm H_2})/N({\rm ^{13}CO})=7\times10^5$ \citep{nag98} and $N({\rm H_2})/N({\rm C^{18}O})=7\times10^6$ \citep{cas95} are adopted in the surface density calculation.}
\label{fig:sden}
\end{figure}

In our calculation, we obtained a mean H$_2$ column density of $5.8\times10^{21}$~cm$^{-2}$ based on $^{12}$CO emission by averaging all pixels with line detection. Similar method produces a mean column density of 3.4, and 11.9$\times10^{21}$~cm$^{-2}$ traced by $^{13}$CO and C$^{18}$O, respectively. We show the surface density map for the three molecular species in Figure~\ref{fig:sden}. All three tracers show a maximum surface density over 500~$M_\sun$~pc$^{-2}$ in the Pelican's Neck region, while the Gulf of Mexico region is optically thick with high surface density only in the C$^{18}$O map. The $3\sigma$ noise at $T_{\rm ex}=10$~K within velocity width of 40~km~s$^{-1}$ correspond to 14, 19, and 146~$M_\odot$~pc$^{-2}$ in $^{12}$CO, $^{13}$CO, and C$^{18}$O map, respectively. Therefore the mass hidden under our detection limit of $^{13}$CO is lower than $5.4\times10^2~M_\odot$, which indicates that the discrepancy in the obtained mass between $^{12}$CO and $^{13}$CO is mainly the result of the different emission area tracing by them. The hidden mass for C$^{18}$O is $1.4\times10^3~M_\odot$ at most, significantly lower than the total mass traced by $^{13}$CO. This means that we have detected over 80\% of mass in our C$^{18}$O observation area.

\citet{bal80} observed a similar field in the NAN complex and estimated the LTE mass as $(3-6)\times10^4~M_\odot$ for a distance of 1~kpc and $^{13}$CO abundance of $N({\rm H_2})/N({\rm ^{13}CO})=1\times10^6$. For the same parameters as we used, it would correspond to (1-2)$\times10^4~M_\odot$. \citet{cam02} obtained a mass of $4.5\times10^4~M_\odot$ for a distance of 580~pc in their near-infrared extinction study covering an area of 6.25~deg$^2$ in the NAN complex. These discrepancy of mass might be due to the dust-to-gas ratio or the $X$ factor.

\subsection{Features in Individual Regions}

\begin{figure}
\centering
\includegraphics[width=0.18\textwidth]{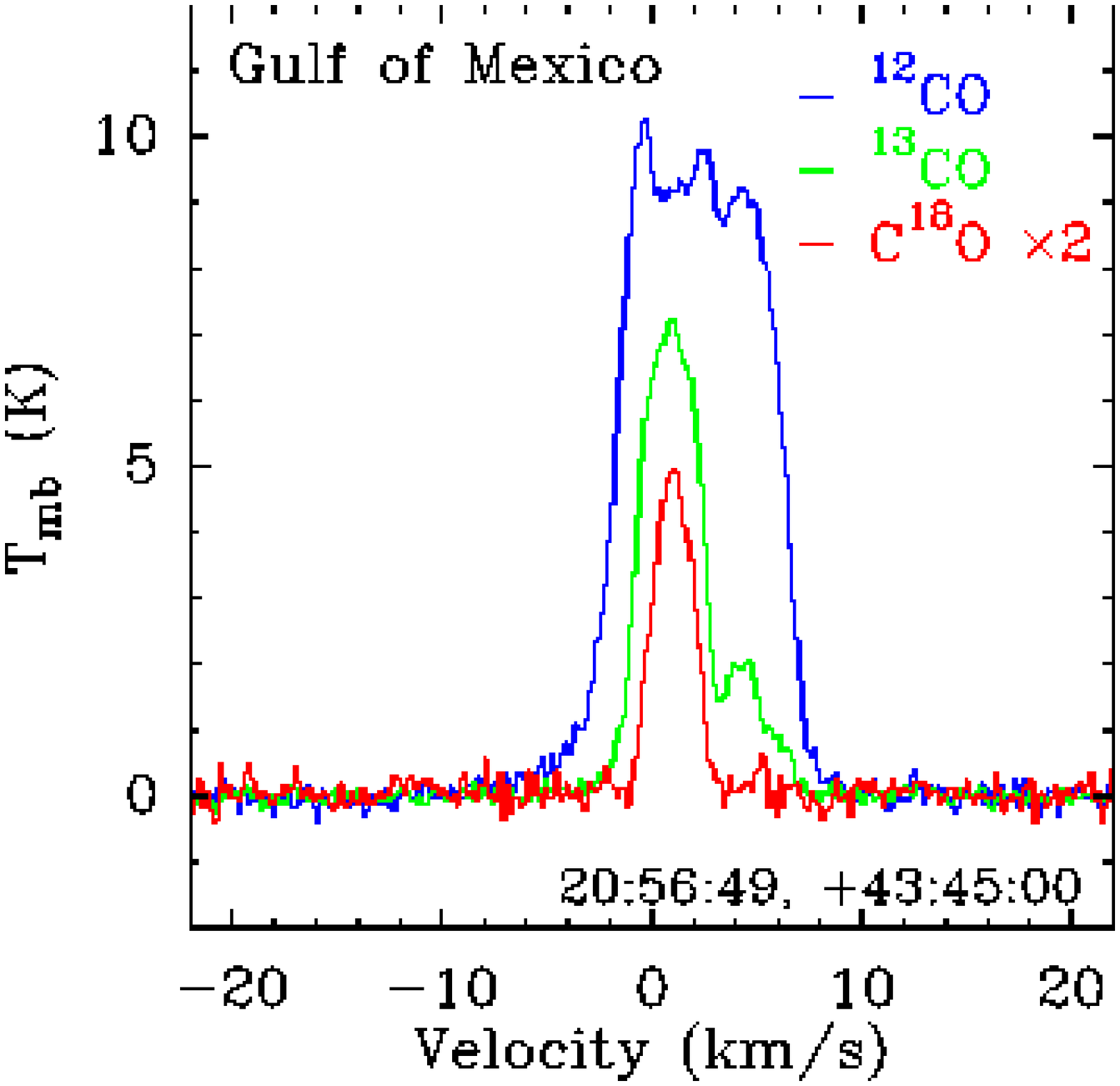}\ %
\includegraphics[width=0.18\textwidth]{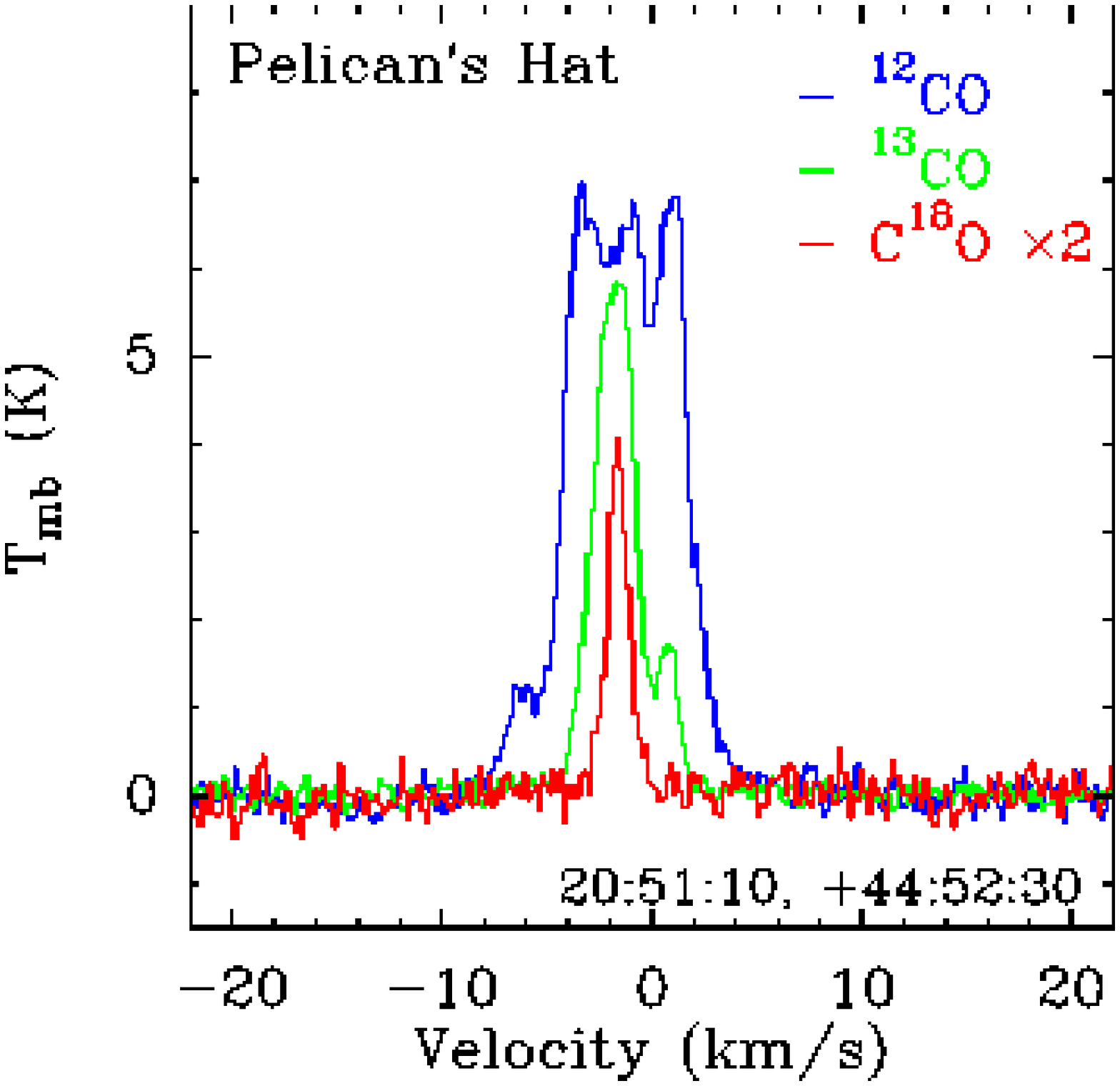}\\
\includegraphics[width=0.18\textwidth]{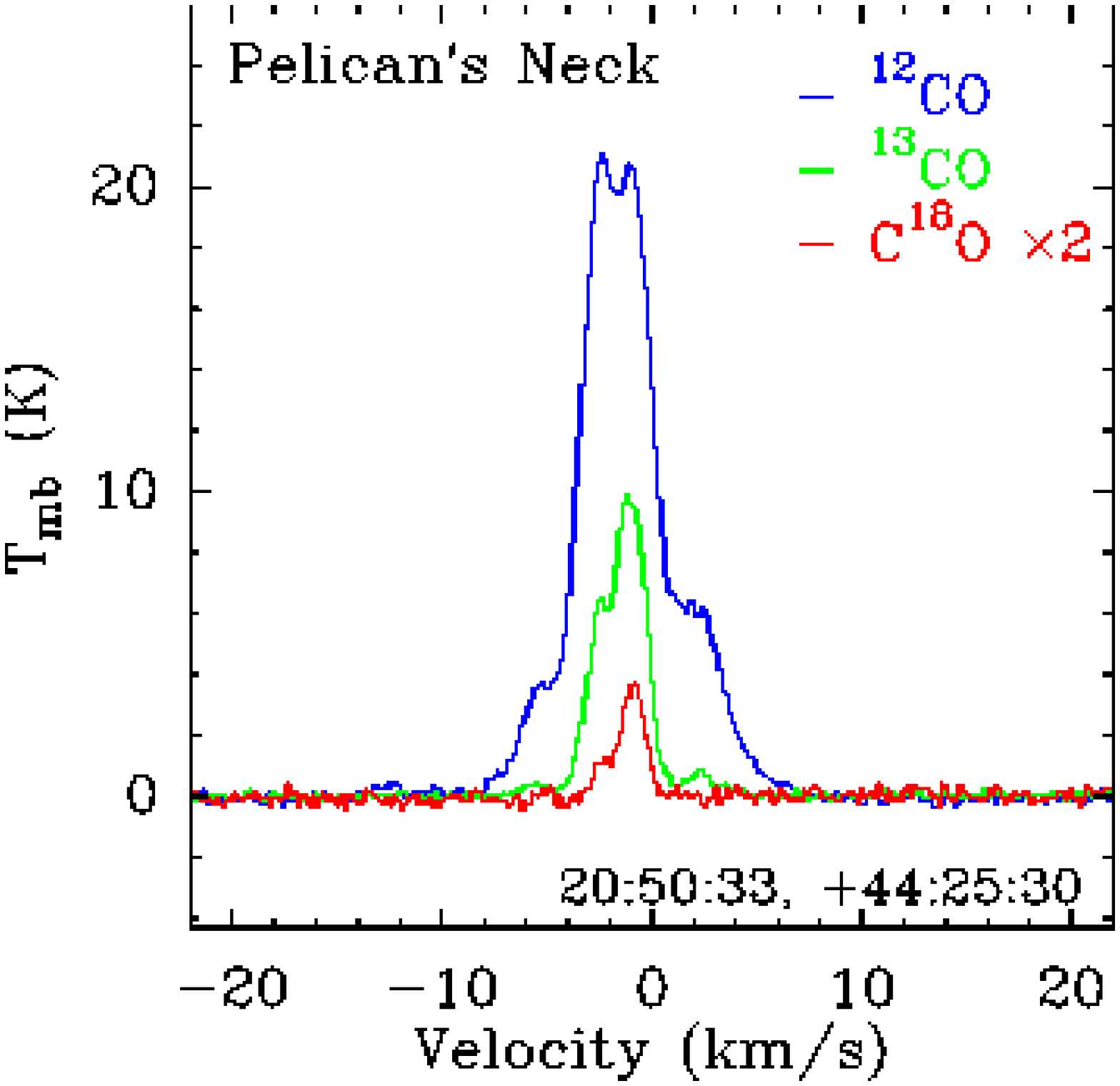}\ %
\includegraphics[width=0.18\textwidth]{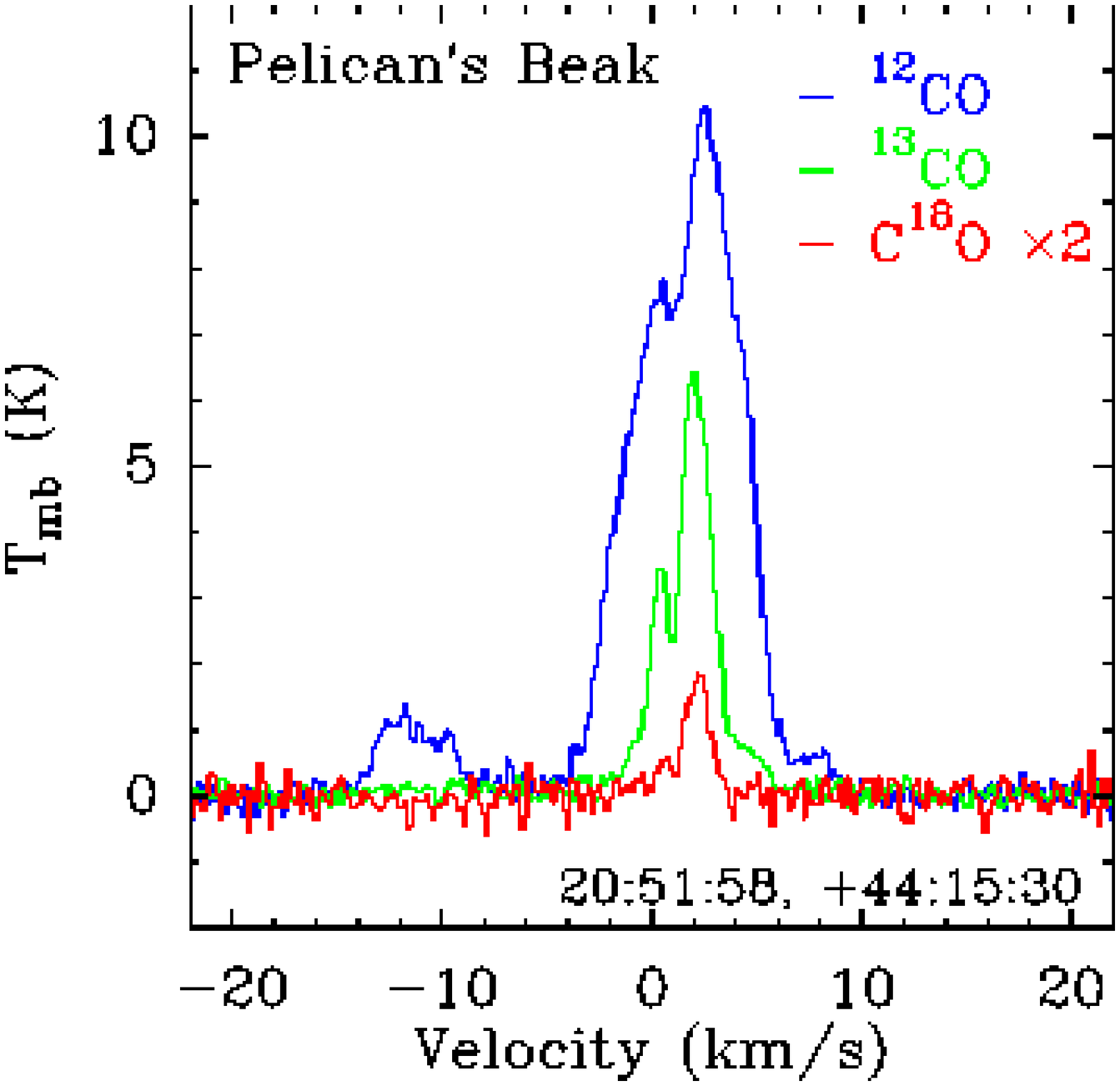}\\
\includegraphics[width=0.18\textwidth]{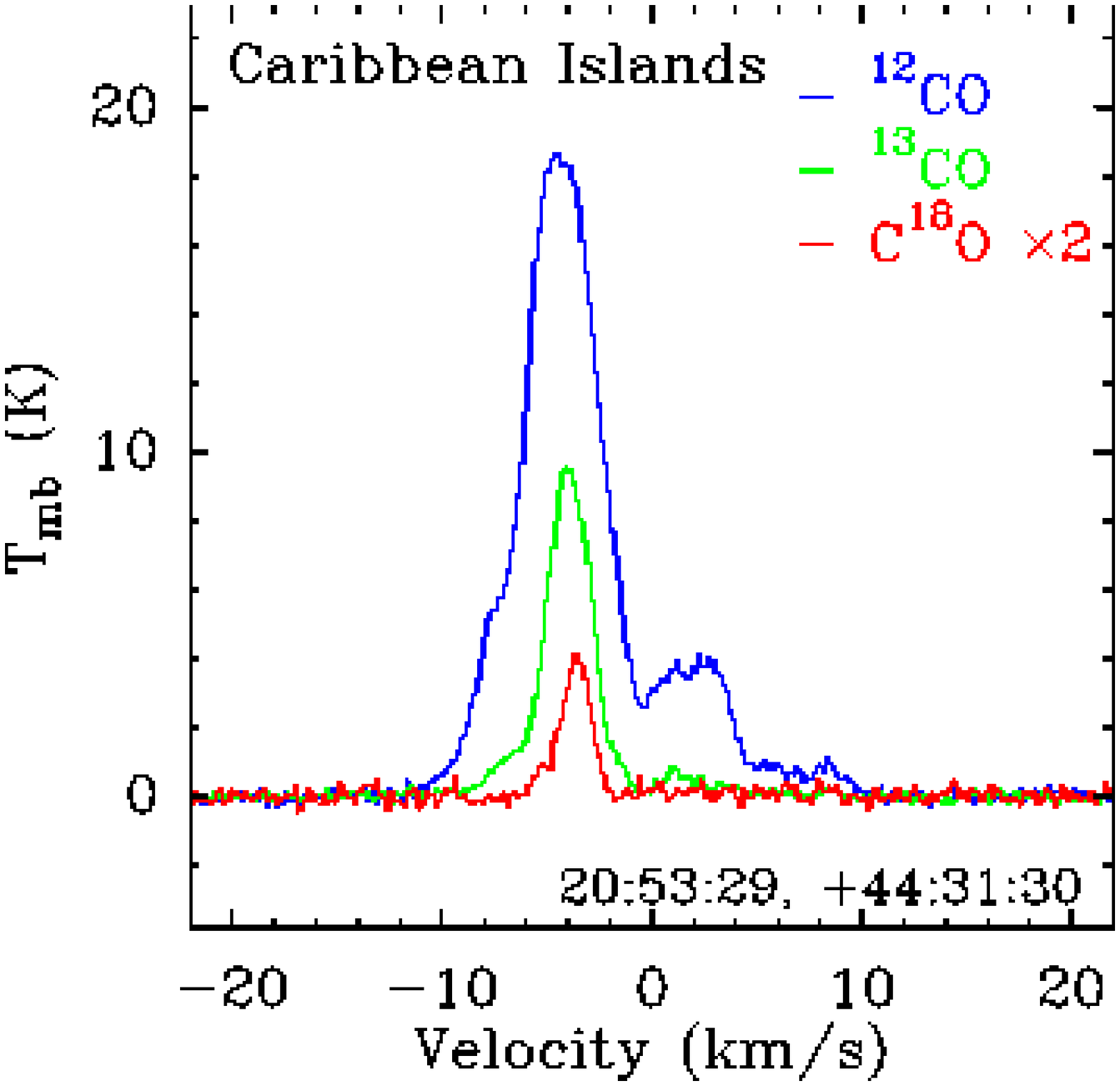}\ %
\includegraphics[width=0.18\textwidth]{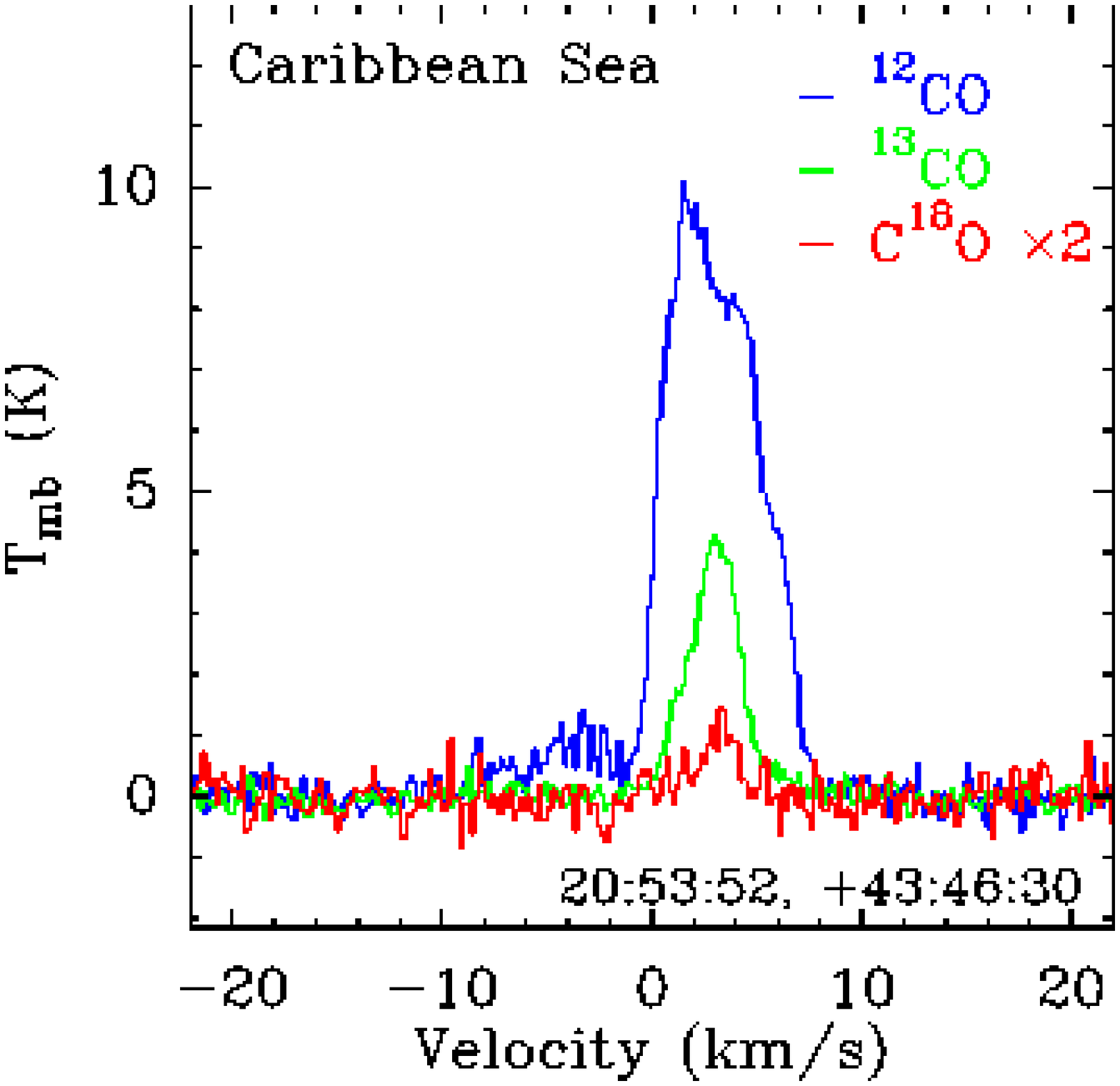}
\caption{Typical spectra in the selected regions with names in the upper left corner. These spectra are averaged within a 2\arcmin$\times$2\arcmin\ box centered at the positions (RA, Dec) marked at the lower right corner.}
\label{fig:spec}
\end{figure}

Several discernible regions and filamentary structures can be identified in our observations. The spectra observed toward the regions are shown in Figure~\ref{fig:spec}. The variety of line profile and intensity ratios indicates distinct kinematic and chemistry environments. Their properties probed by different tracers are summarized in Table~\ref{tbl:region} and details for each region are listed below.

\begin{turnpage}
\begin{deluxetable*}{lcccccccccccccccc}
\tabletypesize{\scriptsize}
\small\addtolength{\tabcolsep}{-5pt}
\tablecolumns{17}
\tablewidth{0pc}
\tablecaption{Properties of regions\label{tbl:region}}
\tablehead{
\colhead{} &
\colhead{} &
\colhead{} &
\multicolumn{3}{c}{$^{12}$CO} &
\colhead{} &
\multicolumn{4}{c}{$^{13}$CO} &
\colhead{} &
\multicolumn{3}{c}{C$^{18}$O} &
\colhead{} &
\colhead{}
\\
\cline{4-6} \cline{8-11} \cline{13-15}
\colhead{Region} &
\colhead{$T_{\rm ex}$} &
\colhead{} &
\colhead{area} & \colhead{$N_{\rm H_2}$} & \colhead{$M$} &
\colhead{} &
\colhead{area} & \colhead{$N_{\rm H_2}$} & \colhead{$M$} &
\colhead{$\Delta v$} &
\colhead{} &
\colhead{area} & \colhead{$N_{\rm H_2}$} & \colhead{$M$} &
\colhead{} &
\colhead{$S({\rm C^{18}O}) \over S({\rm ^{13}CO})$}
\\
\colhead{} &
\colhead{(K)} &
\colhead{} &
\colhead{(pc$^2$)} & \colhead{($10^{22} {\rm cm}^{-2}$)} & \colhead{($10^3M_\odot$)} &
\colhead{} &
\colhead{(pc$^2$)} & \colhead{($10^{22} {\rm cm}^{-2}$)} & \colhead{($10^3M_\odot$)} &
\colhead{(km~s$^{-1}$)} &
\colhead{} &
\colhead{(pc$^2$)} & \colhead{($10^{22} {\rm cm}^{-2}$)} & \colhead{($10^2M_\odot$)} &
\colhead{} &
\colhead{}
}
\startdata
Gulf of Mexico      & 14.2 & & 18.2 & 1.2 & 11.8 & & 5.7 & 1.3 & 5.4 & 5.93 & & 2.0 & 2.5 & 32.0 & & 0.13 \\
Pelican's Hat         & 12.2 & & 16.4 & 0.7 &  4.1 & & 4.6 & 0.6 & 1.7 & 4.37 & & 1.3 & 1.3 & 8.4 &  & 0.14 \\
Pelican's Neck       & 22.8 & &  4.4 & 1.6 &  6.0 & & 3.4 & 1.6 & 3.1& 2.99 & & 0.6 & 2.4 & 9.9 & & 0.07 \\
Pelican's Beak       & 15.9 & &  7.5 & 1.1 &  4.8 & & 2.2 & 0.8 & 1.3 & 2.24 & & 0.3 & 1.1 & 1.3 & & 0.08 \\
Caribbean Islands & 18.7 & & 19.2 & 1.3 & 12.3 & & 2.2 & 1.3 & 5.0 & 2.98 & & 0.3 & 3.6 & 8.5 & & 0.11 \\
Caribbean Sea      & 12.1 & & 41.6 & 0.7 & 8.1 & & 10.2 & 0.4 & 2.5 & 3.86 & & - & - & - & & - \\
\enddata
\tablecomments{The properties of the regions in the NAN complex, including excitation temperature, area within the half maximum contour line, mean column density of H$_2$, and mass of $^{12}$CO, $^{13}$CO, and C$^{18}$O, line width of averaged spectra for $^{13}$CO, and integrated intensity ratio of C$^{18}$O to $^{13}$CO. The column density and mass for $^{12}$CO are derived with a constant CO-to-H$_2$ mass conversion factor, and those for $^{13}$CO, and C$^{18}$O are derived under the LTE assumption. The C$^{18}$O properties in Caribbean Sea is missing because of the low C$^{18}$O detection rate in this region.}
\end{deluxetable*}
\end{turnpage}

\begin{figure}[t]
\centering
\includegraphics[width=0.2\textwidth]{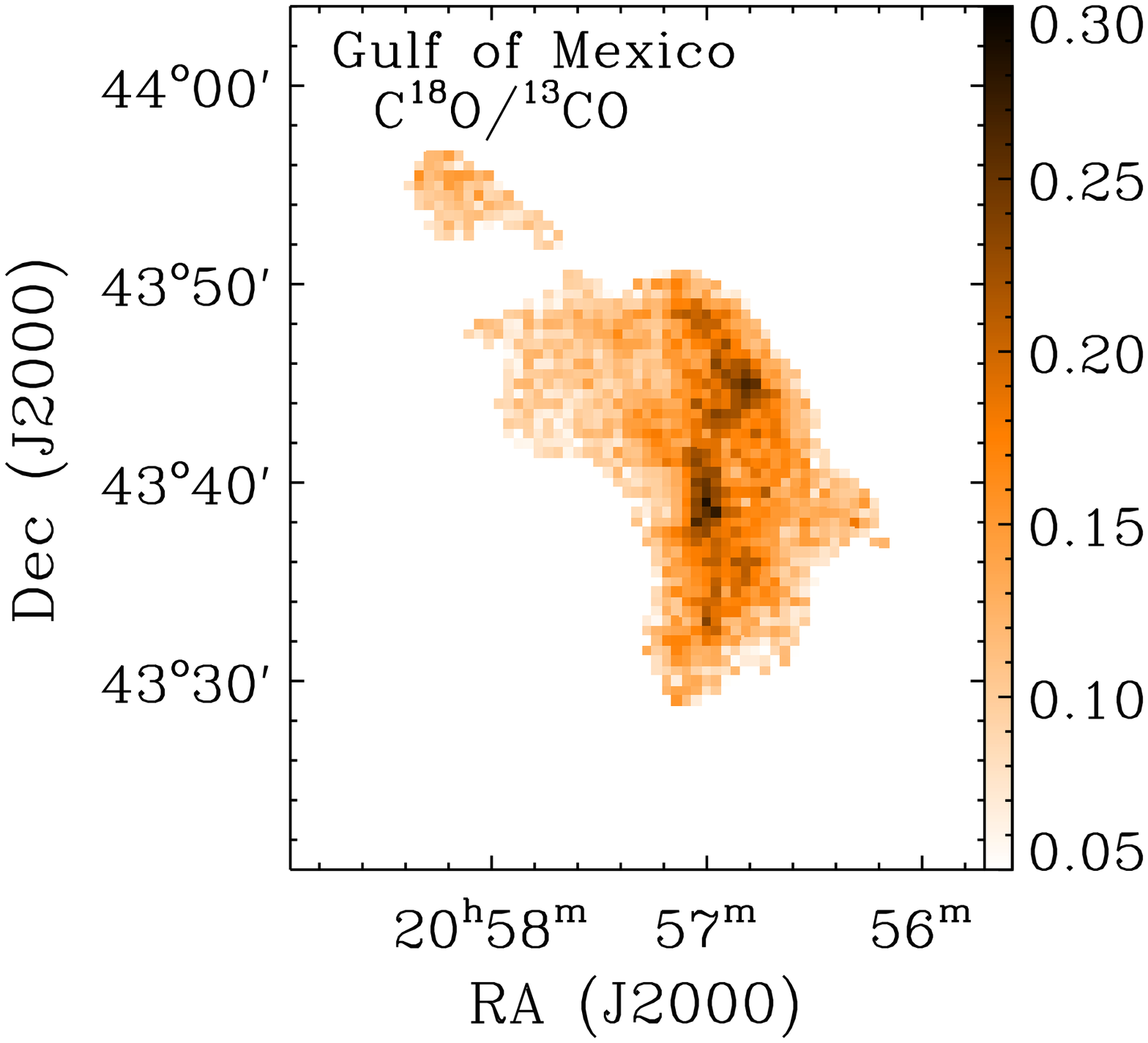}%
\includegraphics[width=0.25\textwidth]{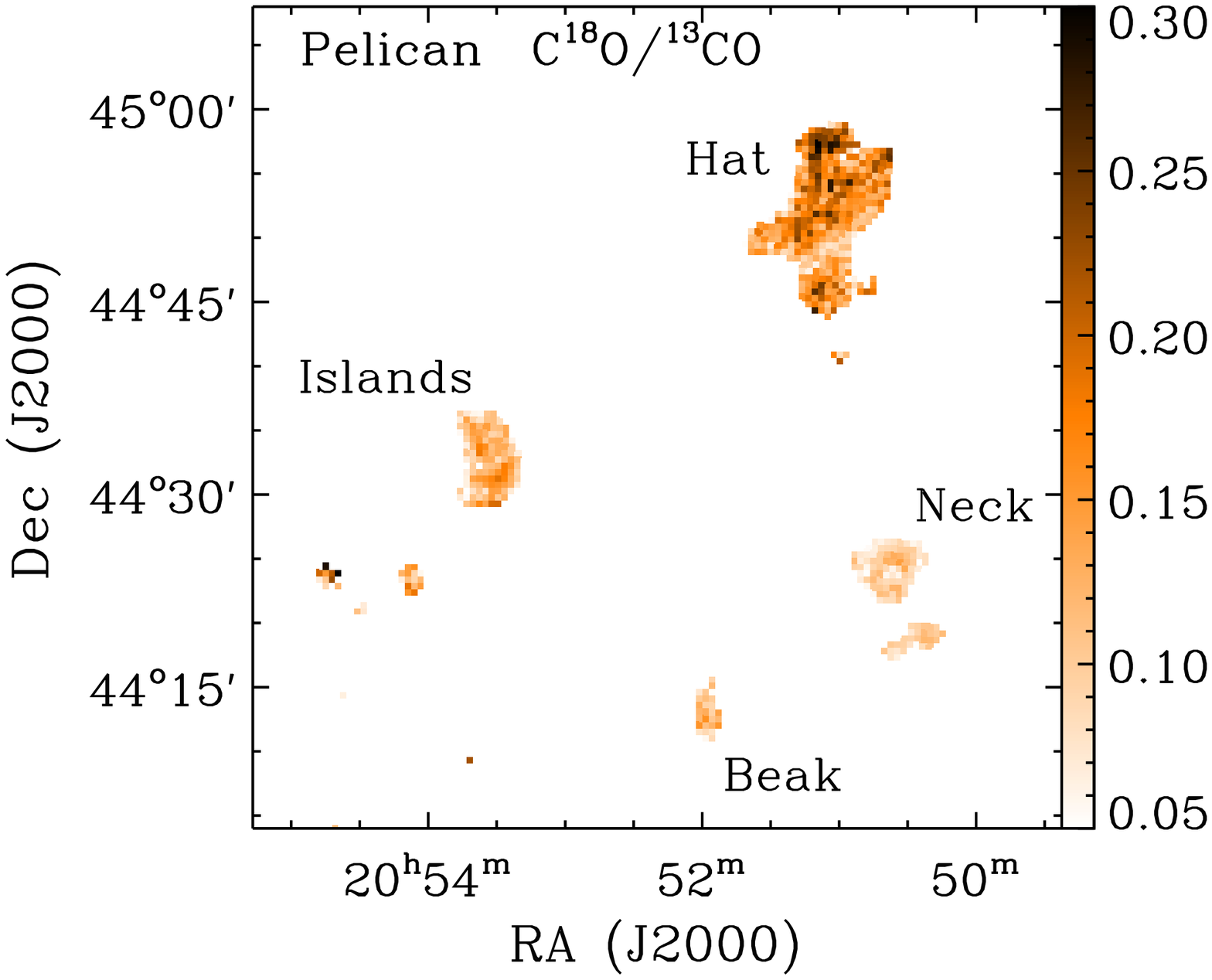}
\caption{Integrated intensity ratio of C$^{18}$O to $^{13}$CO. The two panels show the same region as that in Figure~\ref{fig:18map}. Names of the regions are marked on the map.}
\label{fig:ratio}
\end{figure}

The ``\textbf{Gulf of Mexico}'' (GoM) is the largest and the most massive region with all three line detections in the southeast of the NAN complex. Two major clumps can be found in this region: one in the north (GoM~N) with weak C$^{18}$O emission, and one in the south (GoM~S), with strong C$^{18}$O emission indicating a pair of parallel arcs, which closely matches the morphology of the filamentary dark cloud in Spitzer mid-infrared image \citep{gui09, reb11}. The $^{12}$CO spectra are flat-topped, indicating high opacity at these locations. Under the LTE assumption, we find a low excitation temperature ($\sim$14~K), large line width, and high column density in this region. The intensity ratio in Figure~\ref{fig:ratio}, which also indicates the relative abundance of C$^{18}$O to $^{13}$CO, shows a higher ratio in the GoM~S. \citet{reb11} reported a young stellar objects (YSOs) cluster is associated with the GoM region. A high concentration of T-Tauri type stars \citep{her58} and an association of H$_2$O maser \citep{tou11} are found in the GoM~N. No IRAS point sources are associated with either clump. These evidence suggest active star formation in the GoM region, and the GoM~N region is relatively more evolved than GoM~S.

The ``\textbf{Pelican'’s Hat}'' locates to the north of Pelican's head, and is similar to but smaller than the GoM~S. The excitation temperature in this region is low ($\sim$12~K), and the relative abundance of C$^{18}$O to $^{13}$CO is higher than those in other regions in Pelican nebula. The line emission resembles the morphology of mid-infrared extinction in Spitzer image \citep{gui09, reb11}.

\begin{figure}[b]
\centering
\includegraphics[width=0.25\textwidth]{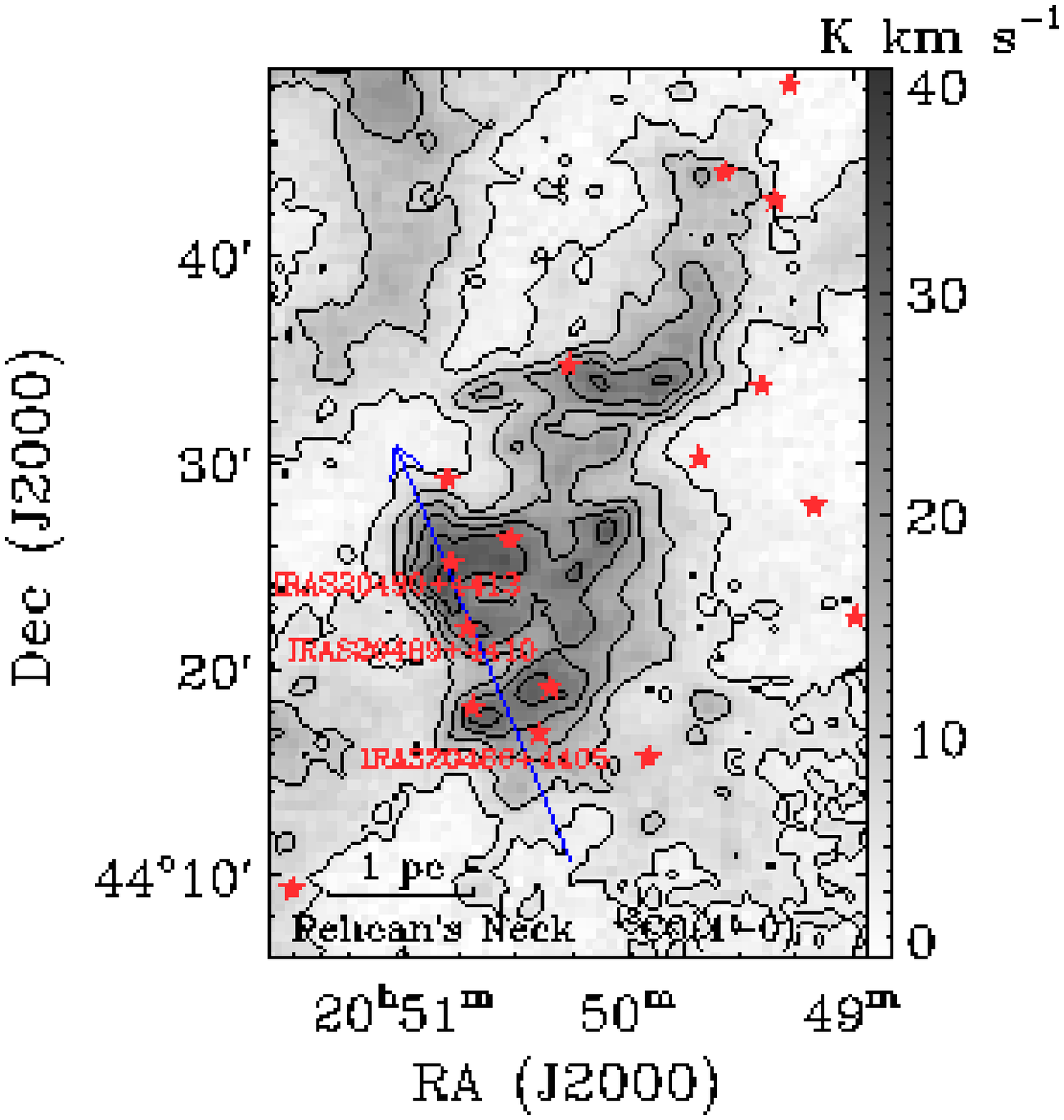}%
\includegraphics[width=0.21\textwidth]{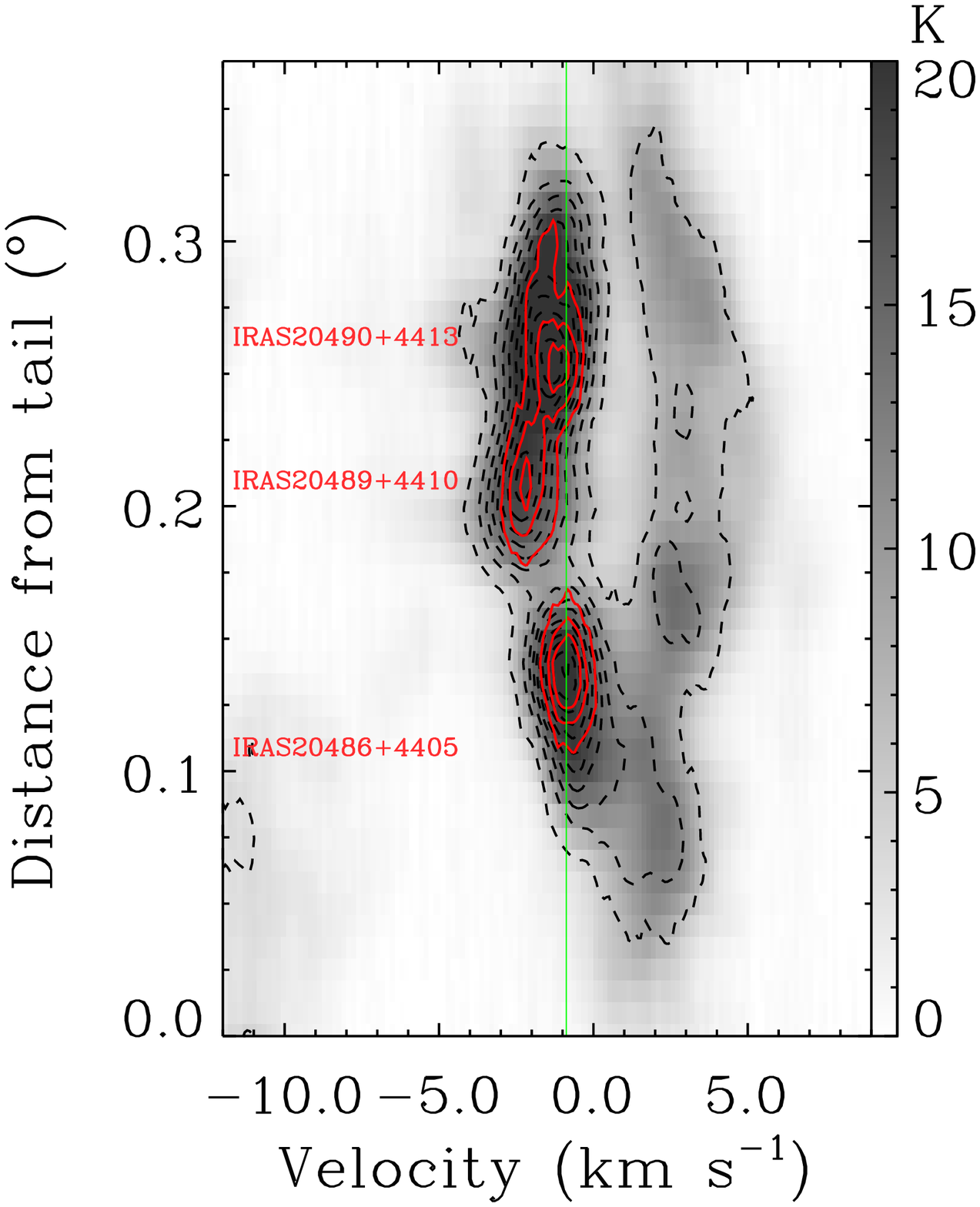}
\caption{Left: the $^{13}$CO integrated intensity map of Pelican's Neck. Red stars indicate the position of IRAS point sources. The names of three massive sources are indicated. The blue arrow indicates the axis of the position-velocity map. Right: position-velocity map along the axis shown in the left panel. The gray-scale background indicates $^{12}$CO, dashed contours indicate $^{13}$CO, and red solid contours indicate C$^{18}$O. The lowest contour is 10$\sigma$ and the contour interval is 15$\sigma$ for $^{13}$CO and 10$\sigma$ for C$^{18}$O. The vertical green line indicates the rest velocity averaged over the whole region. Projected positions of three IRAS sources are marked.}
\label{fig:pvslice}
\end{figure}

The ``\textbf{Pelican'’s Neck}" region to the west of the NAN complex shows the most intense $^{12}$CO emission in our survey. It presents a high excitation temperature of $\sim$23~K, narrow line width, and weak C$^{18}$O emission. The molecular emission shows a bright feature oriented in the north-south direction, with a sharp cut-off towards the east. Several IRAS sources are associated with the peaks on the $^{13}$CO map. A position-velocity slice along the east edge of the Pelican's Neck (as in Figure~\ref{fig:pvslice}) reveals a weak component at $\sim$3~km~s$^{-1}$ that is separate from the molecular clump and forms a cavity near IRAS~20489+4410 and IRAS~20490+4413 in the velocity dimension. Such a structure could be the result of an embedded H~II region.

The molecular emission in Pelican's Neck matches the morphology of the brightest surface brightness region in Spitzer mid-infrared image, and it is at the west edge of the Pelican Cluster, an active star forming cluster of YSOs identified by \citet{reb11}. A clustering of T-Tauri type stars \citep{her58} was found around the molecular clump. These all indicate that the Pelican's Neck is a warm region with active star formation.

The ``\textbf{Pelican'’s Beak}" region is a small elongated region to the southeast of Pelican's Neck. The excitation temperature is intermediate ($\sim$15~K) with weak C$^{18}$O emission. The molecules protrude along a filament to the south at the velocity of 3~km~s$^{-1}$. Its properties may suggest an intermediate stage between the cold dense regions (e.g. GoM, Pelican's Hat) and the warm active regions (e.g. Pelican's Neck, Caribbean Islands).

The ``\textbf{Caribbean Islands}" are several bright clumps extending from the west of GoM and to the east of Pelican's head. The southern half of Caribbean Islands is spatially coincident with the Caribbean Sea. The channel map indicates these ``Islands'' are part of a filamentary structure (see \ref{sbs:fil}). They associate with several highly localized nebulous bright blobs in Spitzers mid-infrared image \citep{reb11}. These clumps show a high excitation temperature, narrow line width, and low relative abundance of C$^{18}$O. These properties indicate a similar situation to that in Pelican's Neck. Together with Pelican's Neck, the north part of Caribbean Islands forms a cavity structure at the position of the Pelican Cluster which can be seen on the $^{13}$CO integrated intensity map.

\begin{figure}
\centering
\includegraphics[width=0.35\textwidth]{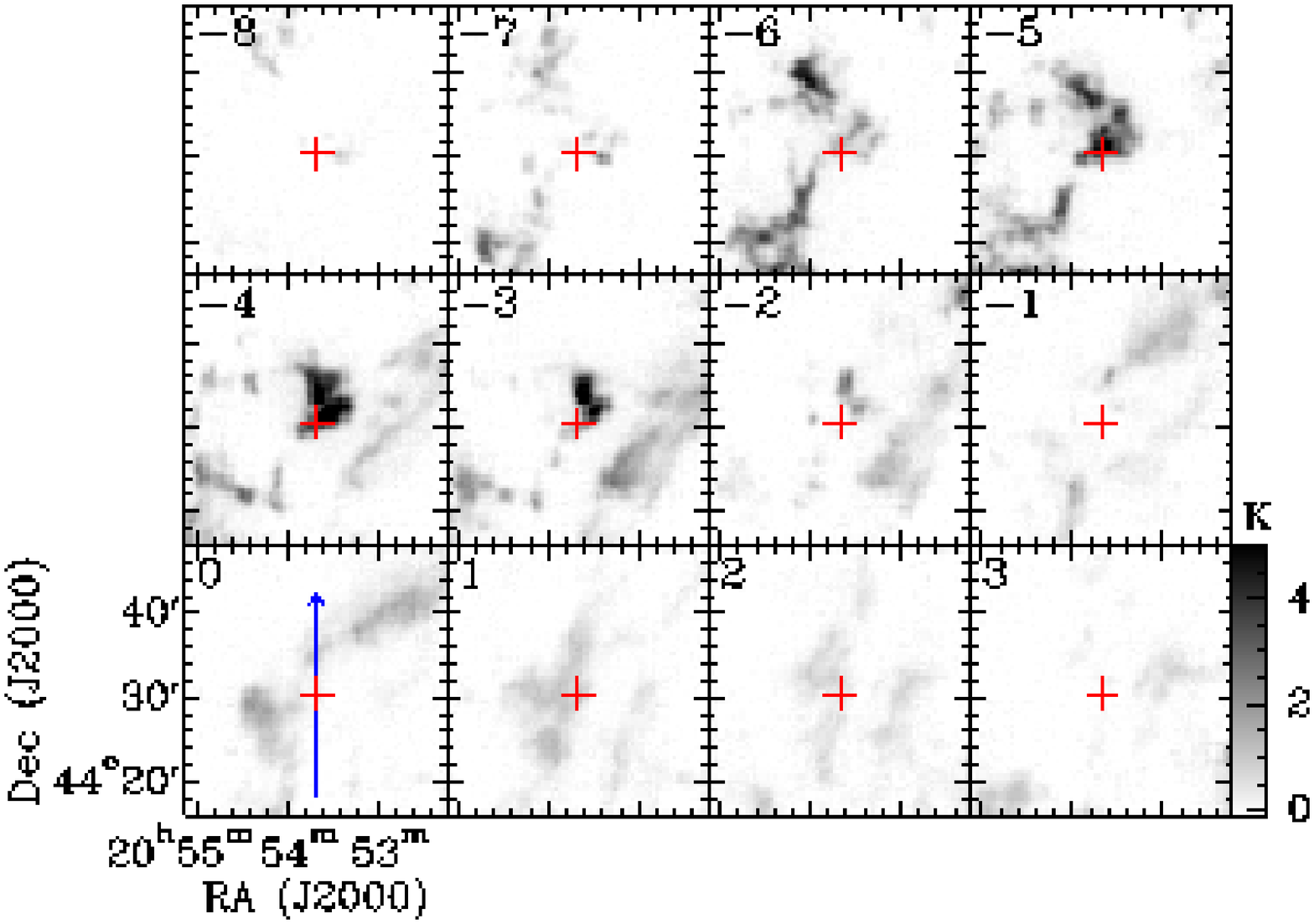}\\
\includegraphics[width=0.35\textwidth]{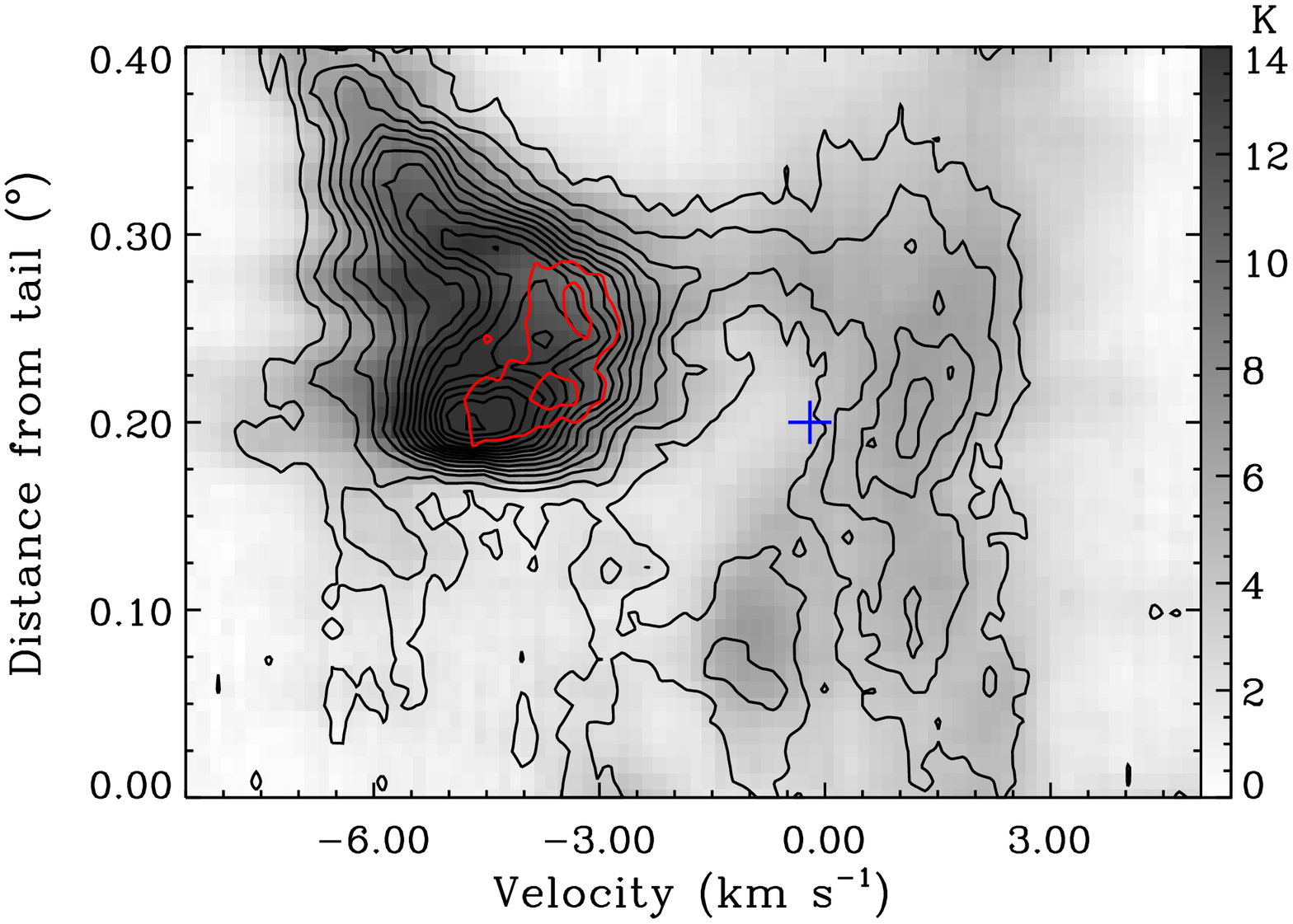}
\caption{Top: channel map of $^{13}$CO near H~II region in the Caribbean Islands region. The cross in each panel marks the position of the H~II region, G085.051$-$0.182, reported by \citet{loc96}. The central velocity of each channel, in km~s$^{-1}$, is marked on the top left corner of each map. The blue arrow is the axis of the position-velocity map. Bottom: position-velocity map along the axis shown in the top panel. The gray-scale background indicates $^{12}$CO, dashed contours indicate $^{13}$CO, and red solid contours indicate C$^{18}$O. The lowest contour is 5$\sigma$ for $^{13}$CO and 10$\sigma$ for C$^{18}$O, with the contour interval of 5$\sigma$. The cross indicates the rest velocity and the position of the H~II region.}
\label{fig:pvhii}
\end{figure}

The molecular cloud in the northernmost part of this region is associated with an H~II regions, G085.051$-$0.182 at $-0.2$~km~s$^{-1}$, identified by \citet{loc96}. Figure~\ref{fig:pvhii} shows that the H~II region is not associated with any dense molecular clumps at its rest velocity. Dense and heated gas with temperature $\sim$27~K appears within the velocity from $-$6 to $-3$~km~s$^{-1}$ near the position of the H~II region, while diffuse clumps are shown in the panels with positive velocities. The position-velocity map shows an incomplete asymmetric molecular shell around the H~II region. It is notable that the densest part of the heated clumps tracing by C$^{18}$O presents a slightly higher velocity, which is closer to the rest velocity of the H~II region than those tracing by $^{12}$CO and $^{13}$CO. These indicate that the H~II region undergoes an asymmetric expansion within the parent molecular cloud.

The ``\textbf{Caribbean Sea}" is a diffuse extended cloud to the west of GoM at the velocity of 3~km~s$^{-1}$ with low excitation temperature and optical depth. $^{12}$CO are detected in a large area, and weak C$^{18}$O emission can only be detected at a few positions. This region shows the lowest column density among all the regions but its total mass is relatively high.

\subsection{Filamentary Structures}\label{sbs:fil}

\begin{figure}
\centering
\includegraphics[width=0.14\textwidth]{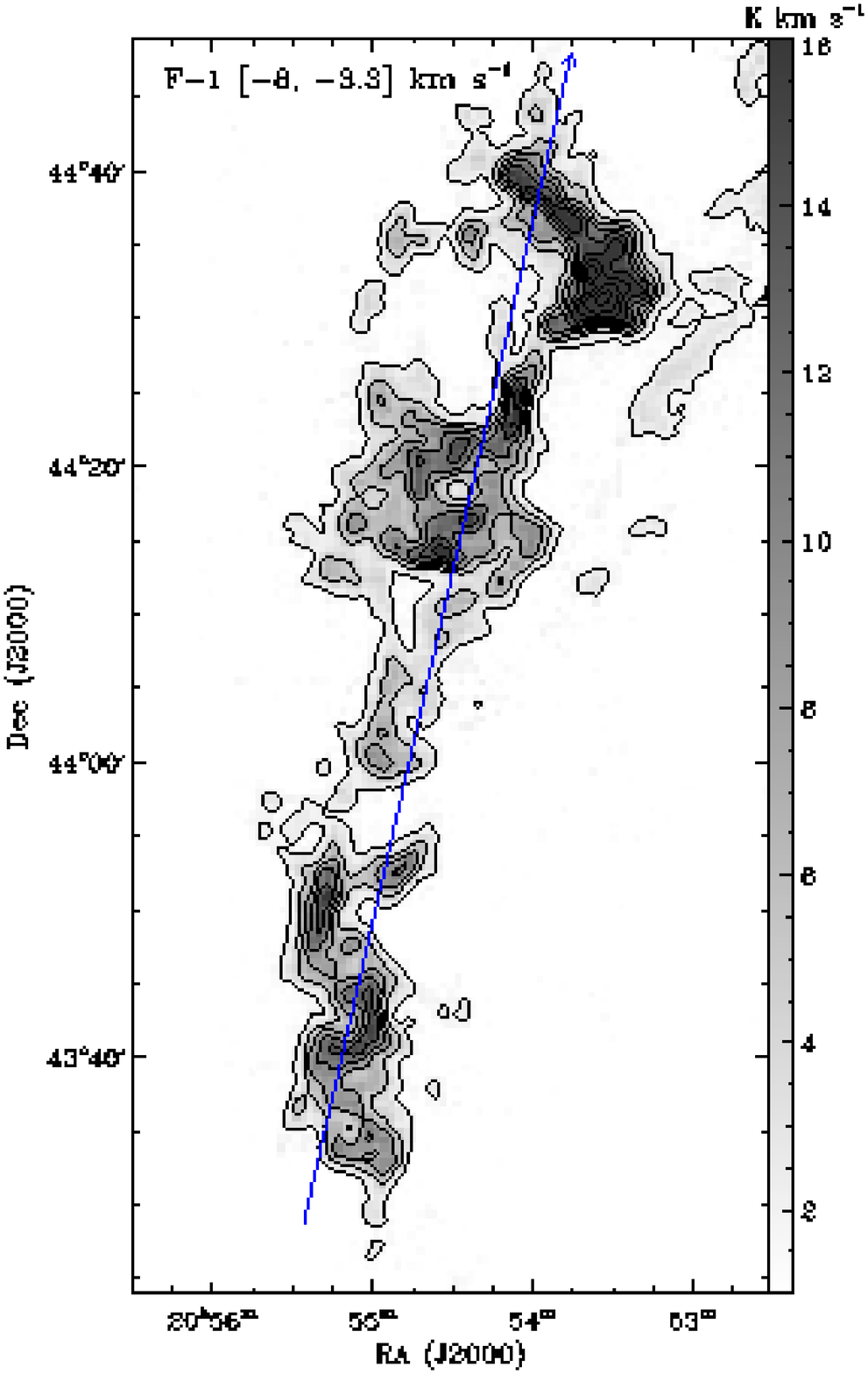}%
\includegraphics[width=0.14\textwidth]{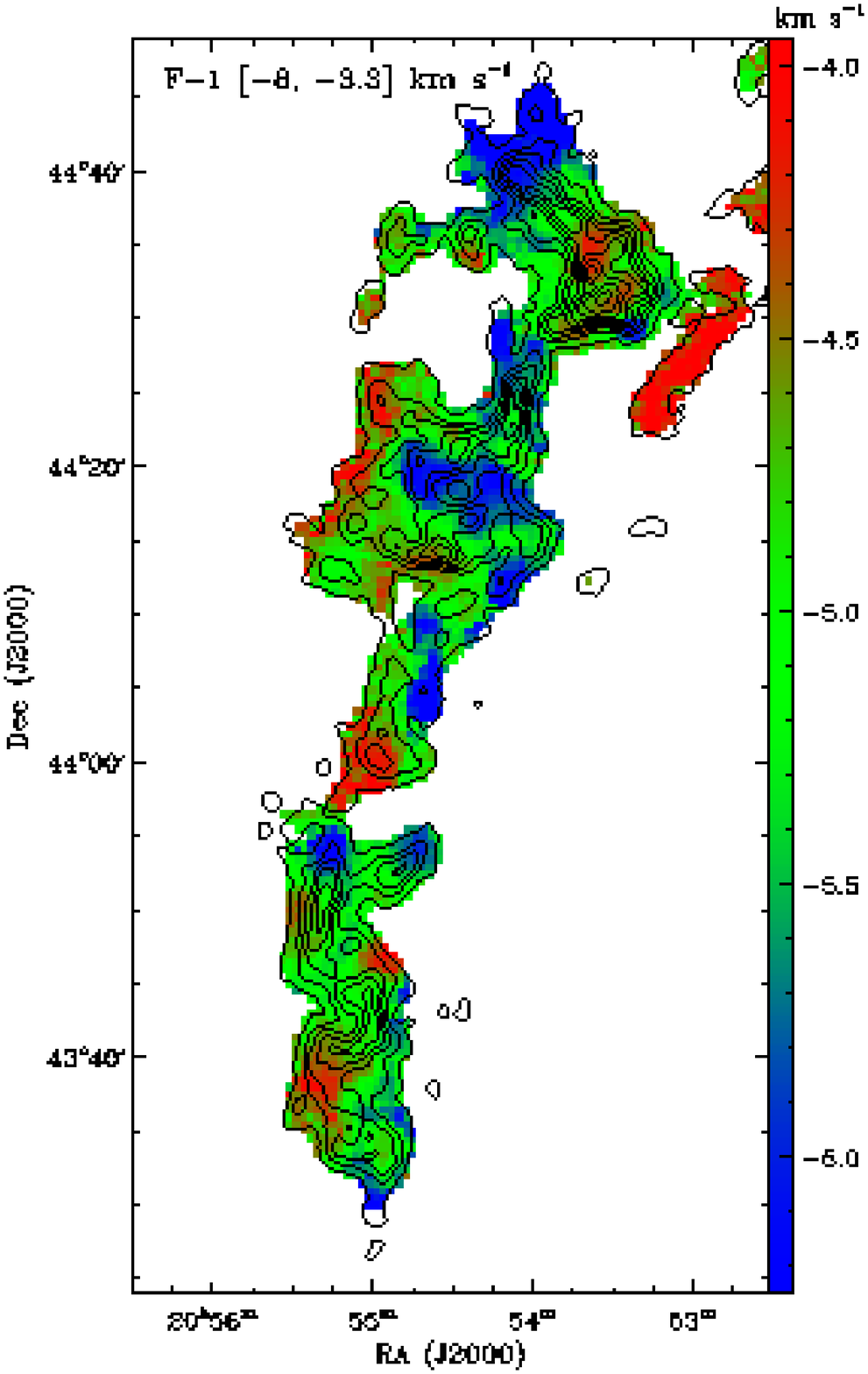}%
\includegraphics[width=0.14\textwidth]{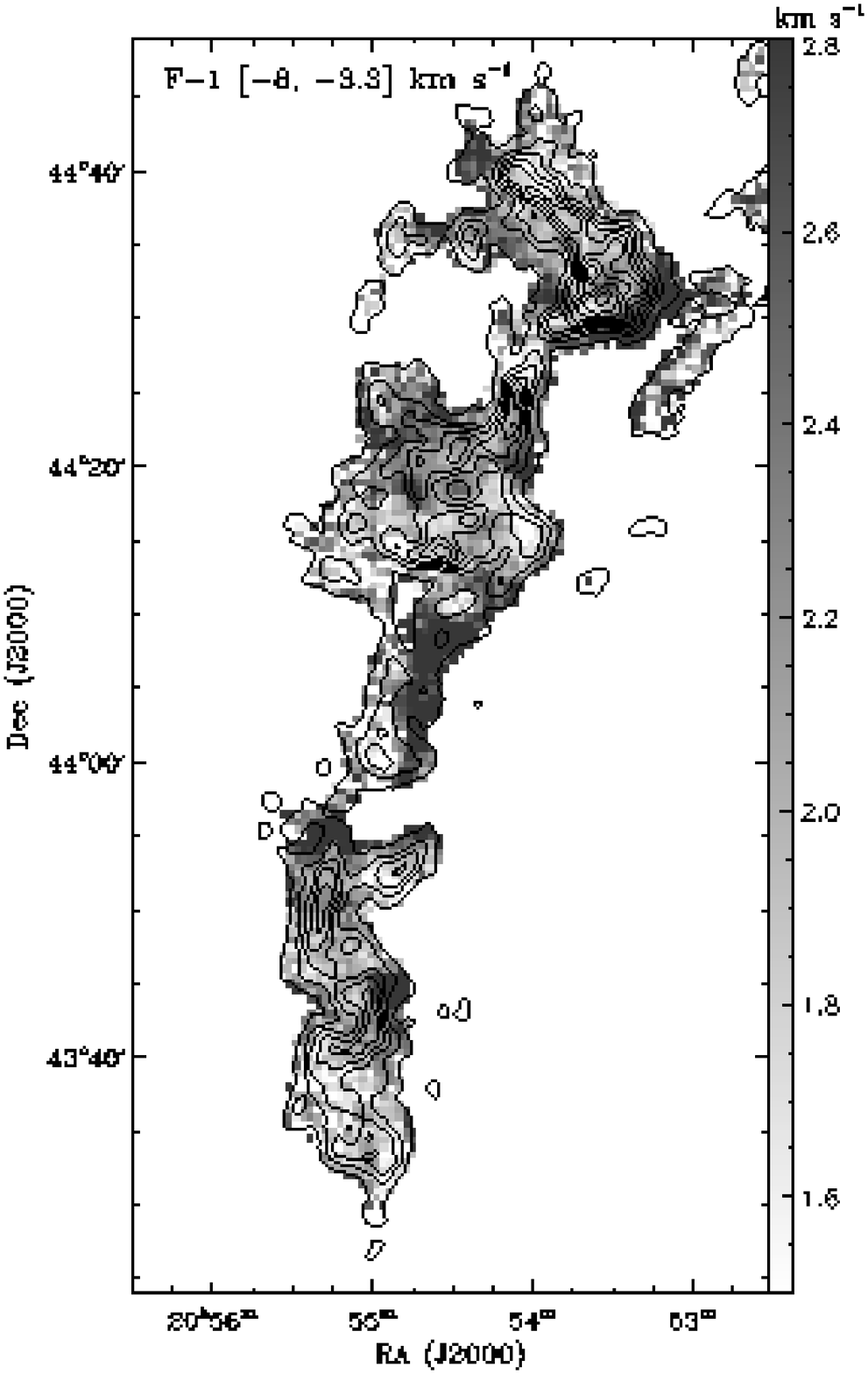}\\
\includegraphics[width=0.14\textwidth]{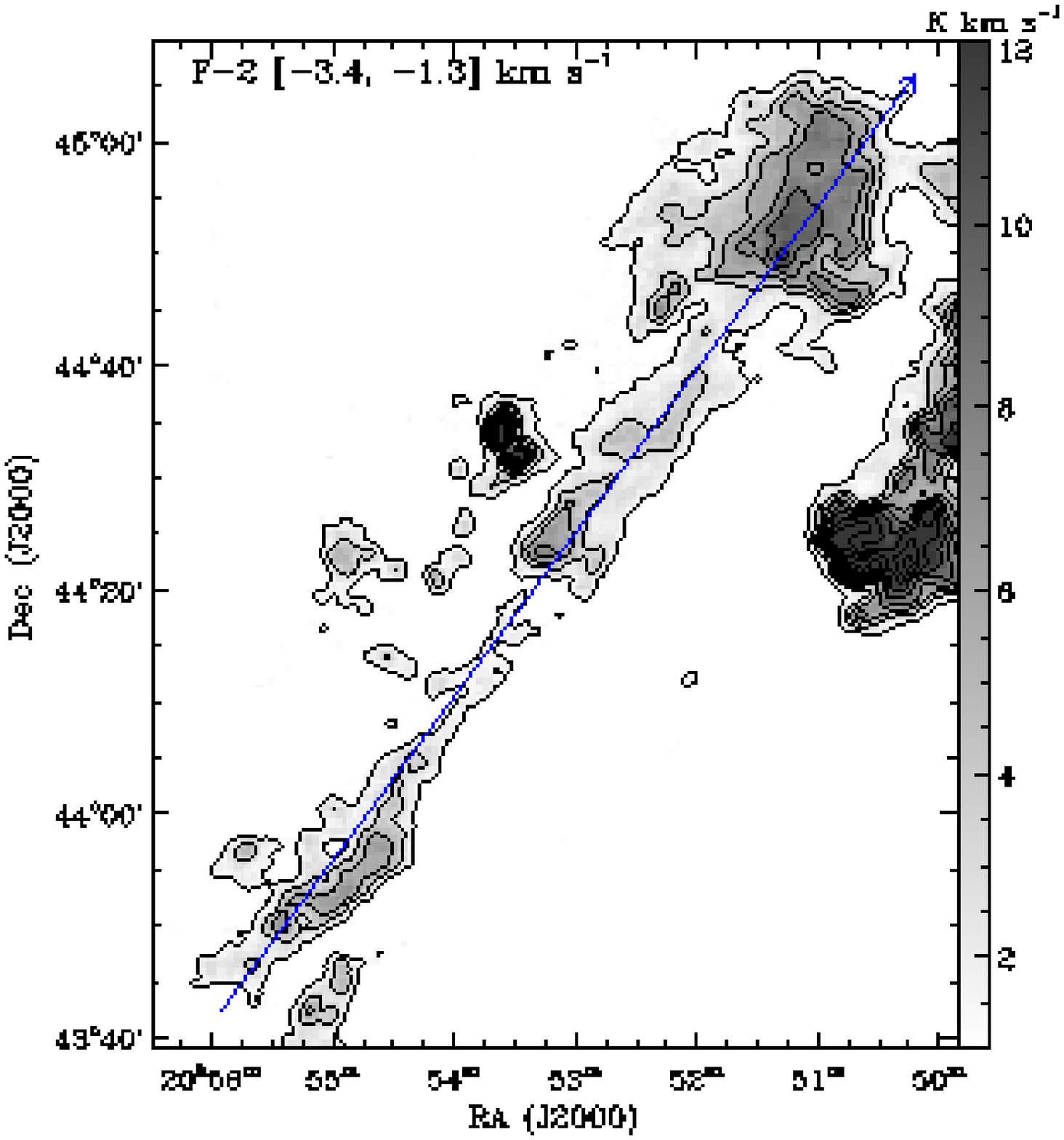}%
\includegraphics[width=0.14\textwidth]{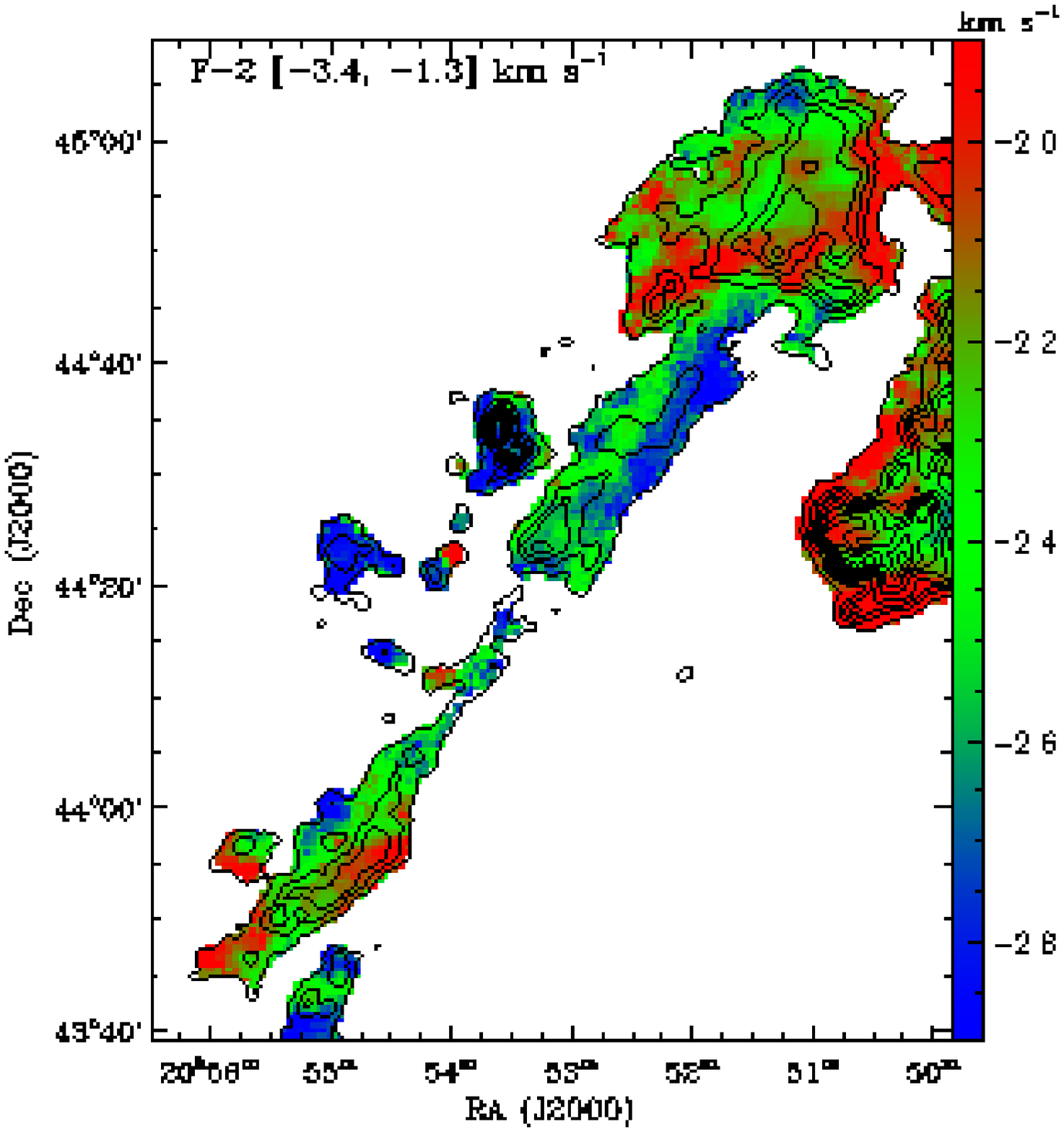}%
\includegraphics[width=0.14\textwidth]{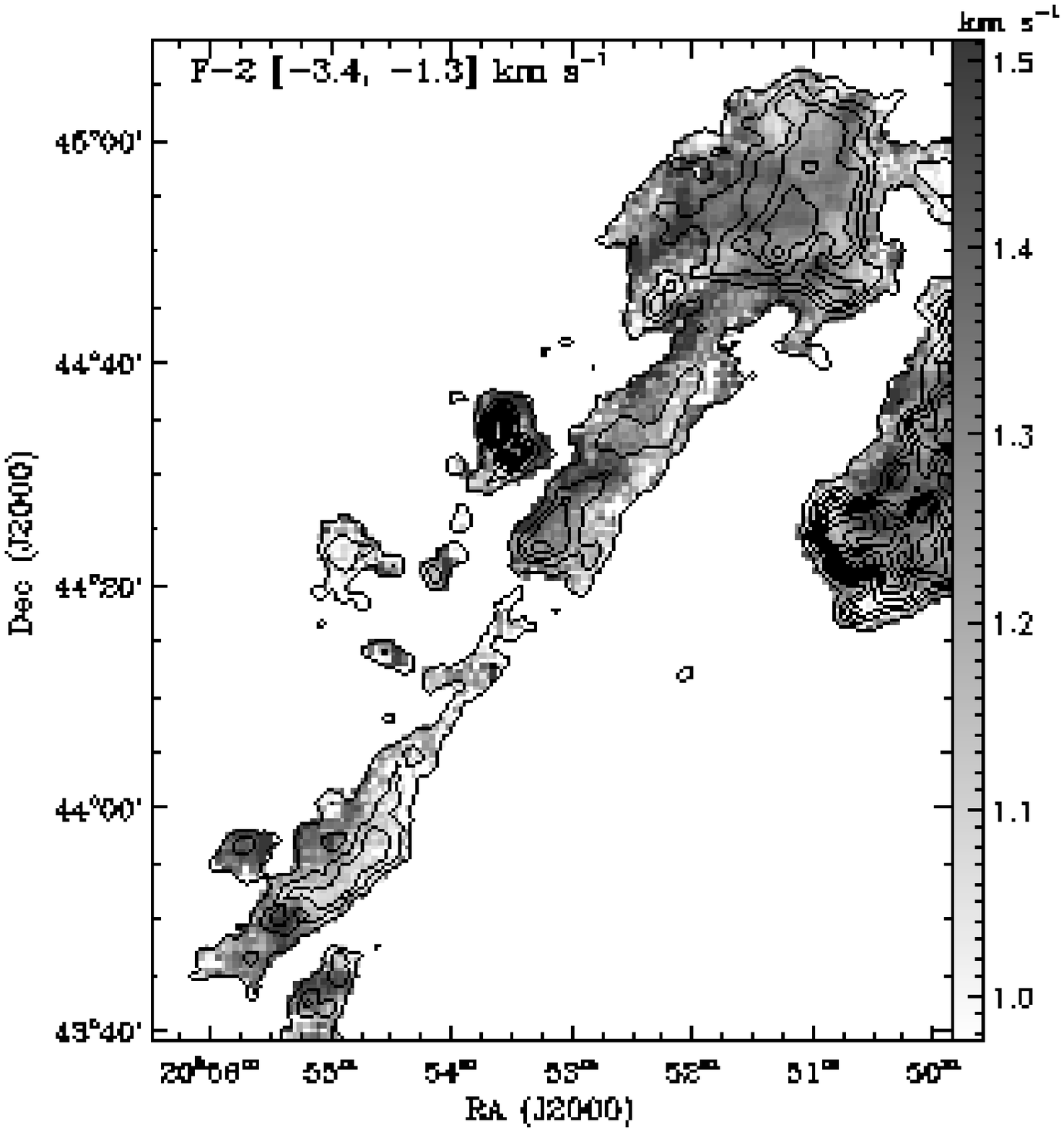}\\
\includegraphics[width=0.14\textwidth]{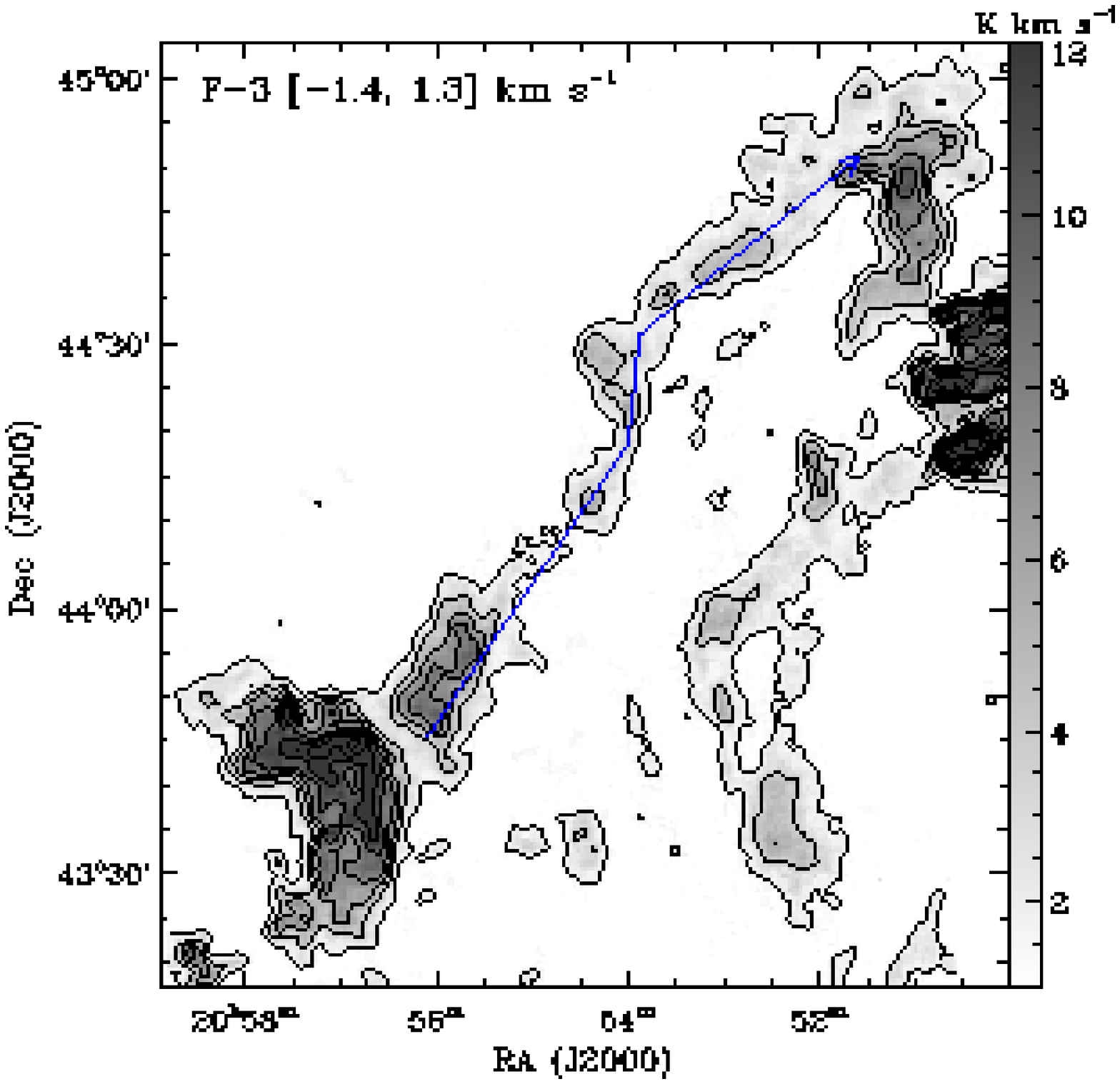}%
\includegraphics[width=0.14\textwidth]{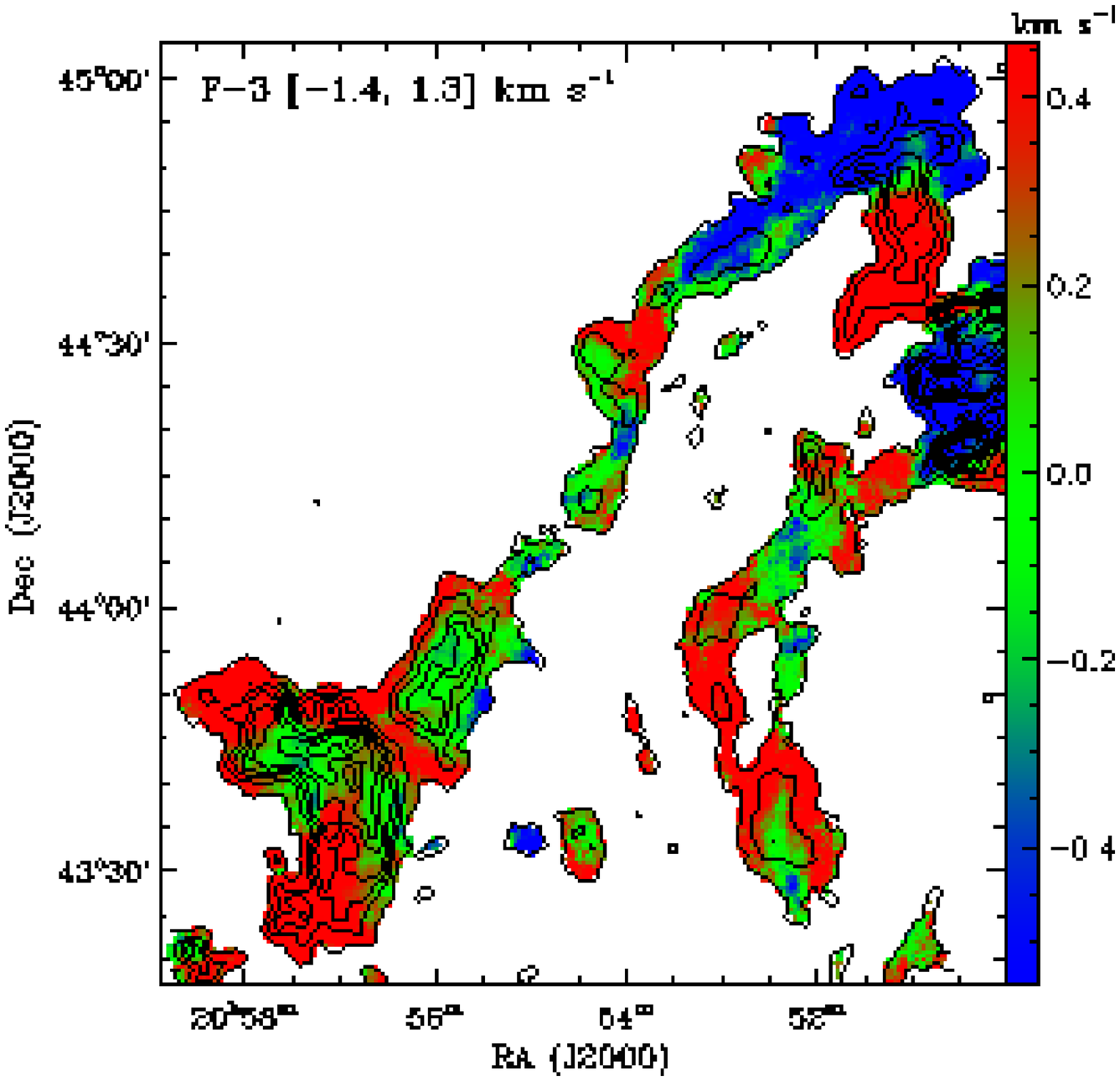}%
\includegraphics[width=0.14\textwidth]{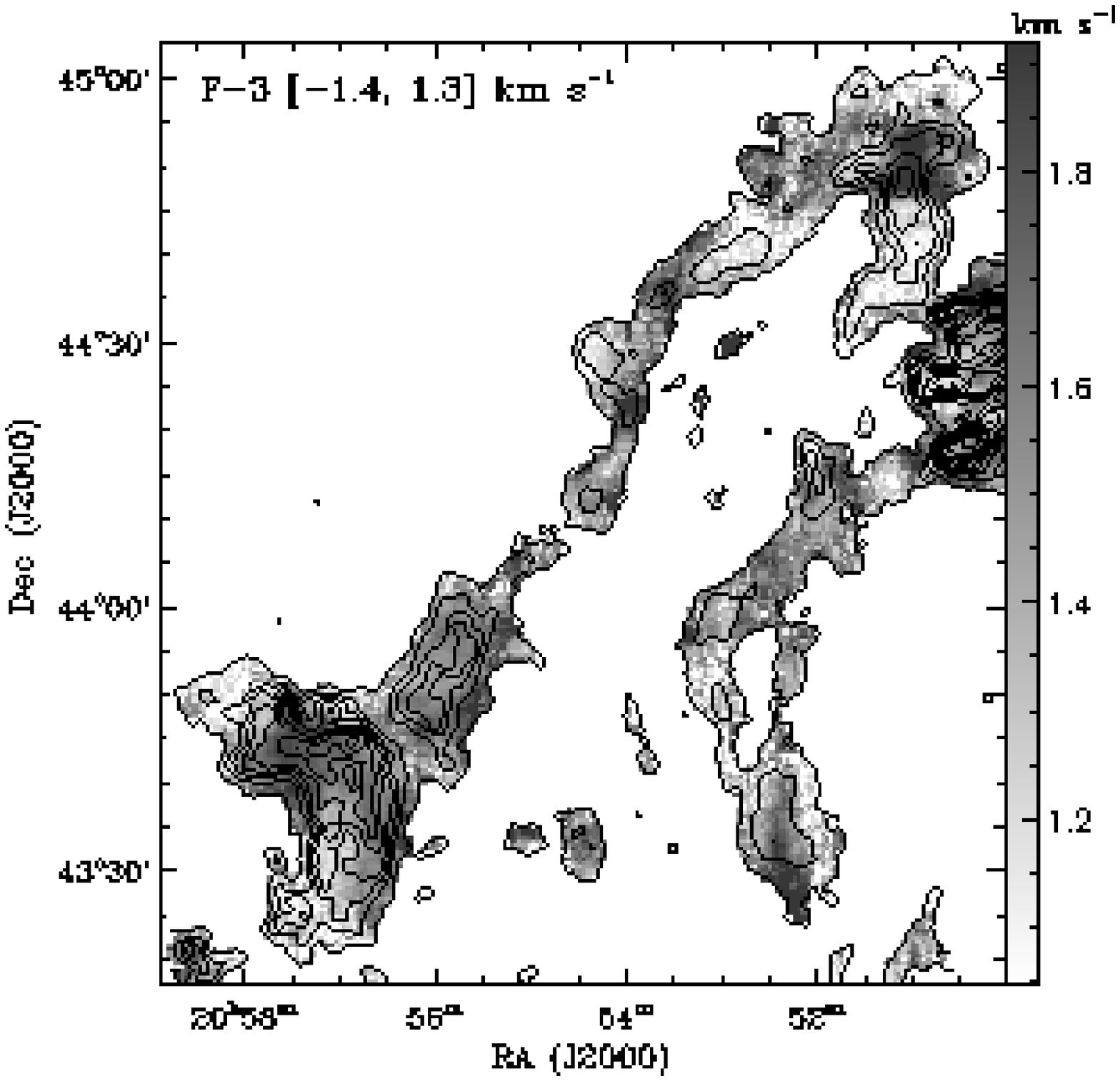}\\
\includegraphics[width=0.14\textwidth]{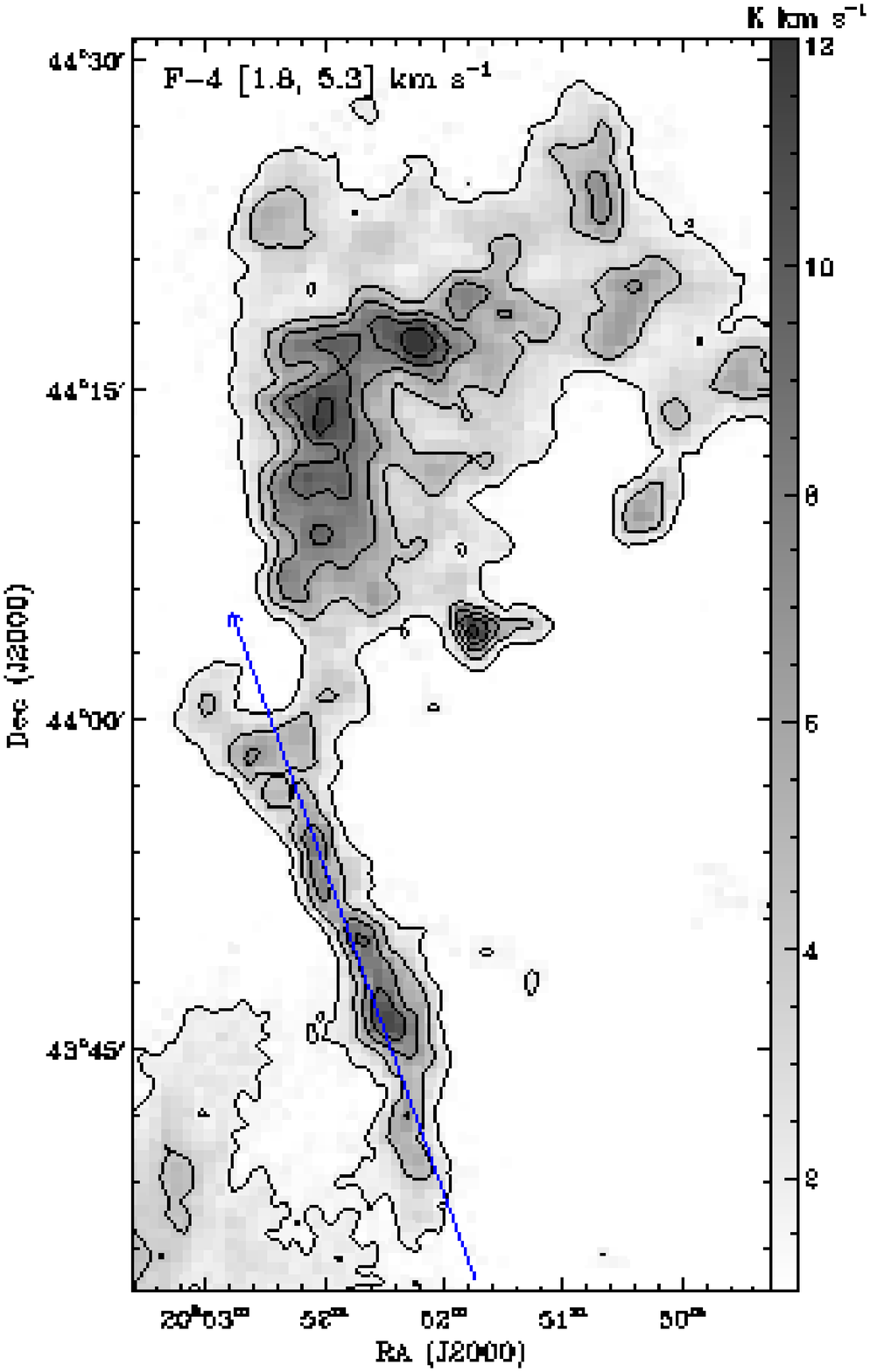}%
\includegraphics[width=0.14\textwidth]{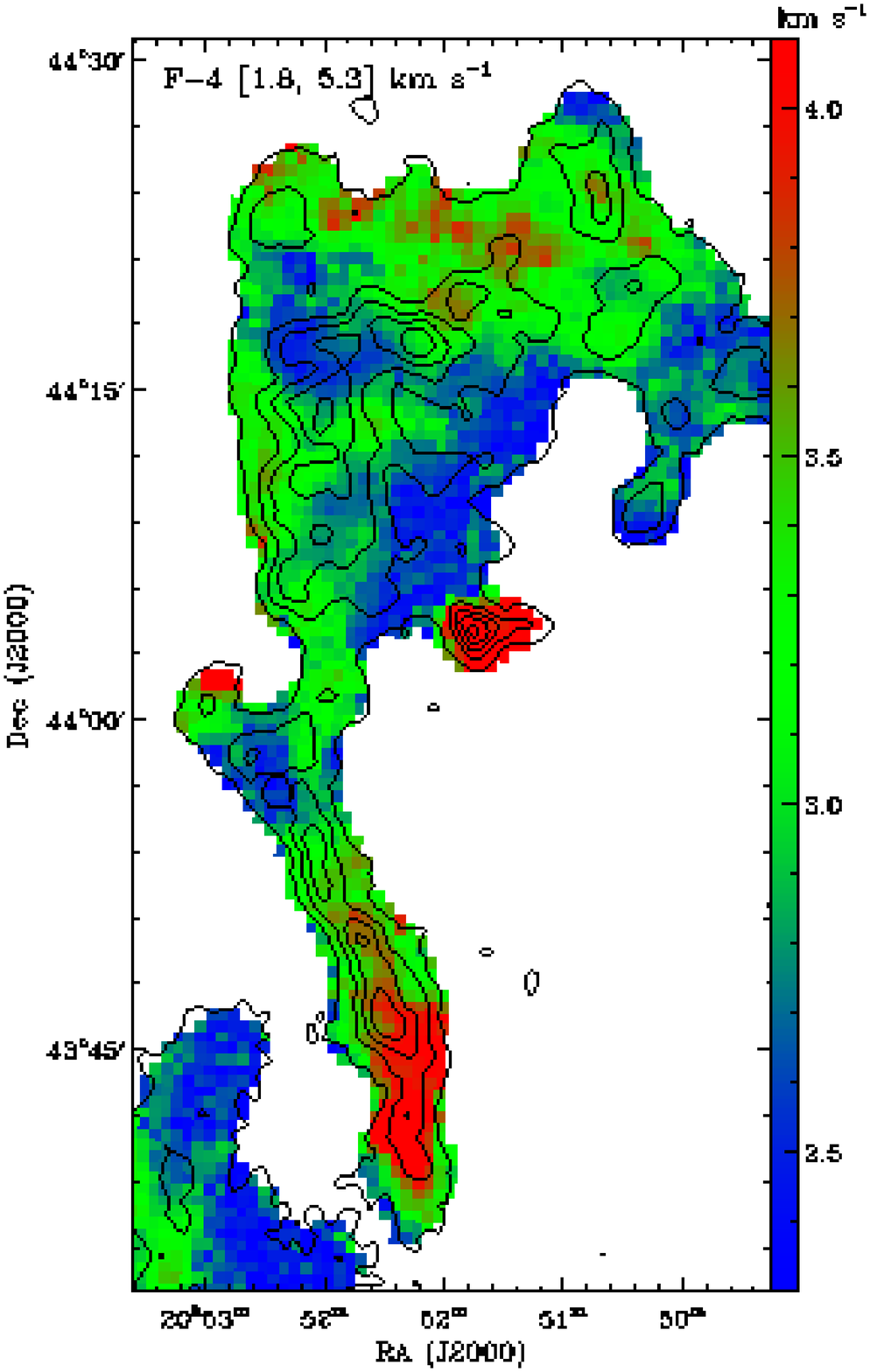}%
\includegraphics[width=0.14\textwidth]{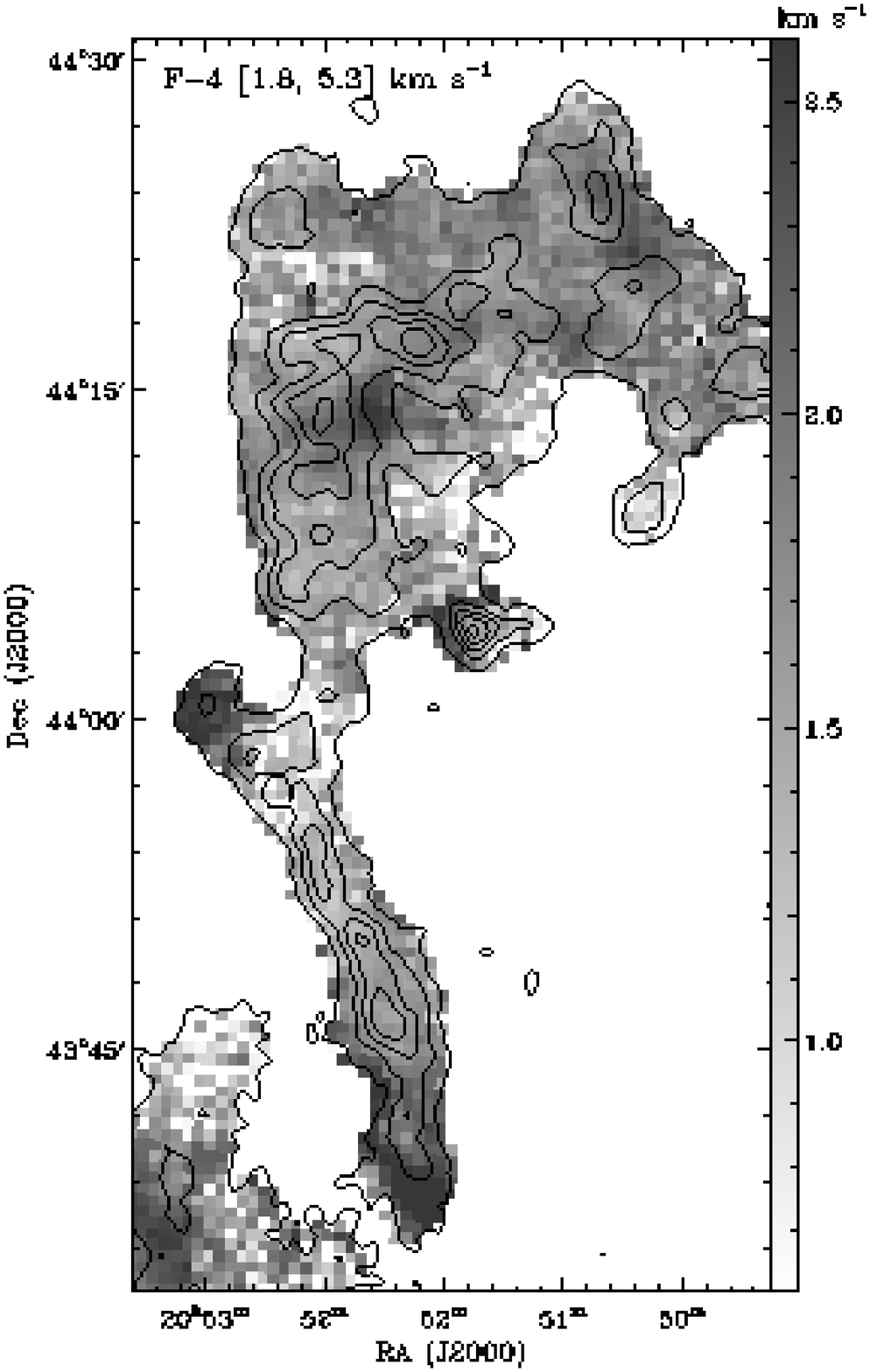}
\caption{$^{13}$CO moment maps of the filaments showing the integrated intensity (left), rest velocity (middle), and line width (right). On each panel, the zeroth moment contour lines are overlaid from 7$\sigma$ with 10$\sigma$ intervals. The name and the velocity range of each filament is marked in the top left corner in each panel.}
\label{fig:filament}
\end{figure}

\begin{figure}
\centering
\scalebox{0.35}{\includegraphics*[3,14][293,557]{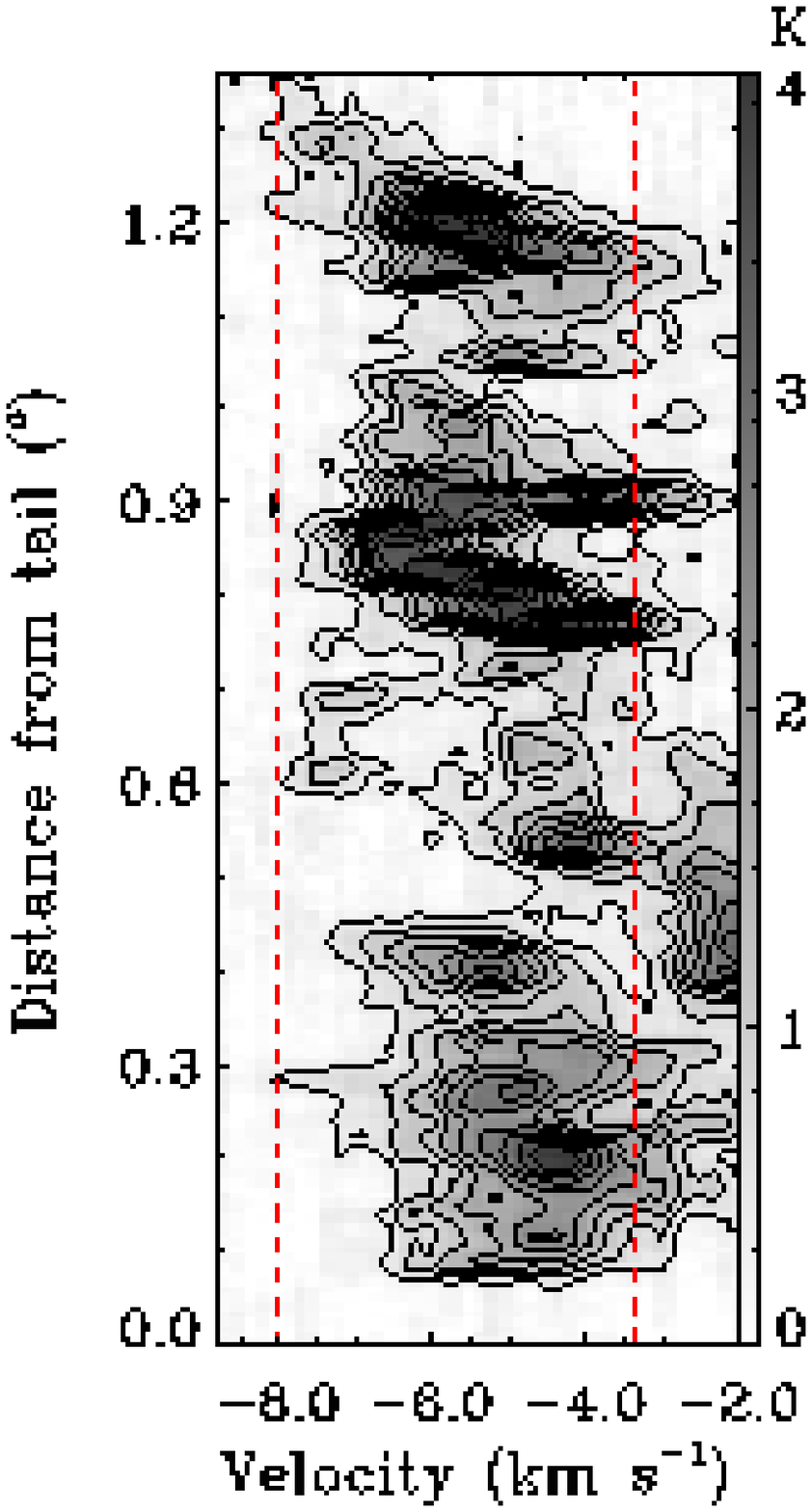}}
\scalebox{0.35}{\includegraphics*[27,14][293,557]{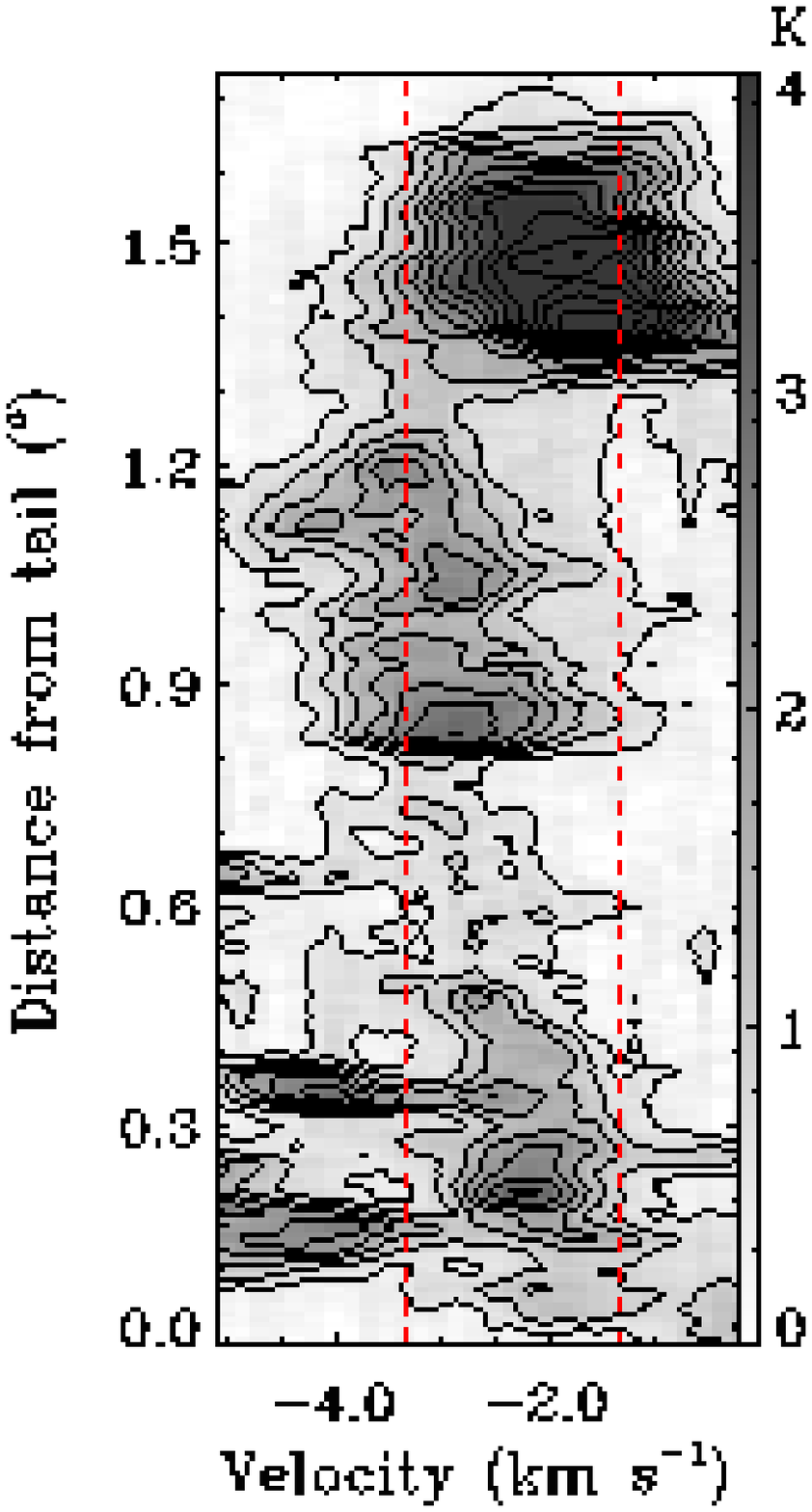}}
\scalebox{0.35}{\includegraphics*[3,14][293,557]{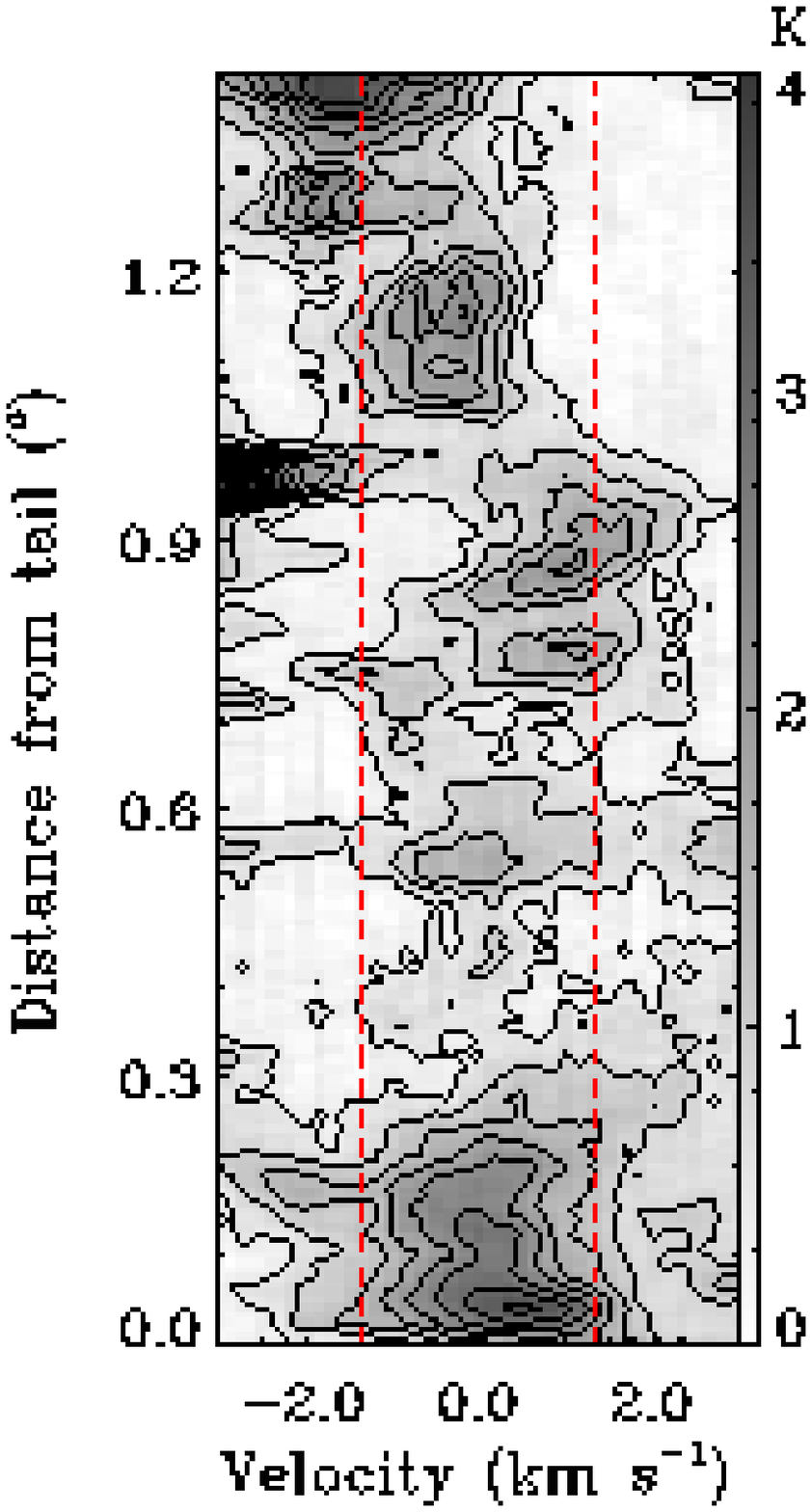}}
\scalebox{0.35}{\includegraphics*[27,14][293,557]{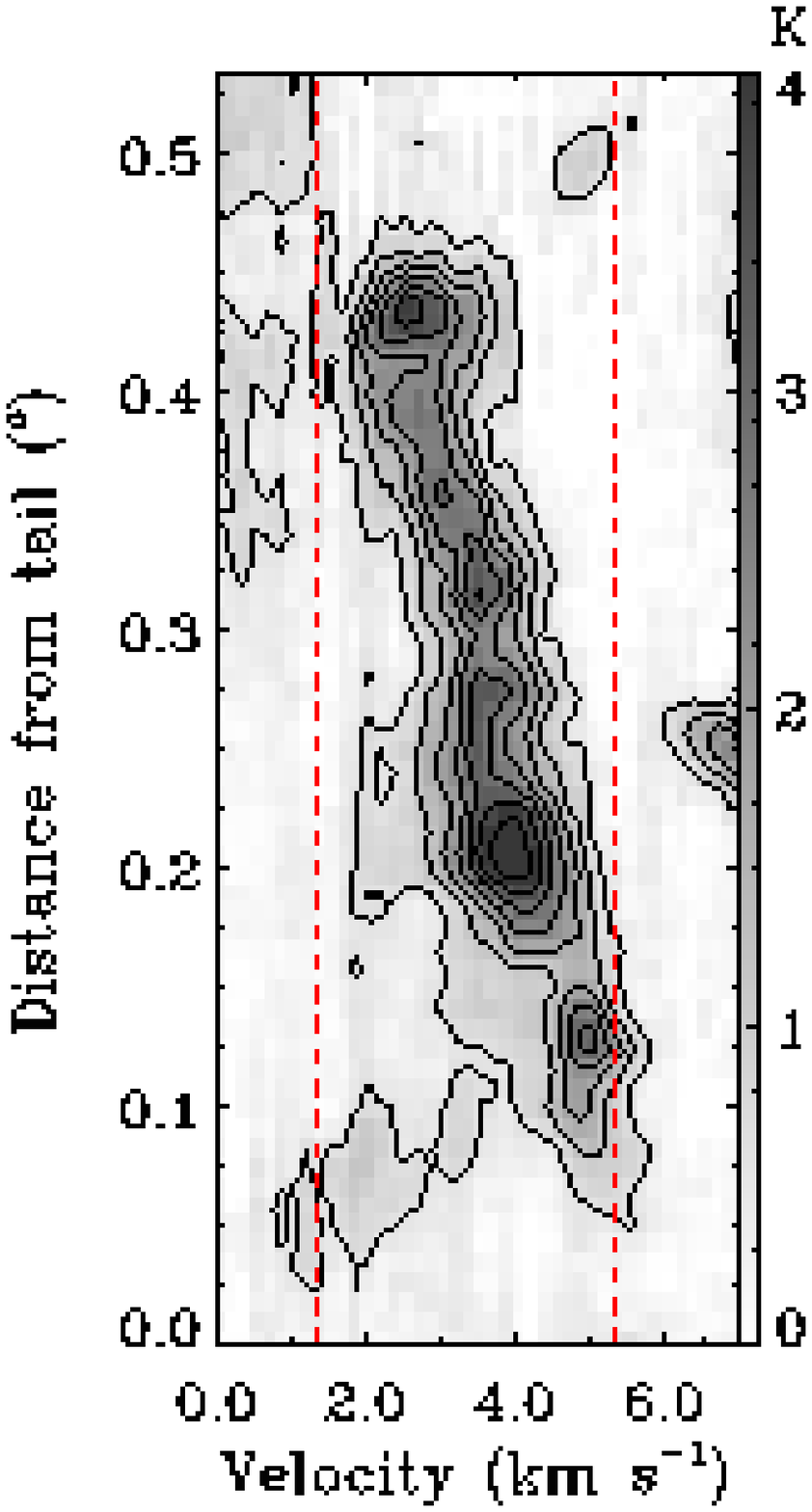}}
\caption{$^{13}$CO position-velocity map of the four filamentary structures along the axes shown in the integrated intensity maps of Figure~\ref{fig:filament}. The lowest contour is 5$\sigma$ and the contour interval is 5$\sigma$ for each panel. Red dashed lines indicate the velocity range of each filaments.}
\label{fig:fpv}
\end{figure}

In our observations with velocity dimensions, we resolve three separate filamentary structures (designated as F-1, F-2, F-3 in ascending velocity order) nearly parallel to each other along the dark lane in the NAN complex. Another filament (F-4) is also resolved near Pelican's Beak region. Figure~\ref{fig:velopeak} shows the positions of the filaments with different color representing their different velocity. Figure~\ref{fig:filament} and \ref{fig:fpv} shows the morphology and velocity structure of these filaments. We found elongated molecular clumps along these filaments. F-1, which contains the bright clumps in Caribbean Islands, presents a complex twisted spatial and velocity structure, with a ring-like structure near $\rm 20^h54^m.5, +44^{\circ}19'$. F-2 and F-3 are discontinuous, and together with the Pelican's Hat region, they form a hub-filament structure \citep{mye09}. The northwest and the southeast section of F-2 show opposite velocity gradient directions. The northwest section of F-3 bends to the east with higher velocity and surrounds the Pelican Cluster. Both F-2 and F-4 show clear velocity gradient along their axes in the position-velocity map.

\begin{figure}
\centering
\includegraphics[width=0.35\textwidth]{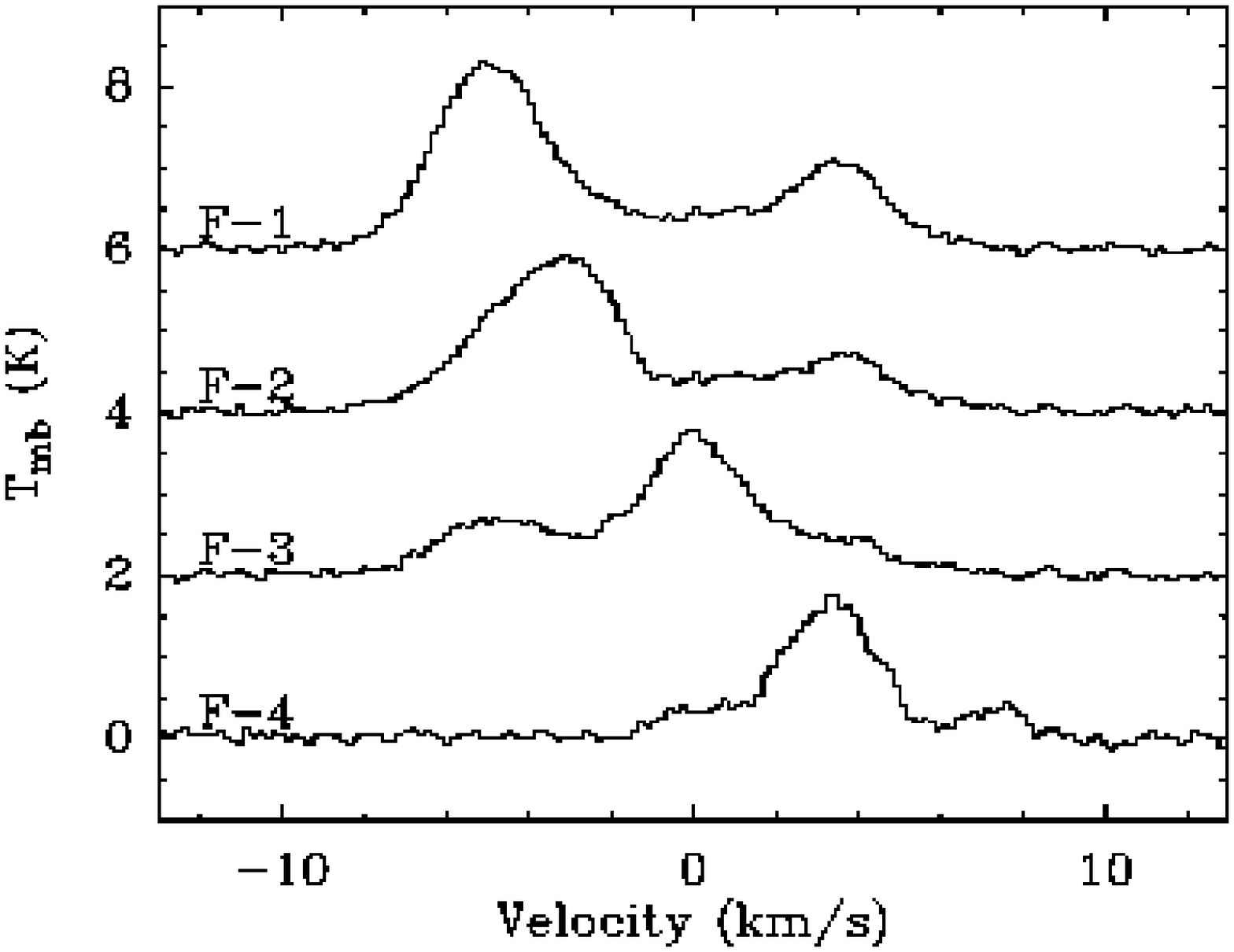}
\caption{Position averaged $^{13}$CO spectra of the filaments. Spectra are moved upwards for clarity. The name of each filament is marked on the left of each spectrum.}
\label{fig:fspec}
\end{figure}

We show the averaged spectra in Figure~\ref{fig:fspec} and some physical properties of the filaments are listed in Table~\ref{tbl:filament}. A typical $^{13}$CO line width of 3.3~km~s$^{-1}$ is shown. These filamentary structures show similar optical depths, while F-1 and F-4 have a higher excitation temperature. We could estimate the mass per unit length by dividing the mass of filaments by their spatial dimension. F-1 shows a higher mass per unit length than that in the other filaments. The twisted structure in F-1 may cause an overestimation of this measurement. A maximum, critical linear mass density needed to stabilize a cylinder structure can be calculated with $(M/l)_{\rm max} = 84(\Delta v)^2M_\odot {\rm pc}^{-1}$ in the turbulent support case, where $\Delta v$ is the line width in unit of km~s$^{-1}$ \citep{jac10}. This means our filaments are gravitationally stable on the assumption of the $^{13}$CO abundance we adopted.

\begin{deluxetable}{cccccc}
\tabletypesize{\scriptsize}
\tablecolumns{6}
\tablewidth{0pc}
\tablecaption{Properties of filaments\label{tbl:filament}}
\tablehead{
\colhead{Filament} &
\colhead{$T_{\rm ex}$} &
\colhead{$\Delta v$($^{13}$CO)} &
\colhead{$\tau$($^{13}$CO)} &
\colhead{$M$} &
\colhead{$M/l$}
\\
\colhead{} &
\colhead{(K)} &
\colhead{(km~s$^{-1}$)} &
\colhead{} &
\colhead{($M_\odot$)} &
\colhead{($M_\odot$ pc$^{-1}$)}
}
\startdata
F-1 & 16 & 3.20 & 0.33 & 1401 & 107 \\
F-2 & 12 & 3.77 & 0.34 & 416 & 30 \\
F-3 & 12 & 3.52 & 0.36 & 487 & 32 \\
F-4 & 17 & 2.75 & 0.26 & 196 & 38 \\
\enddata
\tablecomments{The properties of the filaments in the NAN complex, including excitation temperature, line width of averaged spectra, optical depth of $^{13}$CO, mass, and mass per unit length. These typical values are the results averaged within the 10$\sigma$ contour line of each filament.}
\end{deluxetable}

\subsection{Clump Identification}

\begin{figure}[b]
\centering
\includegraphics[width=0.4\textwidth]{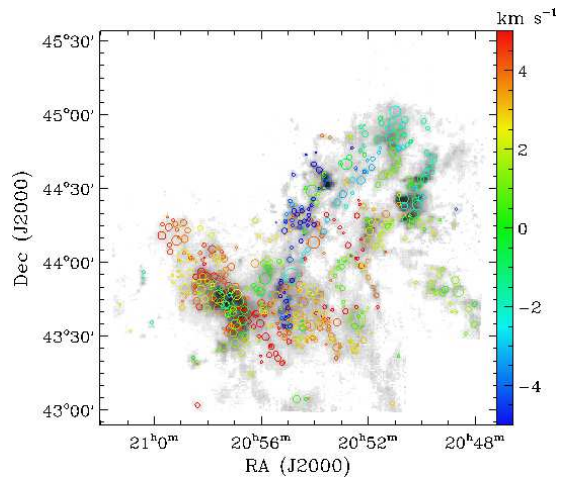}
\caption{Clumps identified in $^{13}$CO data cube. The circles indicate the clump positions on the integrated intensity map of  $^{13}$CO. The colors of the circles represent the velocities the clumps, while the circles are scaled according to the sizes of clumps.}
\label{fig:clump}
\end{figure}

We use the FINDCLUMPS tool in the CUPID package (a library of Starlink package) to identify molecular clumps in the obtained $^{13}$CO FITS cube. The ClumpFind algorithm is applied in the process of identification. The algorithm first contours the data and searches for peaks to locate the clumps, and then follows them down to lower intensities. We set the parameters TLOW=5$\times$RMS and DELTAT=3$\times$RMS, where TLOW determines the lowest level to contour a clump,  and DELTAT represents the gap between contour levels which determines the lowest level at which to resolve merged clumps \citep{wil94a}. The parameters of each clump, such as the position, velocity, size in RA and Dec directions, and one-dimensional velocity dispersion, are directly obtained in this process. The clump size has removed the effect of beam width, and velocity dispersion is also de-convolved from the velocity resolution. The morphology of the clumps are checked by eye within the three-dimension RA-Dec-velocity space to pick out clumps with meaningful structures. We then mark every clump on their velocity channel in the $^{13}$CO cube to confirm the morphology and emission intensity of the molecular gas within the clumps. In addition, clumps with pixels that touch the edge of the data cube are removed. 22 clumps are removed in these checking steps. Eventually, a total of 611 clumps are identified, and the position, velocity, and size of the clumps as illustrated in Figure~\ref{fig:clump} are consistent with the spatial and velocity distribution of the molecular gas.

\begin{figure}
\centering
\includegraphics[width=0.14\textwidth]{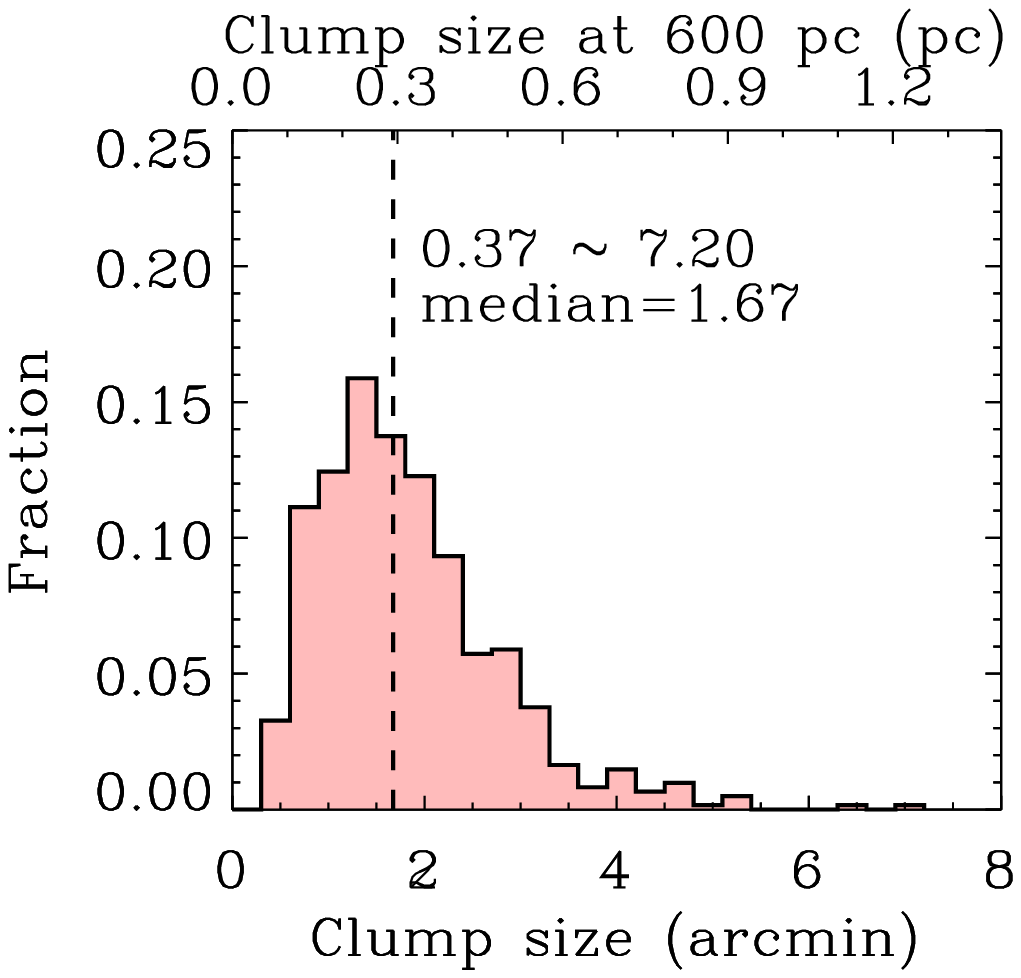}
\includegraphics[width=0.14\textwidth]{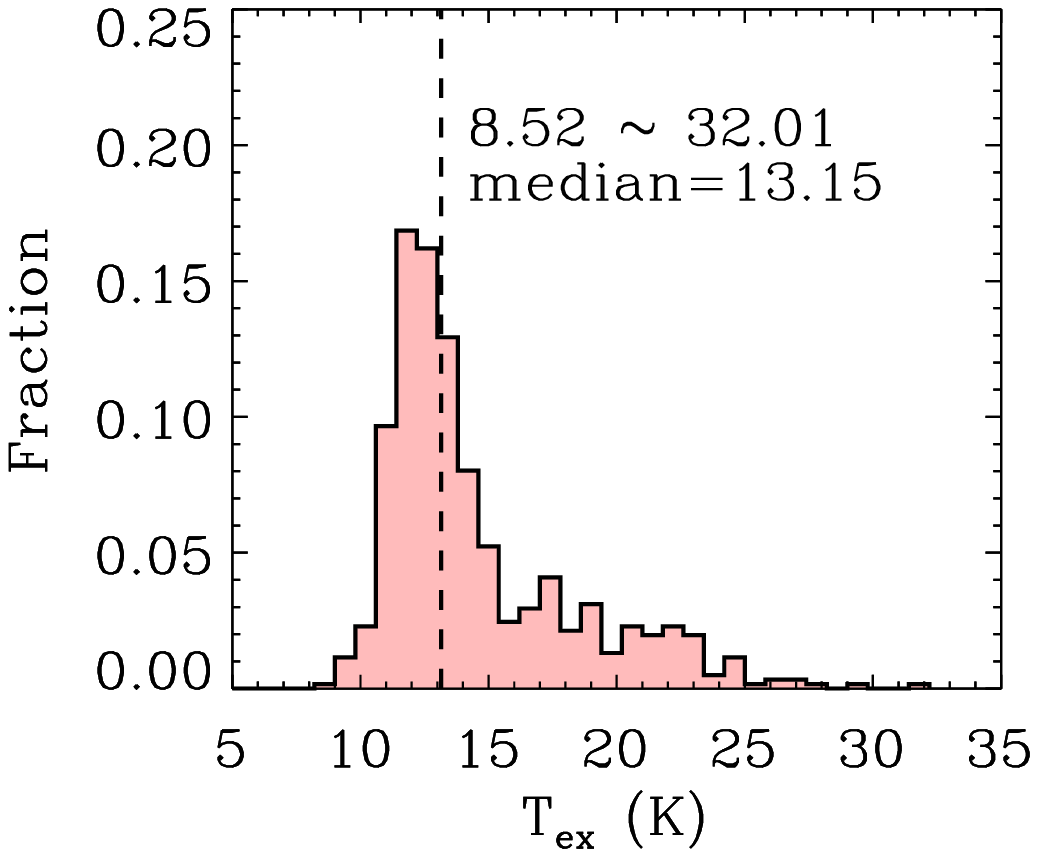}
\includegraphics[width=0.14\textwidth]{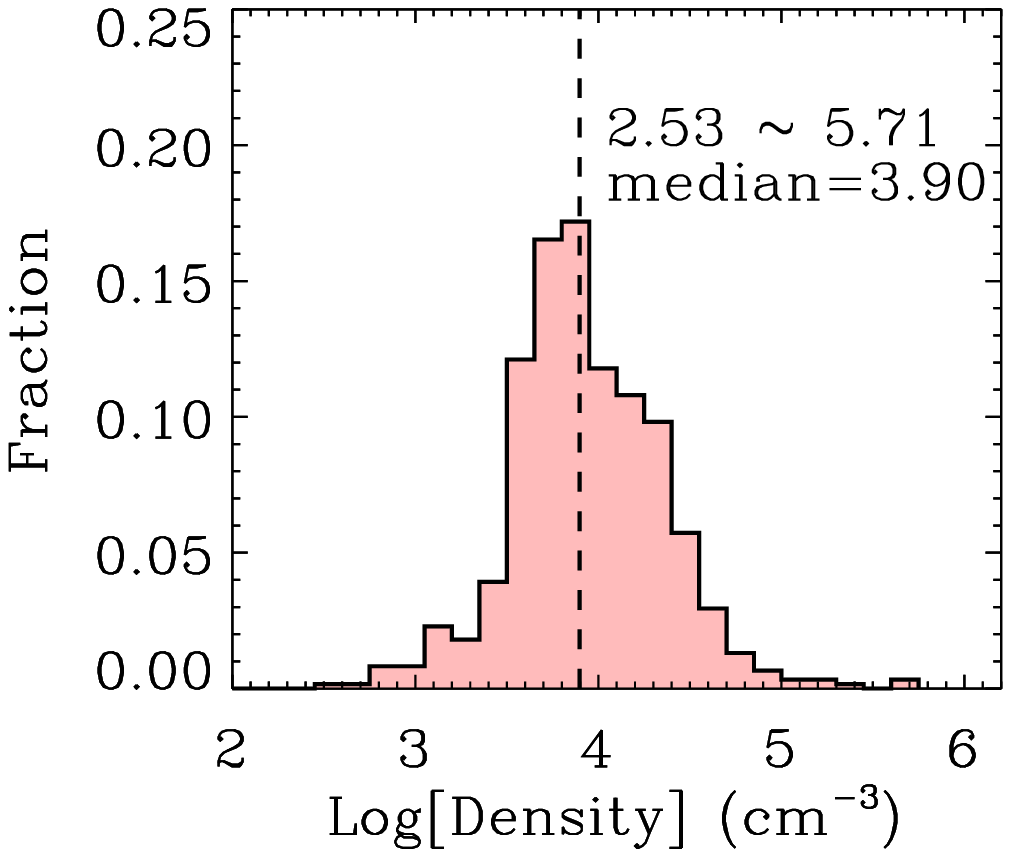}
\caption{Distribution of clump size (left), excitation temperature (middle), and volume density (right). The size is the geometric mean of the size in the R.A. and Dec. direction, and density is derive under the spherical assumption. The range and typical value of each property are marked on each plot.}
\label{fig:clumpstat}
\end{figure}

\begin{figure}
\centering
\includegraphics[width=0.45\textwidth]{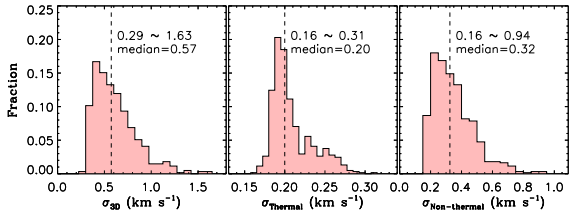}
\caption{Histogram of three-dimensional velocity dispersion (left), and the thermal (middle) and non-thermal (right) one dimensional velocity dispersions. The range and typical value of each velocity dispersion are marked on each panel.}
\label{fig:velodisp}
\end{figure}

We extract the excitation temperature for each clump from their $^{12}$CO datacube under LTE assumption and can then derive the LTE mass. The parameters of the clumps are listed in Table~\ref{tbl:clump}. The clump size are derived from the geometric mean of the clump size in two directions. Figure~\ref{fig:clumpstat} shows the distributions of clump size, excitation temperature, and volume density, which yield typical properties of $\sim$0.3~pc, 13~K, and 8$\times10^3$~cm$^{-3}$, respectively. The left panel in Figure~\ref{fig:velodisp} shows the distribution of three-dimensional velocity dispersion estimated as $\sigma_{v \rm 3D} = \sqrt{3} \sigma_{v \rm 1D}$. The thermal portion in the velocity dispersion is $\sigma_{\rm Thermal} = \sqrt{k T_{\rm kin} / m}$, where $k$ is the Boltzmann constant, $m$ is the mean molecular mass, and $T_{\rm kin}$ is the kinetic temperature equal to the excitation temperature, while the non-thermal portion is $\sigma_{\rm Non-thermal}=\sqrt{\sigma_{v \rm 1D}^2-\sigma_{\rm Thermal}^2}$. The distribution of the thermal and non-thermal velocity dispersion are shown in Figure~\ref{fig:velodisp}. There are 568 (93\%) clumps with $\sigma_{\rm Non-thermal}$ larger than $\sigma_{\rm Thermal}$. The mean ratio of $\sigma_{\rm Non-thermal}$ and $\sigma_{\rm Thermal}$ is 1.57. This suggests that non-thermal broadening mechanisms (e.g., rotation, turbulence, etc) play a dominant role in the clumps.

\begin{deluxetable*}{ccccrrcccccccc}
\tabletypesize{\scriptsize}
\tablecolumns{14}
\tablewidth{0pc}
\tablecaption{Properties of clumps\label{tbl:clump}}
\tablehead{
\colhead{Clump} &
\colhead{R.A.} &
\colhead{Dec.} &
\colhead{Velocity} &
\colhead{$\Delta{\rm R.A.}$} &
\colhead{$\Delta{\rm Dec.}$} &
\colhead{$R$} &
\colhead{$\delta_{v \rm 1D}$} &
\colhead{$T_{\rm peak}$} &
\colhead{$T_{\rm ex}$} &
\colhead{$\Sigma$} &
\colhead{$n_{\rm H_2}$} &
\colhead{$M_{\rm LTE}$} &
\colhead{$\alpha_{\rm Vir}$}
\\
\colhead{} &
\colhead{(J2000)} &
\colhead{(J2000)} &
\colhead{(km~s$^{-1}$)} &
\colhead{(\arcsec)} &
\colhead{(\arcsec)} &
\colhead{(pc)} &
\colhead{(km~s$^{-1}$)} &
\colhead{(K)} &
\colhead{(K)} &
\colhead{($M_\sun \rm pc^{-2}$)} &
\colhead{$10^3\rm cm^{-3}$} &
\colhead{($M_\odot$)} &
\colhead{}
}
\startdata
  1 & 20 48 01.2 & +43 42 59.4 &  +0.01 & 115.9 & 120.7 & 0.17 & 0.35 &  4.40 & 16.67 &  23.9 &  11.2 &  15.6 &  1.52 \\
  2 & 20 48 01.6 & +43 34 43.8 &  +1.13 & 139.2 & 178.4 & 0.23 & 0.35 &  4.57 & 17.73 &  31.9 &  10.1 &  33.2 &  0.98 \\
  3 & 20 48 15.2 & +43 40 59.4 &  +1.50 & 128.9 & 160.3 & 0.21 & 0.24 &  3.62 & 15.32 &  17.0 &   5.2 &  13.1 &  1.07 \\
  4 & 20 48 20.6 & +43 31 04.7 &  +0.99 &  56.5 &  94.2 & 0.10 & 0.28 &  3.27 & 14.66 &   6.9 &   5.6 &   1.8 &  5.00 \\
  5 & 20 48 37.6 & +43 48 11.8 &  +1.01 & 194.6 & 319.4 & 0.36 & 0.56 &  3.73 & 18.72 &  40.1 &   5.7 &  74.4 &  1.74 \\
  6 & 20 48 41.2 & +43 39 38.6 &  +1.67 &  52.0 &  31.2 & 0.06 & 0.21 &  7.21 & 14.41 &   9.8 &  29.7 &   1.6 &  1.80 \\
  7 & 20 48 42.3 & +44 21 38.6 &  $-$4.26 &  47.5 &  65.4 & 0.08 & 0.36 &  5.19 & 17.79 &  15.3 &  26.4 &   3.9 &  3.09 \\
  8 & 20 48 44.2 & +43 40 15.1 &  +2.25 &  32.2 &  79.2 & 0.07 & 0.19 &  6.69 & 13.16 &   7.2 &  10.6 &   1.2 &  2.58 \\
  9 & 20 48 44.8 & +43 52 52.4 &  +1.47 & 169.5 & 193.1 & 0.26 & 0.27 &  3.61 & 16.00 &  17.9 &   3.0 &  14.8 &  1.46 \\
 10 & 20 48 46.3 & +44 15 17.7 &  +1.95 &  45.3 &  80.5 & 0.09 & 0.40 &  6.49 & 24.47 &  38.2 &  53.3 &   9.9 &  1.63 \\
\enddata
\tablecomments{The properties of the clumps in the NAN complex. Columns are clump number, clump position (R.A. and Dec.), rest velocity, clump size in R.A. and Dec. direction, clump radius, one-dimensional velocity dispersion, temperature of emission peak, excitation temperature, surface density, volume density, LTE mass, and virial parameter. The entire table is published in its entirety in the electronic edition. A portion is shown here for guidance regarding its form and content.}
\end{deluxetable*}

\section{Discussion}\label{sec:dis}

\subsection{Comparison with Other Star Formation Regions}

In a typical low-mass star-forming region, the Taurus region, \citet{gol08} gives a LTE column density of $\le 10^{22}$~cm$^{-2}$ for the most dense region based on the data from the Five College Radio Astronomy Observatory (FCRAO) survey \citep{nar08}. This column density is lower than the averaged column density we derived in several dense regions of the NAN complex. \citet{qia12} searched for clumps in the Taurus survey data and derived a typical mean H$_2$ density of $\sim$2000 cm$^{-3}$, lower than the clump density in NAN complex of 8000~cm$^{-3}$. In addition, we found a number of clumps with densities over 10$^4$~cm$^{-3}$, which is hardly seen in the Taurus region. The $^{13}$CO line width (0.4-2.2~km~s$^{-1}$) in NAN complex is also slightly higher than that in Taurus (0.5-1.7~km~s$^{-1}$). These may indicate potential massive stars are forming in some of the dense clumps in the NAN complex.

In an active high-mass star-forming region, such as the Orion Nebula, a survey of the Orion A region by \citet{nag98} yielded an averaged column density similar to our results. Their survey found regions with excitation temperature $\ge$~20~K in most areas, and an even higher temperature of $\ge$~60~K in the Orion~KL region. Meanwhile, the high temperature regions in the NAN complex are limited to those around the Pelican Cluster. In fact, the statistical properties of the clumps we identified in the NAN complex are similar to those of the Planck cold dense core \citep{pla11} of the Orion complex as studied by \citet{liu12}. These results suggest most of the clumps, especially the cold ones, in the NAN complex are in an early evolutionary stage of star formation dominated by a non-thermal environment.

\subsection{Gravitational Stability of the Clumps}

The gravitational stability of clump determines whether the molecular clump could further collapse and form a star cluster. We firstly calculate the escape velocity ($v_{\rm escape}=\sqrt{2GM_{\rm LTE}/R}$) for each clump and compare with its three-dimensional velocity dispersion. The escape velocities range from 0.21 to 2.84~km~s$^{-1}$ with a typical value of 0.64~km~s$^{-1}$. About 493 (72\%) clumps have velocity dispersion smaller than escape velocity, and only 8 (1\%) clumps have velocity dispersion larger than twice the escape velocity. We note that the clumps with high $\sigma_{v \rm 3D}$ to $v_{\rm escape}$ ratios are faint with low emission peaks.

By simply assuming the clumps have a density profile of $\rho(r)=r^{-k}$ with power-law index $k=1$, we could further derive the virial mass using the standard equation (e.g. \citealt{sol87, eva99}): $M_{\rm Vir}=1164R\sigma_{v \rm 1D}^2 [M_\odot]$, where the clump size $R$ is in pc, and three-dimensional velocity dispersion $\sigma_{v \rm 1D}$ is in km~s$^{-1}$. A steeper power-law index of $k$ would result in a lower estimation of virial mass. The virial parameter, defined as the ratio of virial mass to LTE mass: $\alpha_{\rm Vir}=M_{\rm Vir}/M_{\rm LTE}$, describes the competition of internal supporting energy against the gravitational energy. We find a typical virial parameter of 2.5 in our clump sample. The virial masses are comparable to the LTE masses. There are 588 (96\%) clumps with virial mass larger than LTE mass, and 221 (36\%) clumps with virial parameter larger than 3. The clumps with high virial parameter ($\alpha_{\rm Vir}>10$) are all faint ones with emission peak lower than 3.3~K. The clumps with $\alpha_{\rm Vir}<1$ are virialized and could be collapsing, while the clumps with higher $\alpha_{\rm Vir}$ could be in a stable or expanding state unless they are external pressure confined. Alternatively, it is possible that the faint clumps may be transient entities \citep{bal06}.

\begin{figure}
\centering
\includegraphics[width=0.4\textwidth]{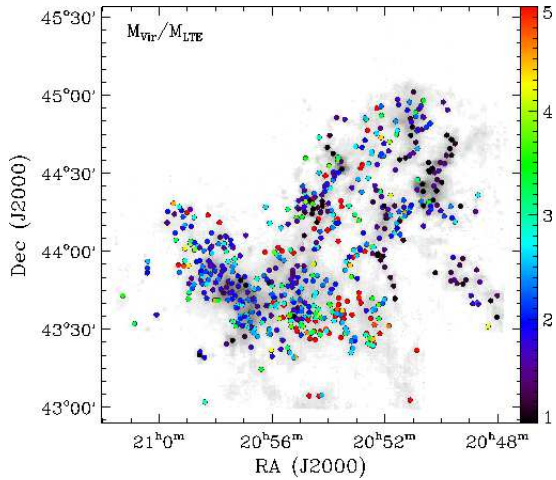}
\caption{Distribution of  clumps with virial parameter coded. The dots represent the clumps overlaid on the integrated intensity map of  $^{13}$CO. The colors of the dots indicate the virial parameter of the clumps.}
\label{fig:vir}
\end{figure}

\begin{figure}
\centering
\includegraphics[width=0.35\textwidth]{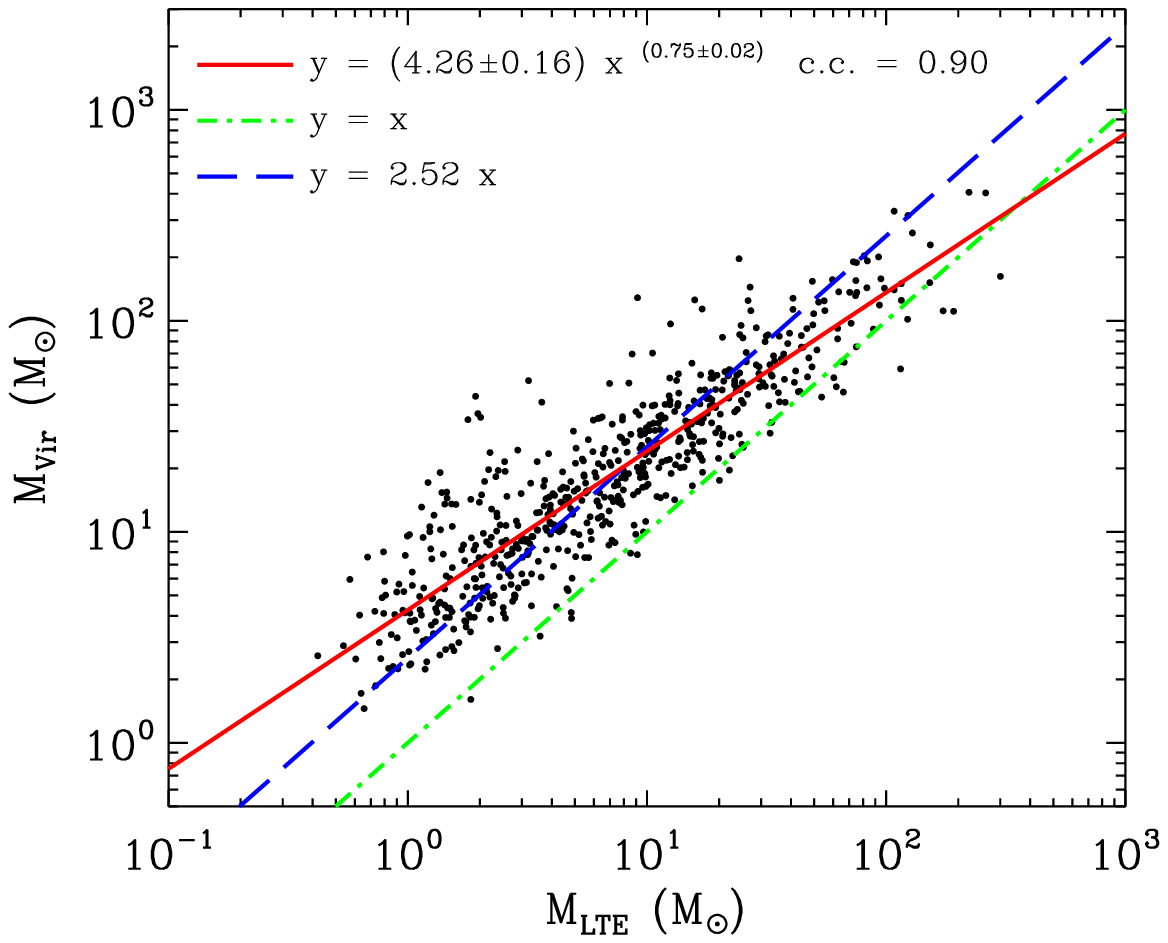}
\includegraphics[width=0.35\textwidth]{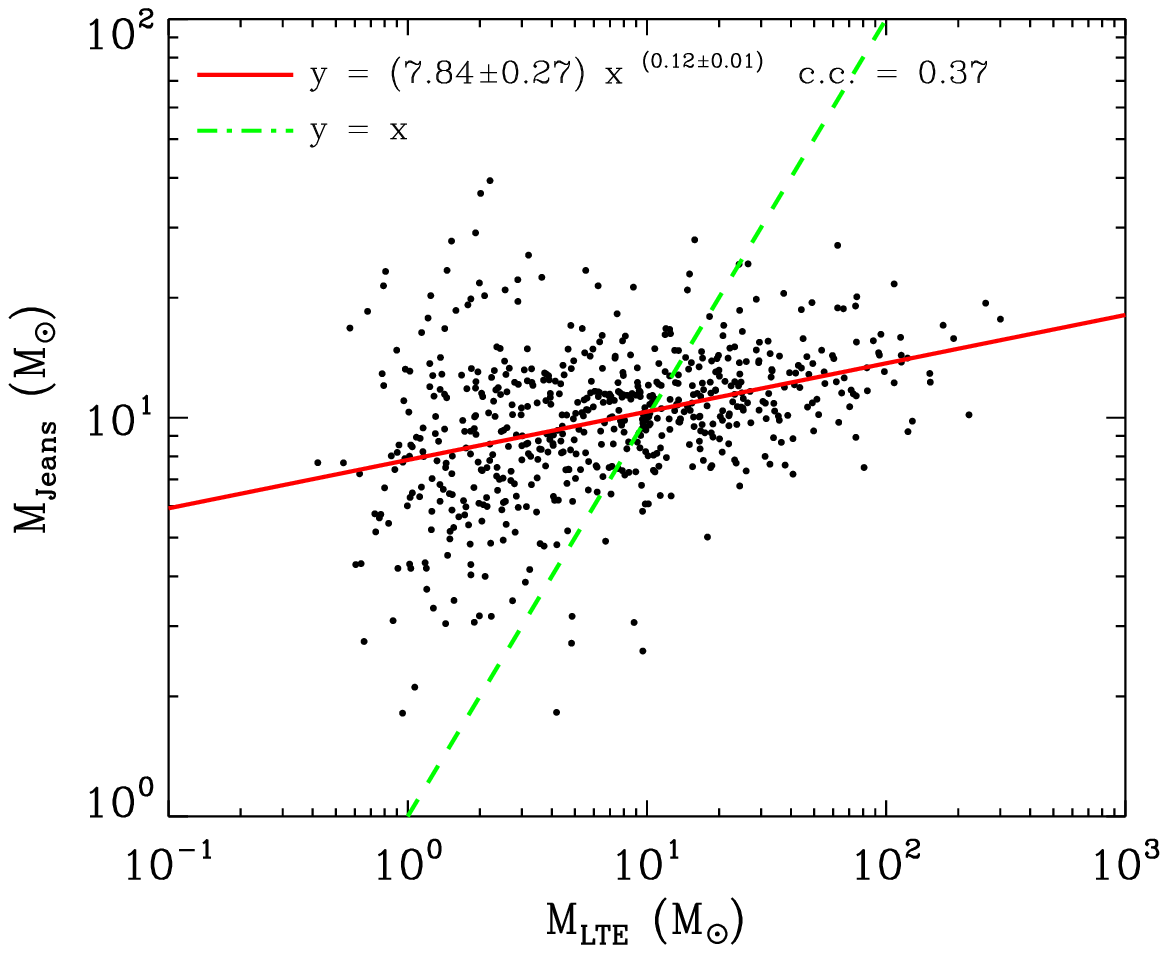}
\caption{Top: Virial mass-LTE mass relation of the clumps. Bottom: Jeans mass-LTE mass relation of the clumps. The dot-dashed green line indicates a mass ratio of 1. The solid red line shows a power-law fit to the relationship. The dashed blue line in the left panel indicates the median mass ratio.}
\label{fig:masscmp}
\end{figure}

Figure~\ref{fig:vir} shows the spacial distribution of clumps with virial parameter coded. It is notable that the clumps close to virial equilibrium associate with dense gas mainly around the Pelican region. The clumps in the Gulf of Mexico region present slightly higher virial parameters than those in the other dense regions, and most of the clumps with weak molecular emissions especially those in the Caribbean Sea region are far from equilibrium state. We compare $M_{\rm Vir}$ and $M_{\rm LTE}$ in the left panel of Figure~\ref{fig:masscmp}. The massive clumps tend to have a lower virial parameter. The mass relationship can be fitted with a power-law of $M_{\rm Vir}/(M_\odot)=(4.26\pm0.16) [M_{\rm LTE}/(M_\odot)]^{(0.75\pm0.02)}$. The power index we obtained is slightly higher than the value in Orion B (0.67) reported by \citet{ike09} and Planck cold clumps (0.61) reported by \citet{liu12}, moreover, significantly higher than the index of pressure-confined clumps ($\alpha_{\rm Vir}\propto M_{\rm LTE}^{-2/3}$) as given by \citet{ber92}. Although our molecular observations reveal several clumps in the NAN complex could be exposed to strong external pressure from ionising radiation and winds of massive stars, the comparability and the high power index of virial and LTE mass suggest that most molecular clumps in the NAN complex are gravitationally bound rather than pressure confined.

We could also derived the Jeans mass with $M_{\rm Jeans}=17.3 {T_{\rm kin}}^{1.5} n^{-0.5} M_\odot$ \citep{gib09} and plot their relationship with LTE mass in the right panel of Figure~\ref{fig:masscmp}. Such relationship could be described with a power-law of $M_{\rm Jeans}/(M_\odot)=(7.82\pm0.27) [M_{\rm LTE}/(M_\odot)]^{(0.12\pm0.01)}$. The flat power index indicates that the LTE masses of most massive clumps are substantially larger than their Jeans mass, suggesting that these clumps will further fragment and may not form individual proto-stars but proto-clusters.

\subsection{Larson Relationship and Mass Function of Clumps}

\begin{figure}
\centering
\includegraphics[width=0.35\textwidth]{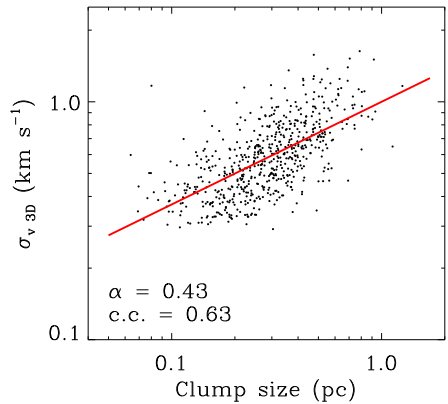}
\caption{Larson relationship for the clumps. The red line indicates a linear fitting to the clump size and velocity dispersion relation.}
\label{fig:larson}
\end{figure}

\citet{lar81} presents a correlation between the velocity dispersion and the region size (range from 0.1 to 100~pc), known as the Larson relationship. The Larson relationship was suggested to exist by several work \citep{leu82,mye83}, but some recent molecular surveys suggest weak or no correlation between line width and size of molecular clouds \citep{oni02,liu12}. Figure~\ref{fig:larson} shows the relationship between size and three dimensional velocity dispersion for our clumps. A fitting to the data gives a correlation of $\sigma_{v \rm 3D}/({\rm km~s^{-1}}) = (1.00\pm0.03) \times [{\rm Size}/({\rm pc})]^{(0.43\pm0.02)}$, with a correlation coefficient of 0.63. The power index is slightly larger than 0.39 given by \citet{lar81}. The correlation is not strong, which might be the result of small dynamic range, and of the scattering of velocity dispersion and clump size (0.06-1.26~pc) we found. The dynamic range is limited by the sensitivity of observations. A uniform survey with sufficient high sensitivity will improve the completeness of less intense clumps with low column density and small size. On the other hand, \citet{liu12} pointed out that turbulence plays a dominant role in shaping the clump structures and density distribution at a large scale, while the small-scale clumps are easily affected by the fluctuations of density and temperature. This will cause a large scattering of line width broadening induced by other factor other than turbulence at small scales. Such scattering of the velocity dispersion may result in a weak or even absent relationship.

\begin{figure}
\centering
\includegraphics[width=0.35\textwidth]{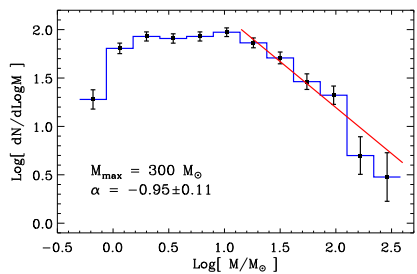}
\caption{Clump mass function (CMF) for the clumps. Red line fit the power-law distribution from 15 to 300~M$_{\odot}$.}
\label{fig:cmf}
\end{figure}

We then study the clump mass function (CMF) in Figure~\ref{fig:cmf} based on the clump mass sample we derived. A power-law distribution of d$N$/d~log$M \propto M^{-\gamma}$ is fitted with our data. Our power index (0.95) is lower than the stellar initial mass function (IMF) of 1.35 give by \citet{sal55}. Several (sub)millimeter continuum studies \citep{tes98,joh06,rei06} and molecular observations \citep{ike09} obtained CMFs which are consistent with the Salpter IMF, while \citet{kra98} reported a flatter power index of 0.6-0.8 in their CO isotopes study of seven molecular clouds. The similarity between CMF and IMF power indices could simply be explained by a constant star formation efficiency unrelated to the mass and self-similar cloud structure, based on a scenario of one-to-one transformation from cores to stars \citep{lad08}. However, such scenario is oversimplified, and ignores the fragmentation in cores whose masses exceed the Jeans mass. Fragmentation in prestellar cores has been observed and discussed by several work \citep{goo07,che10,mau10}. In addition, a simulation by \citet{swi08b} suggested that the obtained IMF is similar to the input CMF even when different fragmentation modes are considered.

\subsection{YSOs in Molecular Cloud}

\begin{figure}
\centering
\includegraphics[width=0.35\textwidth]{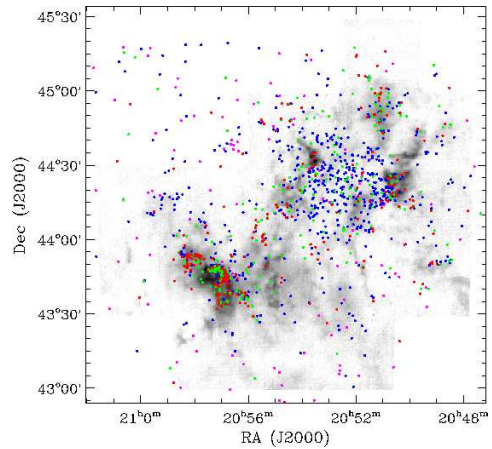}
\caption{YSOs in the NAN complex identified by \citet{reb11}. The gray-scale image is the integrated intensity map of $^{13}$CO. Red dots are Class~I, green are flat, blue are Class~II, and purple are Class~III.}
\label{fig:yso}
\end{figure}

\begin{figure}
\centering
\includegraphics[width=0.22\textwidth]{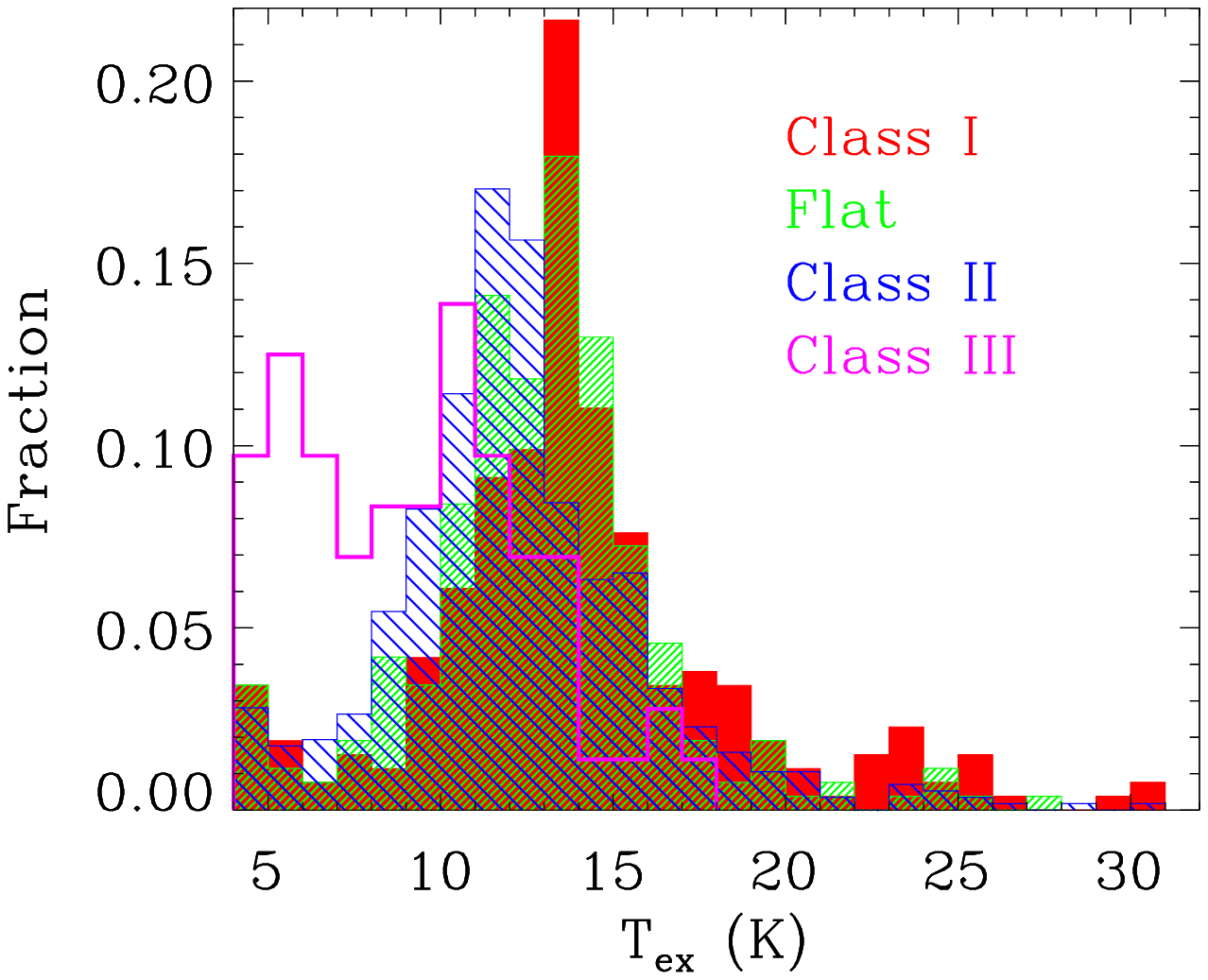}
\includegraphics[width=0.22\textwidth]{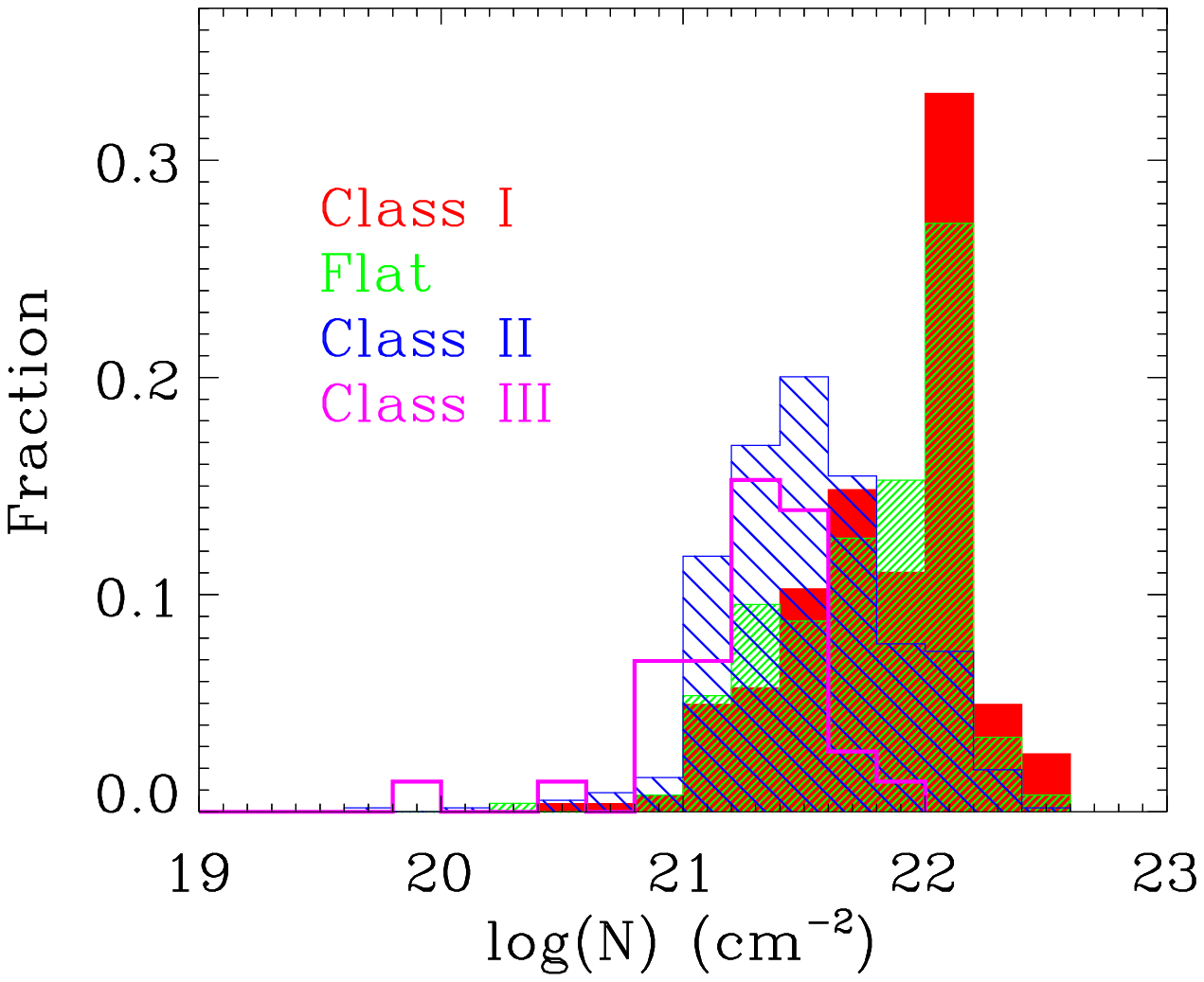}
\caption{The molecular properties (left: excitation temperature; right: column density of H$_2$) associated with YSOs, with the same color code as Figure~\ref{fig:yso}.}
\label{fig:ysostat}
\end{figure}

\citet{cam02} identified nine young stellar clusters in the NAN complex, and \citet{gui09} provided a list of more than 1600 YSOs in their four Infrared Array Camera (IRAC) bands study with the Spitzer Space Telescope. Lately, \citet{reb11} incorporated their Multiband Imaging Photometer for Spitzer (MIPS) observations with earlier archival data, and identified a list of 1286 YSOs in the NAN complex. We compare the distribution of the YSOs from \citet{reb11} with our molecular observations in Figure~\ref{fig:yso}. The Class~I and flat sources are concentrated in cold and dense molecular clouds, especially in the Gulf of Mexico and the Pelican's Hat region, while the Class~II sources are spread across the cloud with low molecular opacity, and only a few YSOs are associated with the diffuse Caribbean Sea region. The molecular properties associated with different classes of YSOs are extracted and studied in Figure~\ref{fig:ysostat}. The histograms indicate that the Class~I and flat sources match the distribution of molecular clouds and prefer a cold dense environment with excitation temperature of $\sim$14~K and column density of $\sim$10$^{22}$~cm$^{-2}$.

Three main YSO clusters are identified from the sample of \citet{reb11}. Two of these with a great fraction of Class~I and flat objects are associated with the molecular cloud of the Gulf of Mexico and the Pelican's Hat region which shows low temperature and high C$^{18}$O abundance. The third cluster, the Pelican Cluster, is surrounded by the Pelican's Neck, the Pelican's Beak, and the Caribbean Islands. Although the Class~II sources constitute a higher fraction in the Pelican Cluster, most of the Class~I and flat objects appear on the east and west edges of the cluster. This distribution is consistent with the molecular distribution in which the molecular gas in the central area is dispersed and surrounded by clouds with higher molecular temperature and low C$^{18}$O abundance. The YSO proportion in the clusters suggests a younger stage of evolution in the most south-eastern and north-western parts of the NAN complex, and an older stage in the center of the Pelican Cluster. If the complex velocity structures in surrounding regions of the Pelican Cluster are indeed the results of feedbacks from the massive cluster members, the cluster may be triggering the star formation in the molecular cloud across a span of over 5~pc and $\sim$10~km~s$^{-1}$.

\begin{figure}
\centering
\includegraphics[width=0.4\textwidth]{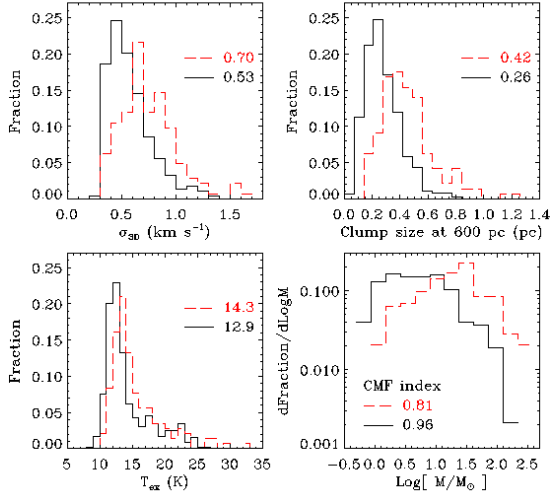}
\caption{The histogram of three-dimensional velocity dispersion (upper left), clump size (upper right), excitation temperature (lower left) and CMF (lower right). Red lines represent the clumps associated with YSOs, while black lines represent the clumps without YSOs. The median values and CMF indices are marked on the plot.}
\label{fig:ysoclump}
\end{figure}

We then compare our clump results with the distribution of YSOs, by separating the clumps spatially associated with YSOs from those containing no YSO. The Class~III YSOs are not considered, as the Class~III catalogue is not complete and their distribution is not associate with molecular cloud. A total of 143 clumps are found to be associated with YSOs. The discrepancies in their physical properties are shown in Figure~\ref{fig:ysoclump}. The clumps associated with YSOs present a higher velocity dispersion, clump size, and excitation temperature, while the discrepancy of the CMF indices is not significant. Further observations with higher signal-to-noise ratio and resolution are needed to extend the limit of mass completeness in CMF comparison.

\section{Summary}\label{sec:sum}

We have presented the PMODLH mapping observations for an area of 4.25 deg$^2$ toward the North American and Pelican Nebulae molecular cloud complex in $^{12}$CO, $^{13}$CO, and C$^{18}$O lines. The main results are listed below:

The molecules distribution is along the dark lane in the southeast-northwest direction. $^{12}$CO emission is bright, extended, while $^{13}$CO and C$^{18}$O emissions are compact. The channel map shows intricate structures within the complex, and filamentary structures are revealed. Position-velocity slice along the full length of the cloud reveals a molecular shell surrounding the W80 H~II region. Gases of two different temperatures are seen in the distribution of excitation temperature.

The surface density map shows several dense clouds with surface density over 500~$M_\sun$~pc$^{-2}$ in the complex. We have derived a total mass of $2.0\times10^4~M_\odot$ ($^{13}$CO) and $6.1\times10^3~M_\odot$ (C$^{18}$O) under the LTE assumption with uniform molecular abundance, and $5.4\times10^4~M_\odot$ with the constant CO-to-H$_2$ factor in the NAN complex. Such a discrepancy in mass may be due to the different extent which the molecules are tracing.

Six regions are discerned in the molecular maps, each with different emission characteristics. Their sizes, column densities, and masses vary with different density tracers. The properties of low temperature, high column density, and high C$^{18}$O abundance found in the Gulf of Mexico, and Pelican's Hat regions indicate a young stage of massive star formation, while the properties of the Pelican's Neck, Pelican's Beak, and Caribbean Islands regions represent a hot, dense, and more evolved environment probably affected by the Pelican Cluster. Only the Caribbean Sea region shows little sign of star formation.

Four filamentary structures are found in the NAN complex. They show complex structures such as a twisted spatial distribution or opposite velocity gradient directions, but these filaments all seem in a gravitationally stable state.

We have identified 611 clumps using the ClumpFind algorithm in the NAN complex, and yield a typical size, excitation temperature, and density of $\sim$0.3~pc, 13~K, and 8$\times10^3$~cm$^{-3}$, respectively. Most of the clumps are non-thermal dominated and in an early evolutionary stage of star formation. The comparison of virial and LTE mass of the clumps indicates that most clumps are gravitationally bound. We obtain a clump mass function index $\gamma=0.95$. The clumps associate with YSOs present more evolved features comparing with those having no association.

\acknowledgements{This work is based on observations made with the Delingha 13.7-m telescope of the Purple Mountain Observatory. The authors appreciate all the staff members of the observatory for their help during the observations. We acknowledge John Bally, the referee of this paper for his valuable comments which helped to considerably improve the quality of the manuscript. This work is supported by the Chinese NSF through grants NSF 11133008, NSF 11073054, NSF 11233007, and the Key Laboratory for Radio Astronomy, CAS.}

\bibliographystyle{aj}
\bibliography{reference}{}

\end{document}